\documentclass[12pt]{article}
\usepackage[top=30truemm,bottom=30truemm,left=25truemm,right=25truemm]{geometry}
\usepackage{amsfonts,amsmath,amssymb,epsf}
\usepackage{amsmath,braket}
\usepackage{physics}
\usepackage{graphicx,hyperref,color}
\usepackage[dvipsnames]{xcolor}
\usepackage{cite}
\usepackage{float}
\hypersetup{
    unicode=false,          
    pdftoolbar=true,        
    pdfmenubar=true,        
    linktocpage=true,       
    pdffitwindow=false,     
    pdfstartview={FitH},    
    pdfnewwindow=true,      
    colorlinks=true,       
    linkcolor=blue,          
    citecolor=blue,        
    filecolor=blue,      
    urlcolor=blue           
}

\numberwithin{equation}{section}									

\newcommand{\de}{\partial}
\newcommand{\be}{\begin{equation}}
\newcommand{\ba}{\begin{eqnarray}}
\newcommand{\ea}{\end{eqnarray}}
\newcommand{\ee}{\end{equation}}

\newcommand{\f}{\frac}
\newcommand{\s}{\sqrt}

\newcommand{\ti}{\tilde}
\newcommand{\ap}{\alpha}

\newcommand{\no}{\nonumber \\}

\newcommand{\bea}{\begin{eqnarray}}
\newcommand{\eea}{\end{eqnarray}}
\newcommand{\bes}{\begin{equation*}}
\newcommand{\beas}{\begin{eqnarray*}}
\newcommand{\eeas}{\end{eqnarray*}}
\newcommand{\bas}{\begin{array*}}
\newcommand{\eas}{\end{array*}}
\newcommand{\ees}{\end{equation*}}
\newcommand{\nn}{\nonumber}
\newcommand{\p}{\partial}
\newcommand{\ep}{\epsilon}

\let\a=\alpha    \let\e=\epsilon   \let\h=\eta \let\k=\kappa \let\l=\lambda  
 \let\p=\phi \let\r=\rho 
 \let\th=\theta     
 \let\D=\Delta \let\G=\Gamma  \let\O=\Omega      
\def\nn{\nonumber}
\def\inf{\infty}

\DeclareMathOperator{\arctanh}{arctanh}

\usepackage{tikz}
\usetikzlibrary{arrows,arrows.meta,intersections, calc,positioning,decorations.pathreplacing,decorations.pathmorphing,shapes}
\usetikzlibrary{patterns}
\usetikzlibrary{decorations.markings}
\usetikzlibrary{knots}

\begin{document}

\begin{titlepage}
\thispagestyle{empty}

\vspace*{-2cm}
\begin{flushright}
YITP-23-07
\\
\end{flushright}

\bigskip

\begin{center}
\noindent{\bf \Large {AdS/BCFT with Brane-Localized Scalar Field}}\\
\vspace{2cm}

Hiroki Kanda$^a$, Masahide Sato$^a$, Yu-ki Suzuki$^a$, 
Tadashi Takayanagi$^{a,b,c}$ and Zixia Wei$^a$
\vspace{1cm}\\

{\it $^a$Center for Gravitational Physics and Quantum Information,\\
Yukawa Institute for Theoretical Physics, Kyoto University, \\
Kitashirakawa Oiwakecho, Sakyo-ku, Kyoto 606-8502, Japan}\\
\vspace{1.5mm}
{\it $^b$Inamori Research Institute for Science,\\
620 Suiginya-cho, Shimogyo-ku,
Kyoto 600-8411 Japan}\\
\vspace{1.5mm}
{\it $^{c}$Kavli Institute for the Physics and Mathematics
 of the Universe (WPI),\\
University of Tokyo, Kashiwa, Chiba 277-8582, Japan}\\

\bigskip \bigskip
\vskip 3em
\end{center}

\begin{abstract}
In this paper, we study the dynamics of end-of-the-world (EOW) branes in AdS with scalar fields localized on the branes as a new class of gravity duals of CFTs on manifolds with boundaries. This allows us to construct explicit solutions dual to boundary RG flows. We also obtain a variety of annulus-like or cone-like shaped EOW branes, which are not possible without the scalar field. We also present a gravity dual of a CFT on a strip with two different boundary conditions due to the scalar potential, where we find the confinement/deconfinement-like transition as a function of temperature and the scalar potential. Finally, we point out that this phase transition is closely related to the measurement-induced phase transition, via a Wick rotation. 
\end{abstract}

\end{titlepage}

\newpage

\tableofcontents

\newpage

\section{Introduction}

A boundary conformal field theory (BCFT) is a conformal field theory (CFT) defined on a manifold with boundaries where half of the conformal symmetries are preserved by the boundaries \cite{Cardy:1984bb,Cardy:2004hm,McAvity:1993ue,McAvity:1995zd}. Gravity duals of boundary conformal field theories (BCFTs) have been actively studied recently as a major generalization of AdS/CFT correspondence \cite{Maldacena:1997re,Gubser:1998bc,Witten:1998qj}, so-called the AdS/BCFT \cite{Karch:2000gx,Takayanagi:2011zk,Fujita:2011fp,Nozaki:2012qd}. Its bottom-up construction is given by cutting out a part of the bulk AdS surrounded by the end-of-the-world brane (EOW brane).  One of the major advantages of AdS/BCFT is that the computation of entanglement entropy in the AdS/BCFT is straightforward, such that the minimal surface in the holographic entanglement entropy \cite{Ryu:2006bv,Ryu:2006ef,Hubeny:2007xt} can end on the end-of-the-world brane as first noted in \cite{Takayanagi:2011zk,Fujita:2011fp} (see also \cite{Miao:2017gyt}).
This was recently employed to analyze black hole evaporation processes to derive the Page curve \cite{Almheiri:2019hni}, where the Island prescription \cite{Penington:2019npb,Almheiri:2019psf} is expected from the brane-world holography \cite{Randall:1999ee,Randall:1999vf,Gubser:1999vj,Karch:2000ct,Giddings:2000mu,Shiromizu:2001jm,Shiromizu:2001ve,Nojiri:2000eb,Nojiri:2000gb,Hawking:2000kj,Koyama:2001rf,Kanno:2002iaa,Emparan:2023dxm}. For further applications of AdS/BCFT to black hole information problem refer to e.g.\cite{Almheiri:2019psy,Rozali:2019day,Chen:2019uhq,Balasubramanian:2020hfs,Geng:2020qvw,Chen:2020uac,Chen:2020hmv,Bousso:2020kmy,Chen:2020jvn,Chen:2020tes,Akal:2020twv,Miyaji:2021lcq,Akal:2021foz,Geng:2021mic,Bhattacharya:2021nqj,Hu:2022ymx,Anous:2022wqh,Kawamoto:2022etl,Bianchi:2022ulu,Akal:2021dqt,Akal:2022qei,Geng:2022dua,Bhattacharjee:2022pcb}. There are many other applications of the idea of AdS/BCFT, ranging from  quantum many-body aspects of chaotic BCFTs \cite{Fujita:2012fp,Ugajin:2013xxa,Erdmenger:2014xya,Erdmenger:2015xpq,Numasawa:2016emc,Seminara:2017hhh,Seminara:2018pmr,Hikida:2018khg,Shimaji:2018czt,Caputa:2019avh,Mezei:2019zyt,Reeves:2021sab,Kusuki:2021gpt, Miyaji:2021ktr, Numasawa:2022cni,Chandra:2022fwi,Kusuki:2022wns,Antonini:2022sfm,Kusuki:2022ozk} and boundary renormalization group flows \cite{Gutperle:2012hy,Estes:2014hka,Kobayashi:2018lil,Sato:2020upl}, to calculating quantum information measures \cite{Chapman:2018bqj,Sato:2019kik,Braccia:2019xxi,Sato:2021ftf,Hernandez:2020nem,Collier:2021ngi,Chalabi:2021jud,Belin:2021nck,Suzuki:2022tan} and building cosmological models \cite{Cooper:2018cmb,Antonini:2019qkt,VanRaamsdonk:2020tlr,VanRaamsdonk:2021qgv,Waddell:2022fbn}.
The basic duality relations and their consistency of AdS/BCFT have also been analyzed in \cite{Suzuki:2021pyw,Omiya:2021olc,Suzuki:2022xwv,Suzuki:2022yru,Izumi:2022opi} and 
string theory realizations have also been studied in \cite{Chiodaroli:2011nr,Chiodaroli:2012vc,Karch:2020iit,Bachas:2020yxv,Simidzija:2020ukv,Ooguri:2020sua,Raamsdonk:2020tin,Uhlemann:2021nhu,Coccia:2021lpp,Karch:2022rvr}. The AdS/BCFT correspondence was generalized in  \cite{Akal:2020wfl,Miao:2020oey,Geng:2022slq,Geng:2022tfc,Ogawa:2022fhy,Lee:2022efh} to construct gravity duals with higher codimensions, named wedge holography.

So far, most of the works on the AdS/BCFT have focused on the case where a single conformally invariant boundary is present in a BCFT. The case where a non-conformal boundary is present or the case where different types of boundaries coexist has not been well studied. The main purpose of this paper is to investigate these problems by analyzing a simple variant of the AdS/BCFT where there is a scalar field $\phi$ localized on the end-of-the-world brane. For example, if we consider a string theory embedding of AdS/BCFT with an orientifold as in \cite{Fujita:2011fp}, we can add D-branes on top of the orientifold, which lead to fields localized on the end-of-the-world brane.  This single scalar field on the brane is expected to be dual to a boundary primary operator in the BCFT. We also expect that some fields in the bulk AdS can be dynamically localized on the brane and these fields can also be treated as brane-localized fields. Also, a brane-localized scalar field plays an important role in the holographic Kondo effect \cite{Erdmenger:2013dpa,Erdmenger:2014xya,Erdmenger:2015xpq,Erdmenger:2015spo,Erdmenger:2020hug}.

Non-trivial profiles of brane-localized scalar fields allow us to construct explicit solutions which describe the boundary RG flows and also those which correspond to the presence of two boundaries with different boundary conditions. This problem was treated in a probe limit in \cite{Suzuki:2022tan} recently. There is another interesting approach to  AdS/BCFT with different boundary conditions by introducing a defect on the EOW brane as discussed in \cite{Miyaji:2022dna,Biswas:2022xfw}.

Analytical results are often available in simple constructions of AdS$_3$/BCFT$_2$ with a brane-localized scalar in three dimensions, because the bulk gravity dynamics is still given by the vacuum Einstein equation in three dimensions, whose local solutions are all equivalent to the pure AdS$_3$. We can even design a large number of solutions by tuning the profile of the potential of the scalar field. As we will see in this paper, this allows us to construct explicit boundary RG solutions when the CFT is defined on an upper half plane, a strip, a round disk, and an annulus as we will explicitly see. Moreover, we will be able to find a solution with the end-of-the-world brane which connects two boundaries with different values of the scalar field, which is dual to a BCFT on a strip with two different boundary conditions. In this strip example, we can construct a connected EOW brane solution using the Poincar\'{e} AdS$_3$ metric which is not possible without the brane-localized scalar field. We will observe that this solution experiences a phase transition which is analogous to the confinement/deconfinement transition in the presence of chemical potential and temperature. Moreover, we will present a connected EOW brane solution for a BCFT on an annulus for the first time, which is again impossible without the scalar field. In this way, our simple models describe a new broader class of boundary conditions in CFTs, which possess rich structures. 

The Wick rotation of the strip example with the scalar field leads to a new time-dependent solution. This can be regarded as a deformation of the solution in \cite{Hartman:2013qma}, which describes a thermalization of a pure state described by a boundary state. In our case with the scalar field, as we will see in this paper, the solution can be more appropriately regarded as the transition matrix instead of a regular quantum state. Accordingly, the holographic entanglement entropy in this class of backgrounds is correctly interpreted as the holographic pseudo entropy. The pseudo entropy is a generalization of entanglement entropy such that it depends on two different quantum states \cite{Nakata:2020luh}.
We will observe that the confinement/deconfinement-like phase transition of the AdS/BCFT with a brane-localized scalar leads to an interesting transition of the time evolution of the entropy from $O(t)$ to $O(\log t)$ and then $O(t^0)$, which looks analogous to the measurement-induced phase transition \cite{Skinner:2018tjl,Li:2018mcv,Li:2019zju}. Moreover, we will suggest a possibility that our time-dependent gravity dual solution with the EOW brane may be interpreted as a non-unitary evolution such as projection measurements of infrared degrees of freedom with the entanglement entropy showing a measurement-induced phase transition. 

This paper is organized as follows. In section two, we present our basic setup of AdS/BCFT with a brane-localized scalar and derive the boundary condition on the end-of-the-world brane. In section three we investigate planar branes in AdS$_3/$BCFT$_2$ with a scalar field, each of which is dual to either the BCFT with a relevant perturbation on an upper half plane or the BCFT on a strip with a marginal perturbation. In section four, we examine round-shaped branes in AdS$_3/$BCFT$_2$, dual to BCFTs on annuli or disks with perturbations. In section five, we study higher dimensional examples. In section six, we consider a finite temperature setup in three dimension and analyze the phase transition between the connected and disconnected solution in the presence of the brane-localized scalar field. In section seven we study transition matrices by choosing the time-dependent profile of the brane-localized scalar field which computes the pseudo entropy. We will show that it exhibits an interesting phase transition. In section eight, we draw conclusions and discuss future problems. In appendix A, a detailed analysis of round-shaped EOW brane solutions is presented.

\section{AdS/BCFT with brane-localized scalar field}

\begin{figure}[t]
    \centering
    \includegraphics[width=.75\textwidth,page=1]{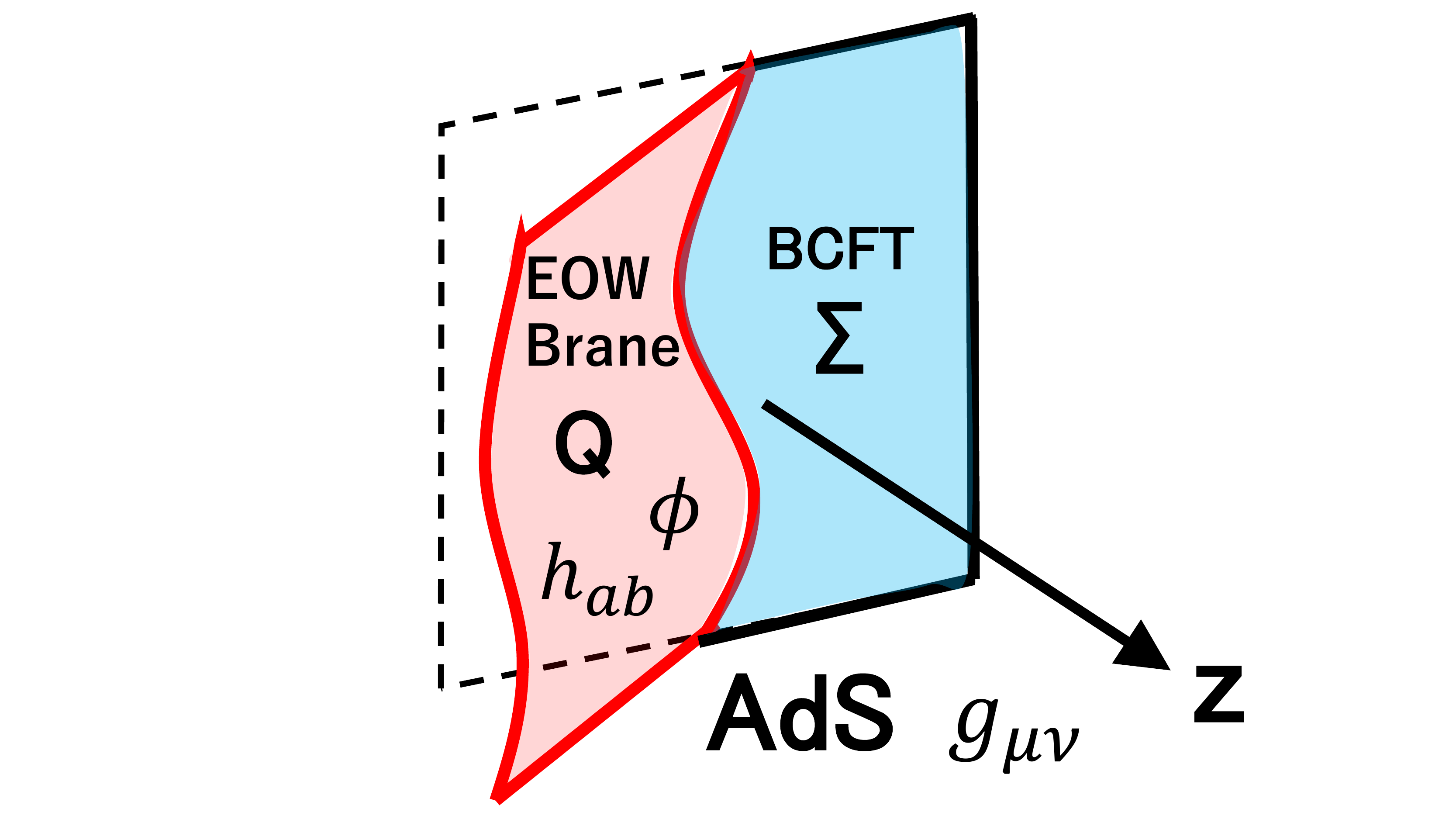}
    \caption{A sketch of our setup of AdS/BCFT.}
    \label{fig:GAdSBCFT}
\end{figure}

In this section, we present a general formulation of Einstein gravity with a scalar field localized on the end-of-the-world brane (EOW brane) $Q$ in AdS$_{d+1}$. This is dual to the $d$-dimensional BCFT  on $\Sigma$ via the
AdS$_{d+1}/$BCFT$_d$. The gravity dual is given by the $d+1$ dimensional region $M$ whose boundary is given by $Q\cup \Sigma$, as depicted in Fig.\ref{fig:GAdSBCFT}.

This model is described by the total action:
\begin{align}
    I&=S_{\text{EH}}+S_{\text{GHY}}+S_{\text{brane}},\nn\\
    I_{\text{EH}}&=-\frac{1}{16\pi G_N}\int_{M} d^{d+1}x\sqrt{g}(R-2\Lambda),\nn\\
    I_{\text{GHY}}&=-\frac{1}{8\pi G_N}\int_\Sigma d^{d}x\sqrt{h} K,\nn\\
    I_{\text{brane}}&=-\frac{1}{8\pi G_N}\int_Q d^{d}x\sqrt{h} (K-h^{ab}\de_a\phi\de_b\phi-V(\phi)), \label{totac}
\end{align}
where $g_{\mu\nu}$ is the bulk metric and 
$h_{ab}$ is the induced metric on the EOW brane $Q$, whose extrinsic curvature is denoted by $K_{ab}$. On the EOW brane we consider a scalar field $\phi$ as a matter field. The first term is the Einstein-Hilbert action with the cosmological constant $\Lambda$ and the second one is the Gibbons-Hawking-York term on the asymptotic AdS boundary $\Sigma$. By setting the AdS radius to be one, we have $\Lambda=-\frac{d(d-1)}{2}$.
The final term consists of  the kinetic term of the scalar field and the potential term in addition to the Gibbons-Hawking term on $Q$. If we set $V=T$ and $\phi=0$, where $T$ is some constant (i.e. the tension of the EOW brane), then this model is reduced to the standard pure gravity model of AdS/BCFT, introduced in \cite{Takayanagi:2011zk,Fujita:2011fp}.
A similar setup with a brane-localized scalar has been analyzed in the context of holographic model Kondo effect \cite{Erdmenger:2013dpa,Erdmenger:2014xya,Erdmenger:2015xpq,Erdmenger:2015spo,Erdmenger:2020hug}.

Now, we derive the boundary condition on the EOW brane $Q$. This is obtained by varying the action with the bulk equation of motion i.e. the Einstein equation $R_{\mu\nu}+dg_{\mu\nu}=0$ imposed:
\begin{align}
    \delta I_{\text{EH}}&=-\frac{1}{16\pi G}\int_{Q\cup\Sigma} d^{d}x\left(\sqrt{h}(K_{ab}-K h_{ab})\delta h^{ab}-2
    \delta(\sqrt{h}K)\right),\nn\\
    \delta I_{\text{GHY}}&=-\frac{1}{8\pi G}\int_{Q\cup\Sigma} d^{d}x\delta(\sqrt{h} K),\nn\\
    \delta I_{\text{brane}}&=\frac{1}{8\pi G}\int_{Q} d^{d}x\sqrt{h} \left(-\frac{1}{2}(h^{cd}\de_c\phi\de_d\phi+V)h_{ab}+\de_a\phi\de_b\phi\right)\delta h^{ab}.
\end{align}
We obtain the boundary condition by setting the total variation to be vanishing:
\be
\delta I=-\frac{1}{16\pi G}\int_Q d^d x\sqrt{h}\left(K_{ab}-Kh_{ab}+(h^{cd}\de_c\phi\de_d\phi+V)h_{ab}-2\de_a\phi\de_b\phi\right)\delta h^{ab}=0.
\ee
Particularly, we consider the dynamical gravity on the brane and we choose the following Neumann-type boundary condition:
\be
K_{ab}-Kh_{ab}=-\left(h^{cd}\de_c\phi\de_d\phi+V(\phi)\right)h_{ab}+2\de_a\phi\de_b\phi.\label{Nbc}
\ee
The equation of motion for the scalar field reads
\be
-2\de_a(\s{h}h^{ab}\de_b\phi)+\s{h}V'(\phi)=0. \label{sceom}
\ee
It is possible to show that this scalar field equation follows from (\ref{Nbc}) as the right hand side is the energy stress tensor of the scalar field, whose conservation law gives (\ref{sceom}).

\section{\texorpdfstring{AdS$_3$/BCFT$_2$}{AdS3/BCFT2} with planar branes}

In this section, we consider a class of AdS$_3/$BCFT$_2$ setup where the EOW brane takes a planar shape.

\begin{figure}[H]
    \centering
    \includegraphics[width=.75\textwidth,page=1]{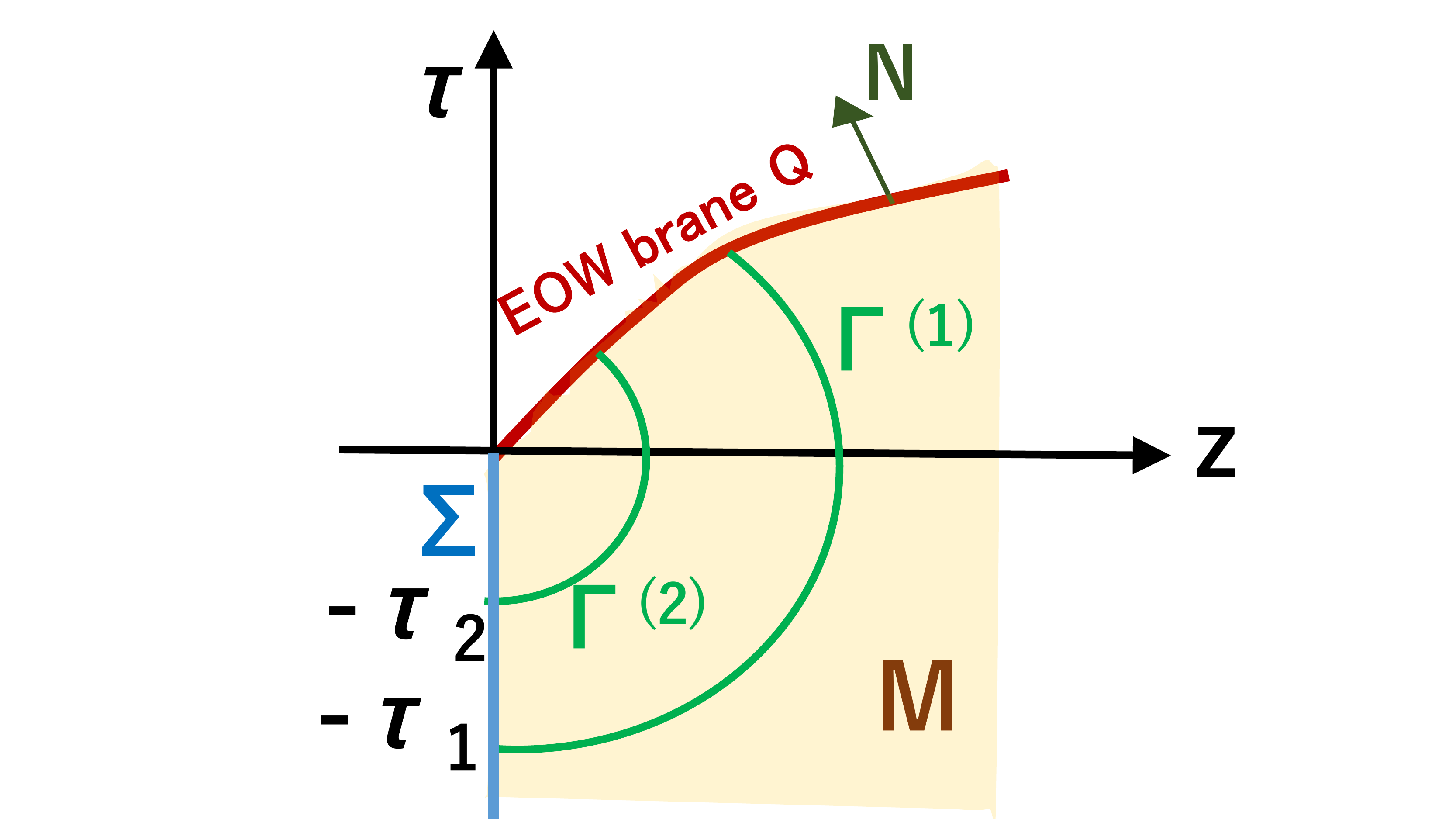}
    \caption{The setup of AdS$_3/$BCFT$_2$ with a planar EOW brane. The red curve shows the end-of-the-world-brane $Q$.  The green curves describe the geodesics $\Gamma^{(1,2)}$ which compute the holographic entanglement entropy.}
    \label{fig:AdSBCFT}
\end{figure}

We assume the bulk metric is the Poincar\'{e} AdS$_3$:
\ba
ds^2=\frac{dz^2+d\tau^2+dx^2}{z^2}.
\ea
We specify the shape of the EOW brane by 
\ba
z=z(\tau),
\ea
assuming the translationally invariance in the $x$ direction. This is depicted in Fig.\ref{fig:AdSBCFT}.
The scalar field also depends on $\tau$ 
\ba
\phi=\phi(\tau).
\ea
The induced metric reads 
\ba
ds^2=\frac{(1+\dot{z}(\tau)^2)d\tau^2+dx^2}{z(\tau)^2},
\ea
where we set $\dot{z}=\de_\tau z$.
The normal vector of the EOW brane reads 
\ba
(N^z,N^\tau,N^x)=\frac{z}{\s{1+\dot{z}^2}}\left(-1,\dot{z},0\right).
\ea
This leads to the extrinsic curvatures 
\ba
&& K_{\tau\tau}-h_{\tau\tau}K=-\frac{\s{1+\dot{z}^2}}{z^2},\no
&& K_{xx}-h_{zz}K=-\frac{1+\dot{z}^2+z\ddot{z}}{z^2(1+\dot{z}^2)^{3/2}}.
\ea
By plugging them into (\ref{Nbc}), the equations are summarized into the following two ones:
\ba
&& \dot{\phi}^2=\frac{\ddot{z}}{2z\s{1+\dot{z}^2}},\label{EOMsca}\\
&& V(\phi)=\frac{z^2\dot{\phi}^2}{1+\dot{z}^2}+\frac{1}{\s{1+\dot{z}^2}}. 
\ea

Note that for any profile of EOW brane $z=z(\tau)$, which satisfies $\ddot{z}\geq 0$, we can find a solution $\phi=\phi(\tau)$. This also fixes the form of potential $V(\phi)$. Thus we can design the potential $V(\phi)$ such that we can obtain a desired solution of the EOW brane.

\subsection{Vacuum solution}
If we set $\phi=0$, then we obtain the standard vacuum solution in AdS/BCFT (see \cite{Takayanagi:2011zk,Fujita:2011fp}):
\ba
z=\lambda\tau,\ \ \ \ \  V=\frac{1}{\s{1+\lambda^2}}(=\mbox{tension of the brane}: T).
\ea
The entanglement entropy is a useful quantity which characterize the degrees of freedom in BCFTs. The entanglement entropy is defined by 
the von Neumann entropy $S_A=-\mbox{Tr}[\rho_A\log\rho_A]$ for the reduced density matrix $\rho_A$, obtained by tracing out the region outside of the subsystem $A$. In AdS/CFT, the entanglement entropy can be computed by the area of minimal area surface $\Gamma_A$ which ends on the boundary of the subsystem $A$ by 
\ba
S_A=\frac{\mbox{Area}(\Gamma_A)}{4G_N}. \label{HEE}
\ea
In AdS/BCFT, there is one more important aspect: the minimal surface $\Gamma_A$ can end on the-end-of-the-world-brane \cite{Takayanagi:2011zk,Fujita:2011fp,Miao:2017gyt}. 

In Poincar\'{e} AdS$_3$, we can easily find that the minimal surface $\Gamma_A$ is given by an arc of a circle as shown in Fig.\ref{fig:AdSBCFT}. If we choose the subsystem $A$ to be the interval 
$-\tau_0\leq \tau\leq 0$, then the holographic entanglement entropy reads (we put the UV cut off $z\geq \ep$)
\ba
S_A&=&\frac{1}{4G_N}\left[\log\frac{2\tau_0}{\ep}+\log\mbox{arcsinh}\left(\frac{1}{\lambda}\right)\right]\no
&=&\frac{c}{6}\log\frac{2\tau_0}{\ep}+S_{\rm bdy},
\ea
where we employed the well-known relation to the central charge $c$ of the dual two dimensional CFT: $c=\frac{3}{2G_N}$ \cite{Brown:1986nw}. By comparing with the CFT result \cite{Calabrese:2004eu}, the boundary entropy (or the logarithm of the $g$-function) is found to be \cite{Takayanagi:2011zk,Fujita:2011fp}
\ba
S_{\rm bdy}=\frac{c}{6}\log\mbox{arcsinh}\left(\frac{1}{\lambda}\right)=\frac{c}{6}\log\s{\frac{1+T}{1-T}}.
\label{bdyent}
\ea
This vanishes at $\lambda=\infty$ or equally at $T=0$ and gets divergent  at $\lambda=0$ or equally at $T=1$.

The boundary entropy has the remarkable property that it is monotonically decreasing under the boundary RG flow \cite{Affleck:1991tk,Friedan:2003yc}, so called $g$-theorem.
We will soon later see explicit holographic examples of the boundary RG flow.

\subsection{Boundary RG flow solutions}

We expect that the boundary RG will be described as follows:
at $\phi=0$ we have UV fixed point with a tachyon mass $V(\phi)\simeq -A\phi^2\ (A>0)$, which will flow into the IR fixed point at $\phi=\phi_*$ by a relevant perturbation such that $V(\phi)\simeq B(\phi-\phi_*)^2\ \  (B>0)$.

At the fixed points, $V$ is equal to the tension $T$ of the EOW brane and is related to the boundary entropy as in (\ref{bdyent}).  We can introduce the boundary entropy (or the logarithm of the $g$-function) in the middle of RG flow by regarding $\dot{z}$ as $\lambda$. This allows us to show the boundary entropy monotonically decreasing as can be derived
by noting that $\ddot{z}$ is non-negative as required by the equation of motion (\ref{EOMsca}).
In more general setups, the holographic $g$-theorem was shown in \cite{Takayanagi:2011zk,Fujita:2011fp} assuming the null energy condition. 

\begin{figure}[H]
    \centering
     \includegraphics[width=.3\textwidth,page=1]{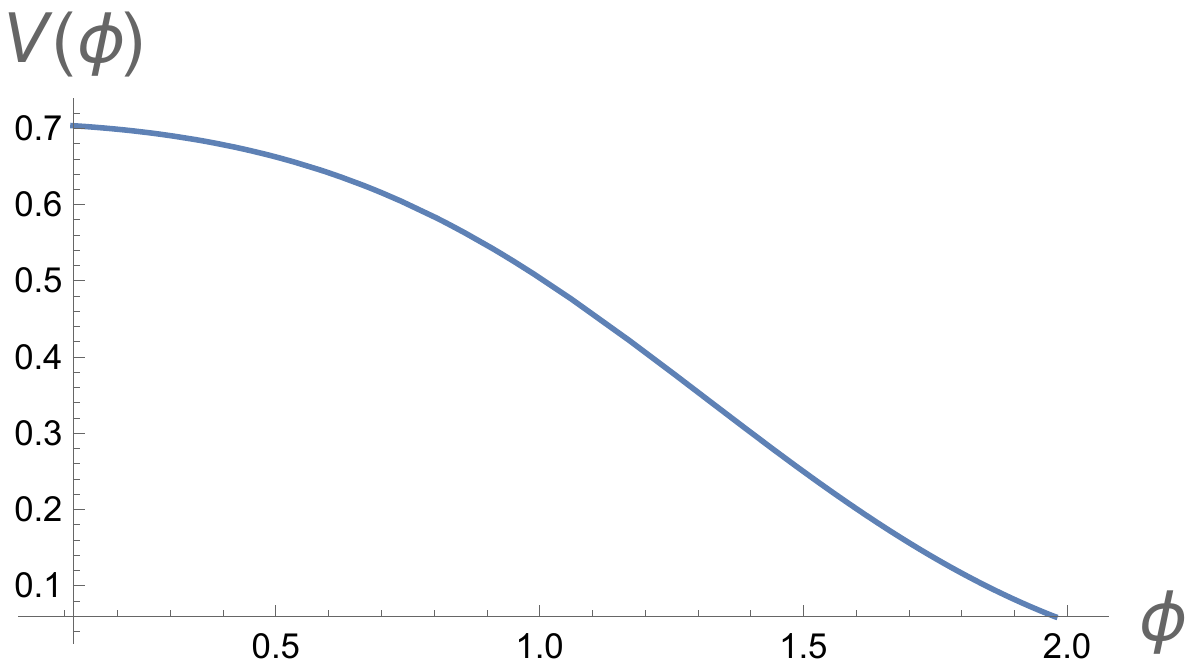}
    \includegraphics[width=.3\textwidth,page=1]{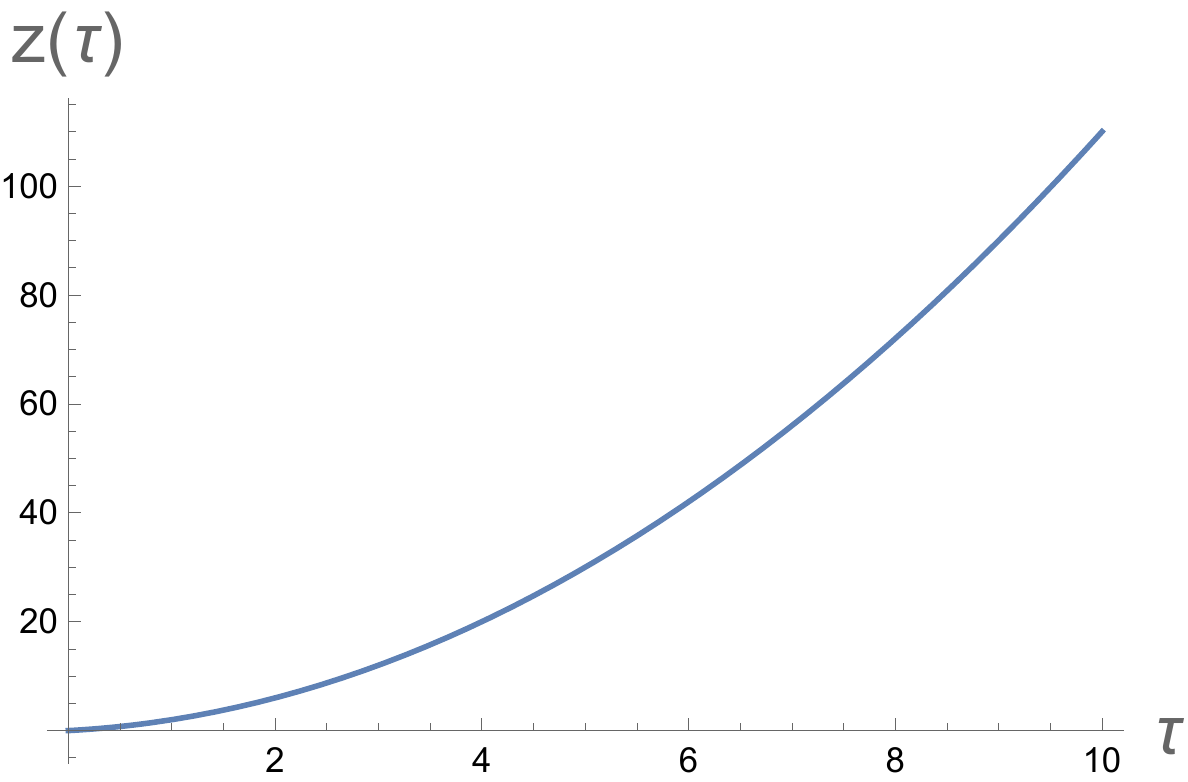}
     \includegraphics[width=.3\textwidth,page=1]{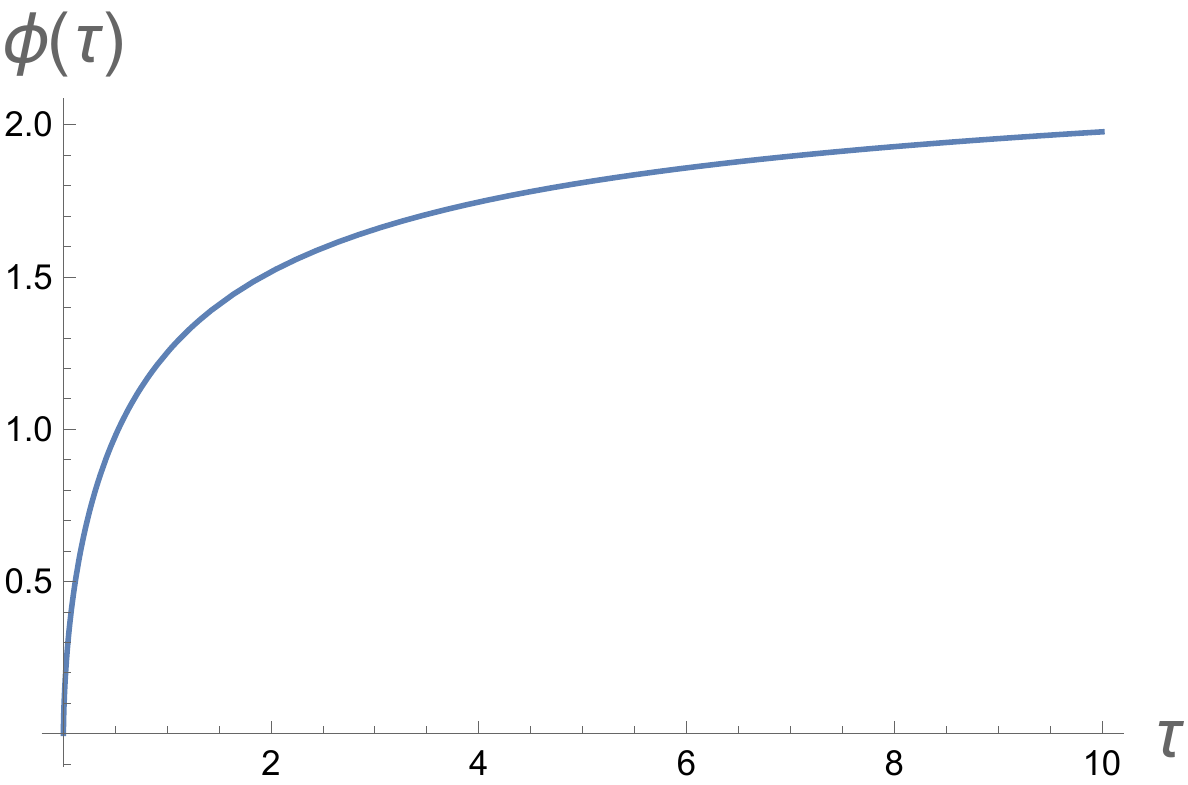}
    \caption{The profile of $V(\phi)$ (left), $z(\tau)$ (middle) and $\phi(\tau)$ (right) for $\ap=\lambda=1$. We find $\phi_*\simeq 2.5.$}
    \label{fig:analyticalplot}
\end{figure}

To construct explicit examples of holographic boundary RG flow, we note that we can design the potential $V$ such that it solves the equations of motion for a desired profile of $z(\tau)$ as we mentioned before. As the first example, we choose $z(\tau)$ to be 
\ba
z(\tau)=\lambda\tau+\ap\tau^2.
\ea
This has the UV fixed point at $\phi=\tau=0$, where $V=T_{\rm UV}=\frac{1}{\s{1+\lambda^2}}$, and the IR fixed point is given by the limit $\tau\to \infty$, where we have $V=T_{\rm IR}=0$. Note that it satisfies 
$T_{\rm UV}>T_{\rm IR}$ and thus $S^{\rm (UV)}_{\rm bdy}>S^{\rm (IR)}_{\rm bdy}$.
The potential reads
\ba
V=\f{1}{\s{1+(\lambda+2\ap\tau)^2}}+\frac{\ap(\lambda\tau+\ap\tau^2)}{\left(1+(\lambda+2\ap\tau)^2\right)^{3/2}}.
\ea
We can find the behavior at $\tau=0$:
\ba
&& \phi\simeq 2\s{\frac{\ap}{\lambda\s{1+\lambda^2}}}\s{\tau},\no
&& V(\phi)\simeq \frac{1}{\s{1+\lambda^2}}-\frac{\lambda^2}{4(1+\lambda^2)}\phi^2.
\ea
The full plot is shown in Fig.\ref{fig:analyticalplot}.

Another example is
\ba
z(\tau)=\frac{\a \tau+\mu \tau^2}{1+\tau}\ \ \ (\a\leq\mu)
\ea
This also has the UV fixed point at $\phi=\tau=0$, where $V=T_{\rm UV}=\frac{1}{\sqrt{1+\a^2}}$, and IR fixed point is also given by the limit $\tau\to\infty$, where we have $V=T_{\rm IR}=\frac{1}{\sqrt{1+\mu^2}}$.  Again it satisfies 
$T_{\rm UV}>T_{\rm IR}$ and thus $S^{\rm (UV)}_{\rm bdy}>S^{\rm (IR)}_{\rm bdy}$.
The potential reads
\ba
V=(1+\tau)^2\frac{-\left(\a^2 (\tau -1)\right)+\a \mu \tau  (\tau +5)+\tau  \left(\tau  \left(\mu^2 (\tau  (\tau +4)+5)+\tau  (\tau +4)+6\right)+4\right)+1}{ \left(\a^2+2 \a \mu \tau  (\tau +2)+\tau  (\tau +2) \left(\left(\mu^2+1\right) \tau  (\tau +2)+2\right)+1\right)^{3/2}}
\ea
we can find the behavior at $\tau=0$:
\ba
&&\phi\simeq 2\sqrt{\frac{\mu-\a}{\a\sqrt{1+\a^2}}}\sqrt{\tau},\\
&&V(\phi)\simeq\frac{1}{\sqrt{1+\a^2}}-\frac{4\a^2}{1+\a^2}\phi^2
\ea

\begin{figure}[H]
    \centering
     \includegraphics[width=.3\textwidth,page=1]{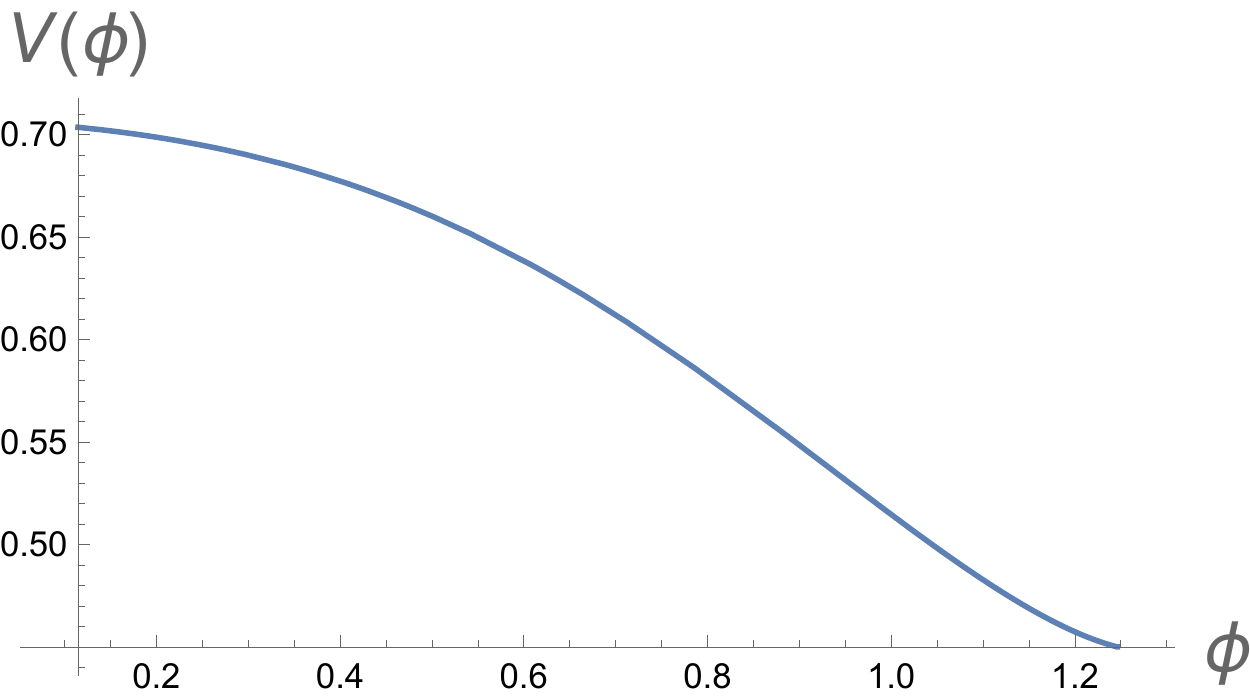}
    \includegraphics[width=.3\textwidth,page=1]{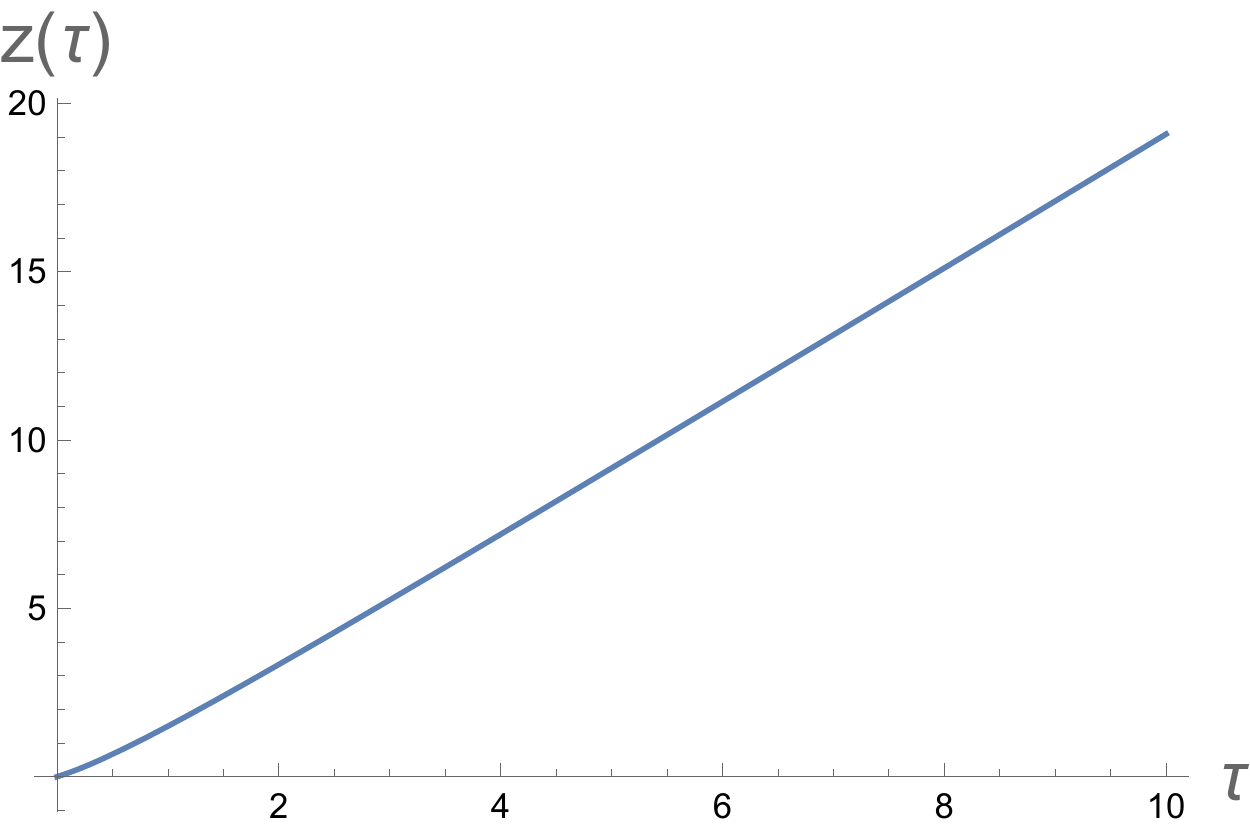}
     \includegraphics[width=.3\textwidth,page=1]{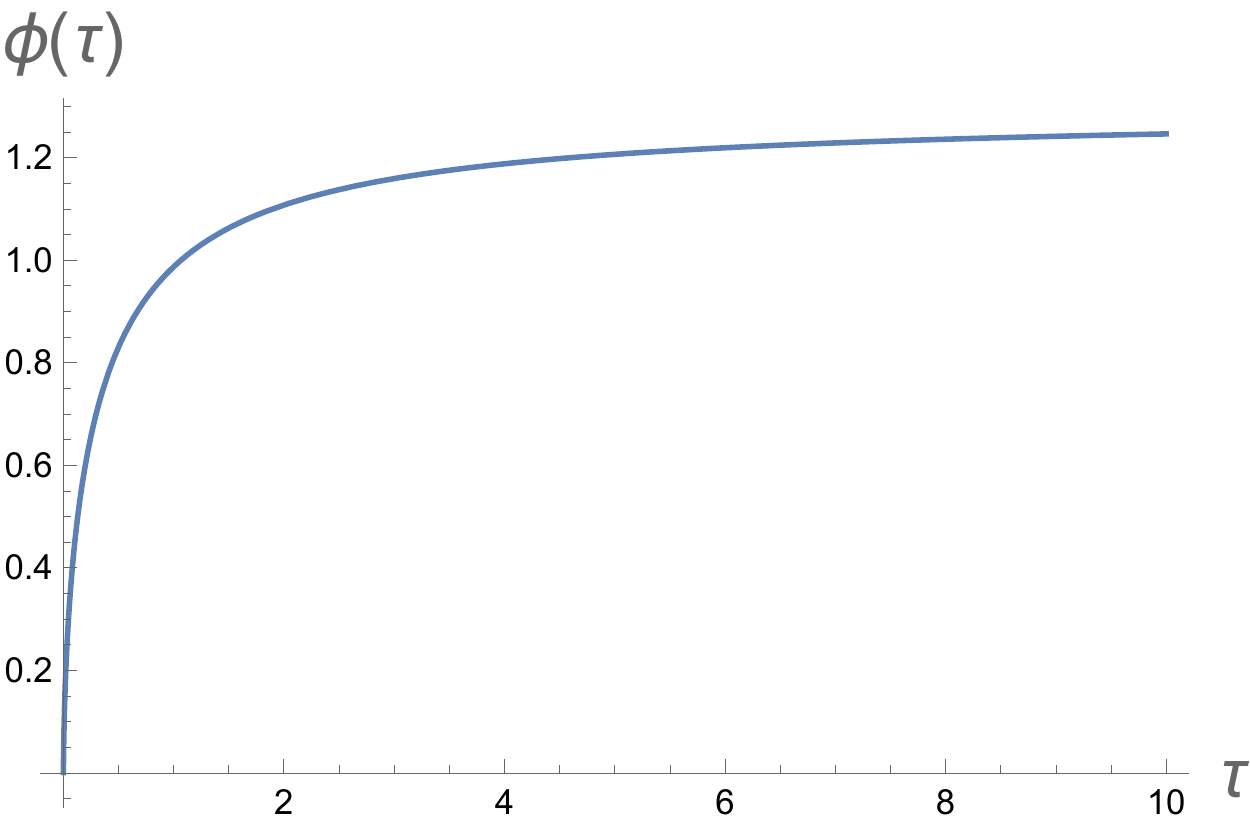}
    \caption{The profile of $V(\phi)$ (left), $z(\tau)$ (middle) and $\phi(\tau)$ (right) for $\a=1,\mu=2$}
    \label{fig:RGplot}
\end{figure}

The full plot is shown in Fig.\ref{fig:RGplot}

It is also helpful to evaluate the holographic entanglement entropy. We choose subsystem $A$ as an interval $-\tau_0\leq \tau\leq 0$ and the coordinates of two endpoints of the interval as
\ba
P:(z,\tau,x)=(\epsilon,-\tau_0,x),\ \ \ Q:(z,\tau,x)=(\epsilon,0,x).
\ea
By increasing the value of $\tau_0$, we can probe from UV to IR degrees of freedom. Thus the flow of $\tau_0$ is interpreted as the RG flow.

\begin{figure}[t]\label{fig:squareEE}
   \centering
    \includegraphics[width=.4\textwidth,page=1]{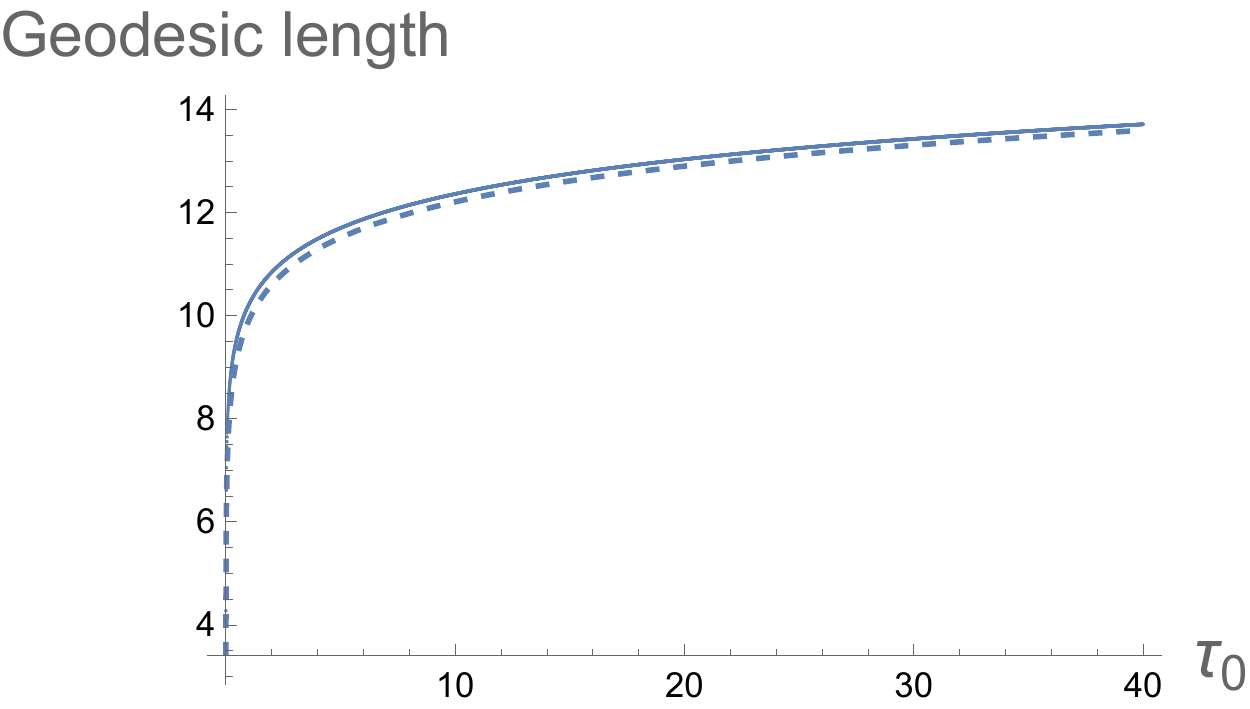}
     \includegraphics[width=.4\textwidth,page=1]{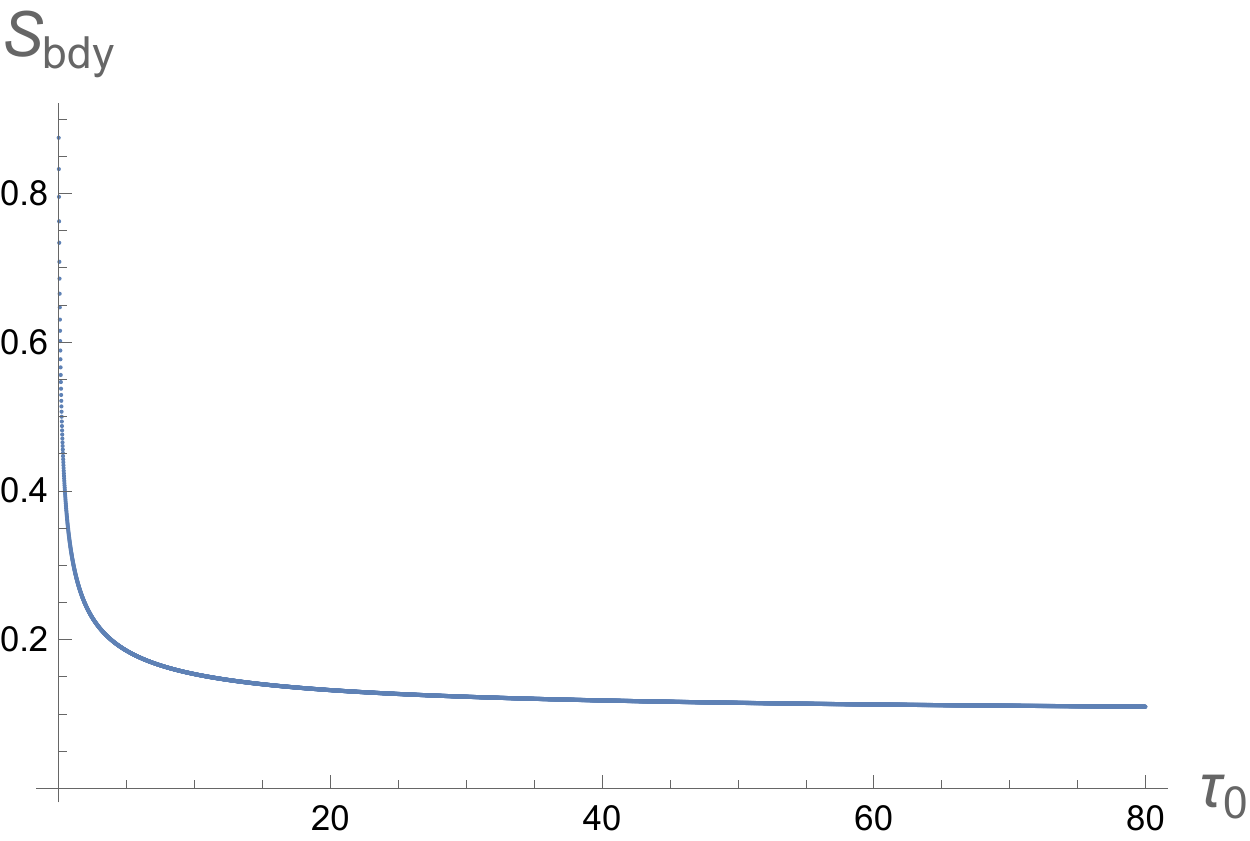}
    \caption{Left figure shows geodesic length (solid line) of $z(\tau)=\tau+\tau^2$ case and plot of $log(2\tau_0/\epsilon)$ (dashed line). Right figure shows the difference between solid line and dashed line.}
\end{figure}

\begin{figure}[t]\label{fig:fractionEE}
   \centering
    \includegraphics[width=.4\textwidth,page=1]{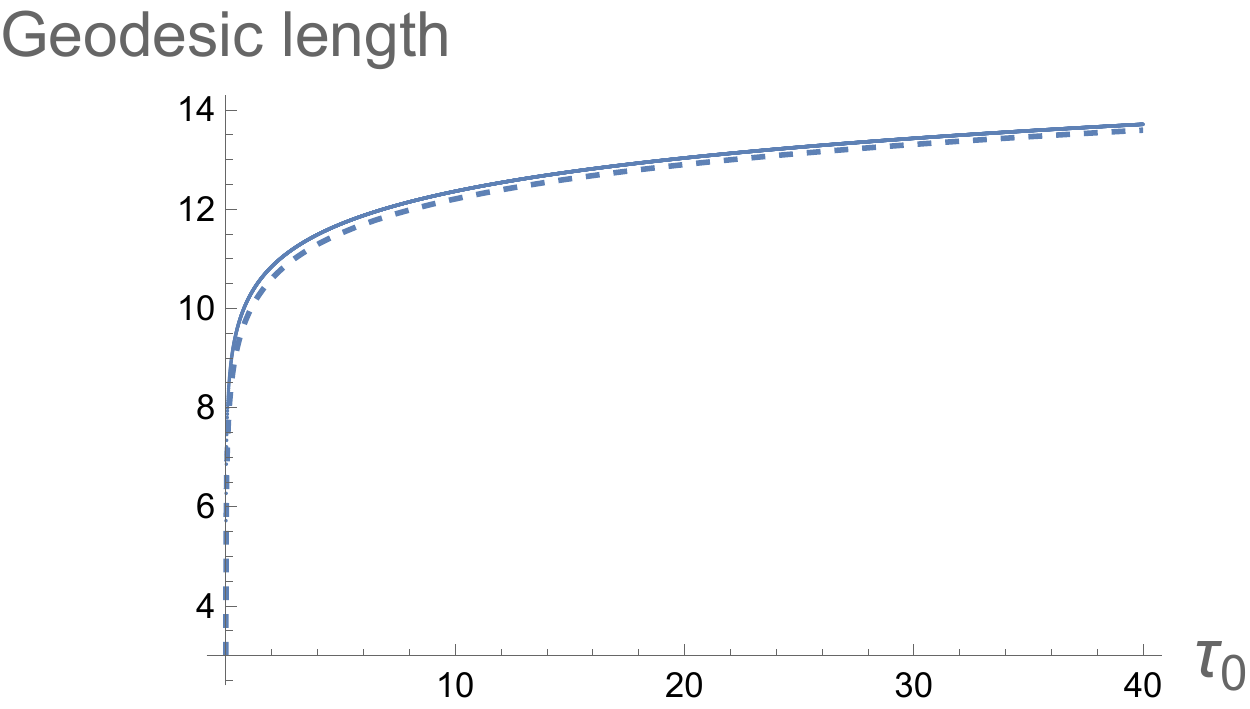}
     \includegraphics[width=.4\textwidth,page=1]{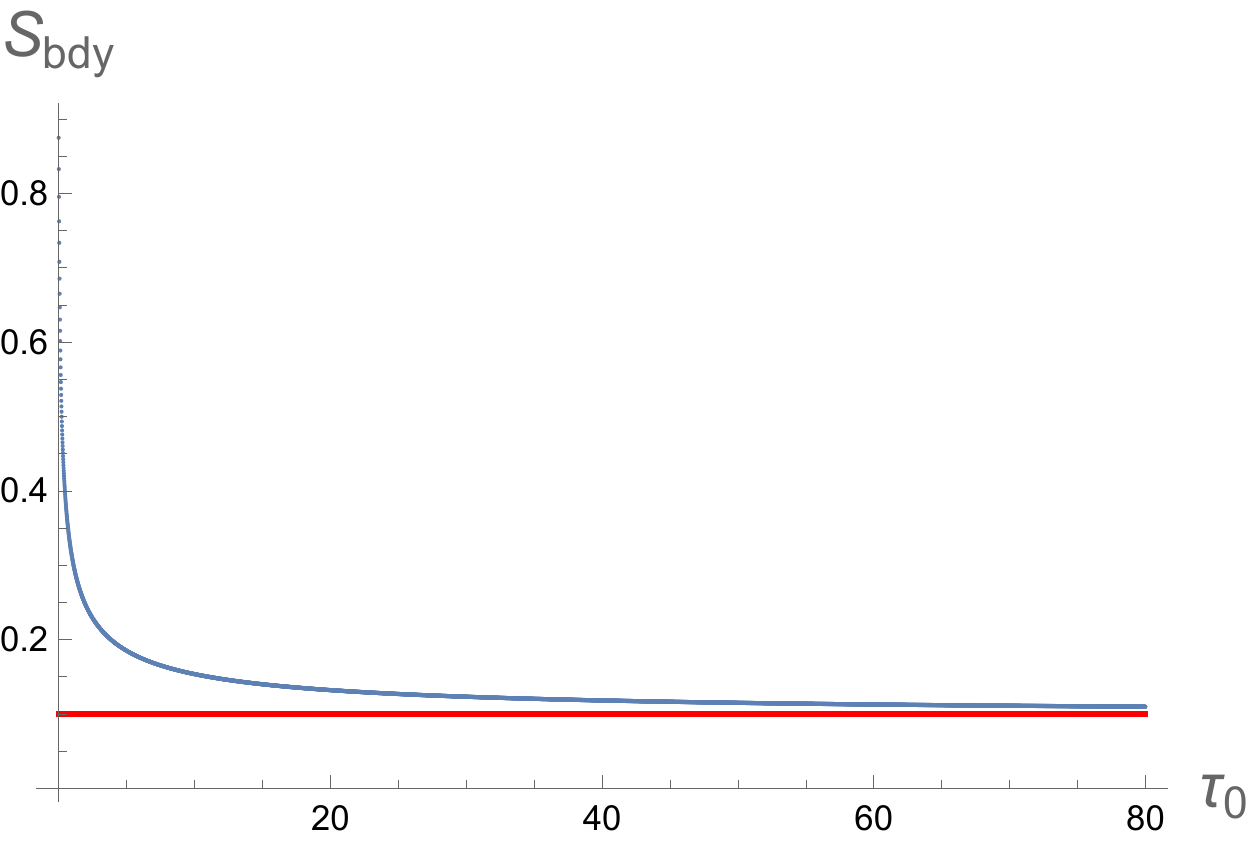}
    \caption{Left figure shows geodesic length (solid line) of $z(\tau)=\frac{\tau+10\tau^2}{1+\tau}$ case and plot of $log(2\tau_0/\epsilon)$ (dashed line). Right figure shows the difference between solid line and dashed line.}
\end{figure}

To calculate the entanglement entropy, we would like to identify the geodesic $\Gamma_A$ which ends on the EOW brane. So we write the endpoint on the brane as $(z,\tau,x)=(z(\tau),\tau,x)$. 
Because of the symmetry of the interval under $x\to -x$, it is sufficient to assume that the surface begins from $Q$ and ends on the EOW brane (refer to Fig.\ref{fig:AdSBCFT}).
It is helpful to embed Euclidean AdS$_3$ spacetime in $4$-dimensional Minkowski space
\ba
ds^2=-dX_0^2+dX_3^2+dX_1^2+dX_2^2
\ea
with the following embedding equation
\ba
X_0^2=X_1^2+X_2^2+X_3^2+1
\ea
where $X_\mu$ satisfies
\ba
X_0=\frac{z}{2}\qty(1+\frac{1+x^2+\tau^2}{z^2}),\ X_1=\frac{\tau}{z},\ X_2=\frac{z}{2}\qty(1-\frac{1-x^2-\tau^2}{z^2}),\ X_3=\frac{x}{z}
\ea
By using this embedding, we can evaluate geodesic length between $(z,\tau,x)$ and $(z',\tau',x')$ as 
\ba\label{eq:geodesiclengthofAds}
\cosh^{-1}(X_0 X'_0+X_3 X'_3-X_1 X'_1-X_2 X'_2)
\ea
where $X$ and $X'$ are embedding of $(z,\tau,x)$ and $(z',\tau',x')$ respectively.
By using this formula, we get the geodesic length between P and the EOW brane
\ba
\cosh^{-1}\qty(\frac{\tau ^2+2 \tau _0 \tau +\tau _0^2+z(\tau )^2}{2 z(\tau )\epsilon}+\mathcal{O}(\epsilon)).
\ea
In order to find the minimum value of the quantity above, we need to minimize the first term in the bracket.
After choosing the profile $z(\tau)$, the $\tau$-dependent term can be minimized by a numerical calculation and we can evaluate the behavior of boundary entropy by subtracting $\log(2\tau_0/\epsilon)$ from the minimum 
geodesic length.
We chose the two different cases: $(i)$ $z(\tau)=\tau+\tau^2$ and $(ii)$ $z(\tau)=\frac{\tau+10\tau^2}{1+\tau}$ and study their behaviors
(we set $\epsilon=10^{-4}$).
In the case $(i)$, since we have $z(\tau)\approx \tau$, 
in the UV limit $\tau_0\to 0$, the boundary entropy takes the value $S^{\rm (UV)}_{\rm bdy}=\mathrm{arcsinh}(1)\approx 0.88$. On the other hand, in the IR limit $\tau_0\to \infty$ we find $S^{\rm (IR)}_{\rm bdy}=0$.
We can confirm this from the plot in right panel of Fig.\ref{fig:squareEE}.
Similarly, in the case $(ii)$, boundary entropy should satisfy 
$S^{\rm (UV)}_{\rm bdy}=\mathrm{arcsinh}(1)\approx 0.88$ and 
$S^{\rm (IR)}_{\rm bdy}=\mathrm{arcsinh}(1/10)\approx 0.10$ in the UV and IR limit, respectively. Again we can confirm these from the right panel of Fig.\ref{fig:fractionEE}. These behaviors are consistent with the $g$-theorem.


\subsection{BCFT on a strip}\label{sec:bdyJanus}

\begin{figure}[H]
    \centering
    \includegraphics[width=.5\textwidth,page=1]{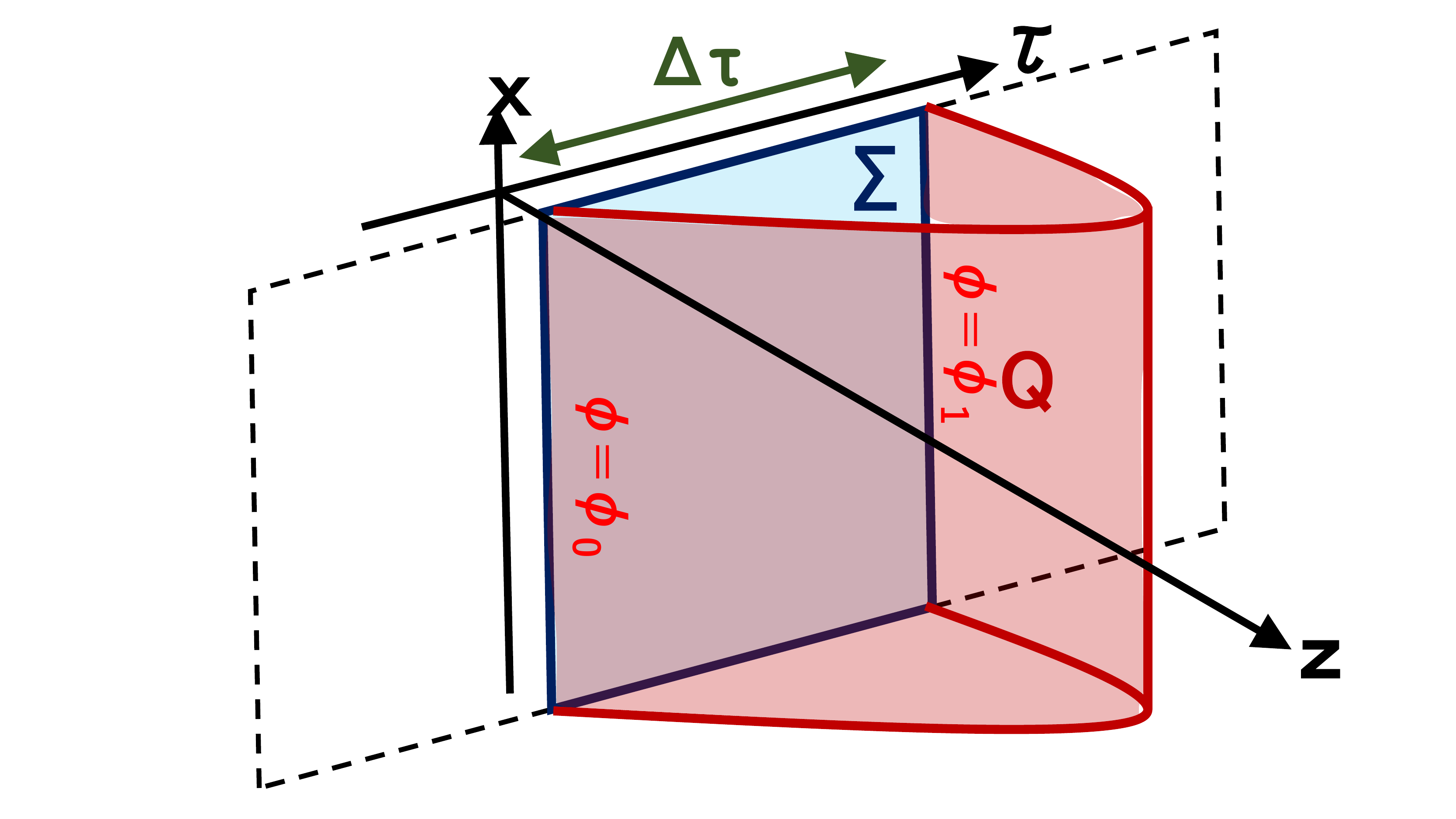}
    \caption{The setup of AdS$_3/$BCFT$_2$ dual to a BCFT on a strip $\Sigma$.}
    \label{fig:AdSBCFTST}
\end{figure}

Now we turn to the AdS$_3/$BCFT$_2$ setup which is dual to a BCFT on a strip $\Sigma$. In this case, assuming that the EOW brane $Q$ is a connected surface, $Q$ takes a half cylinder shape as depicted in Fig.\ref{fig:AdSBCFTST}.
We again assume the translational invariance along $x$ and write the profile of $Q$ as $z=z(\tau)$. Choosing the sign convention of the normal vector as $N^z>0$, we obtain the following equation of motion 
\ba
&& \dot{\phi}^2=-\frac{\ddot{z}}{2z\s{1+\dot{z}^2}},\no
&& V(\phi)=\frac{z^2\dot{\phi}^2}{1+\dot{z}^2}-\frac{1}{\s{1+\dot{z}^2}}.
\ea
By eliminating $\phi$, we obtain
\ba
-V(\phi)=\frac{z\ddot{z}+2+2\dot{z}^2}{2(1+\dot{z}^2)^{3/2}}.
\ea

For simplicity, we set $V(\phi)=0$, i.e. the massless scalar and zero tension. In this case we have $z\ddot{z}+2+2\dot{z}^2=0$, which is solved by setting $y=z^3$, leading to $\ddot{y}=-6y^{1/3}$.
In the end we find that $z=z(\tau)$ is obatined by solving
\ba
1+\dot{z}^2=\frac{z_0^4}{z^4},
\ea
where $z_0$ is an integration constant which gives the turning point.
We obtain
\ba
&& \ddot{z}=-\frac{2z_0^4}{z^5},\no
&& \dot{\phi}^2=\frac{z_0^2}{z^4}.\label{EOMsctd}
\ea

\begin{figure}[H]
    \centering
    \includegraphics[width=.5\textwidth,page=1]{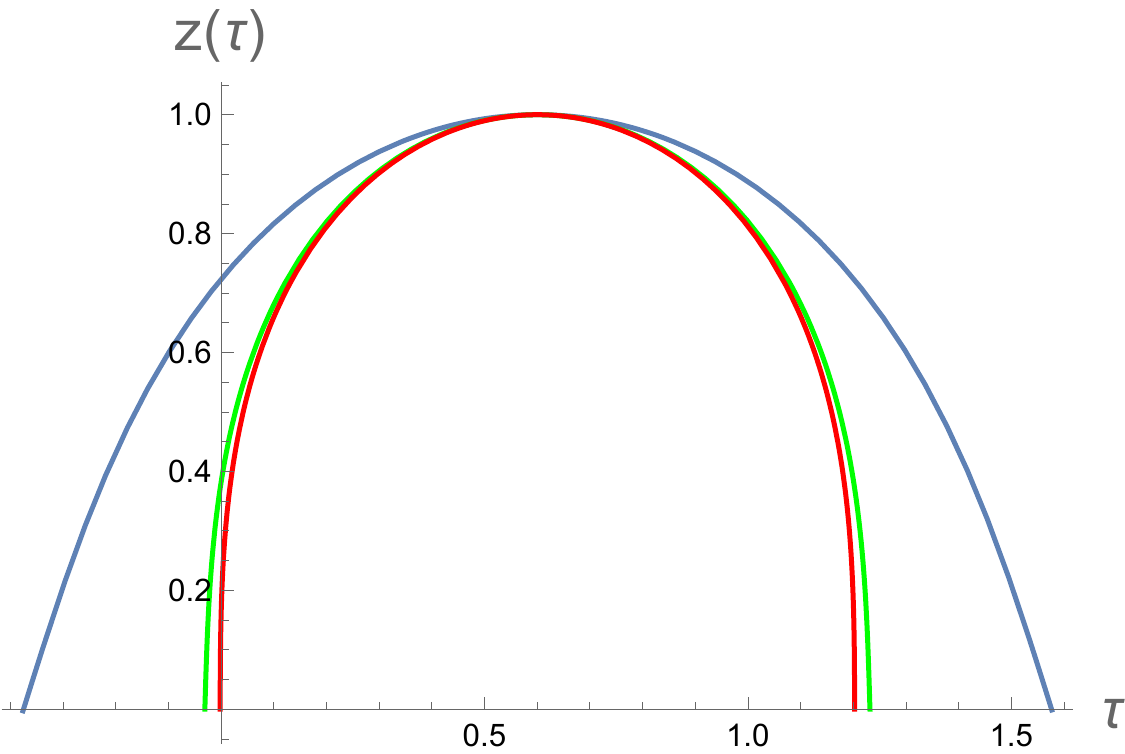}
    \caption{Blue, green, red one represents the profile of EOW brane for $T=-1/3,-1/30,-1/300$ respectively. In this plot we set $z_0=1.$}\label{fig:plotST}
\end{figure}

Near $z=0$, they behave like
\ba
&& z\simeq (3z_0^2 \tau)^{\frac{1}{3}},\no
&& \phi\simeq \phi_0+\left(\frac{3\tau}{z_0}\right)^{\frac{1}{3}}.
\ea
We write the two end points as $\tau=0$ and $\tau=\Delta\tau$. The full solution of $z=z(\tau)$ is plotted in Fig.\ref{fig:plotST}.

It is also easy to estimate the shift:
\ba
&&\Delta \tau=2\int^{z_0}_0 \frac{z^2dz}{\s{z_0^4-z^4}}=(2E[-1]-2K[-1])z_0\simeq 1.2 z_0\no
&& \Delta \phi=\phi_1-\phi_0=2\int^{z_0}_0 \frac{z_0 dz}{\s{z_0^4-z^4}}=2K[-1]\simeq 
2.6.  \label{xxpola}
\ea
 
 It is also intriguing to note that $z$ dependence of $\phi$
 on $Q$, which is asymptotically AdS$_2$, agrees with the standard asymptotic behavior of scalar field dual to a marginal deformation: 
 \ba
 \phi\sim z^{1-\Delta}J+z^{\Delta}\expval{O},\ \ \ \Delta=1,
 \ea
 in the boundary limit  $z\to 0$. Here, $O$ is the marginal boundary primary operator in the BCFT and $J$ is its source.
 This agrees with the standard dictionary of AdS/CFT \cite{Gubser:1998bc,Witten:1998qj,Klebanov:1999tb}.
 It is intriguing to remember that without any scalar field, a connected solution dual to the BCFT on a strip is not available in the Poincar\'{e} AdS$_3$ but is only possible in the thermal AdS$_3$ geometry. In the presence of brane-localized scalar, we are able to construct such a solution in the Poincar\'{e} AdS$_3$. Even though we only constructed the solution with a special value of $\Delta\phi$, solutions for other values of $\Delta\Phi$ will be constructed in section \ref{sec:phasetr}, using either the thermal AdS$_3$ or BTZ spacetime.

\begin{figure}[H]
    \centering
    \includegraphics[width=.5\textwidth,page=1]{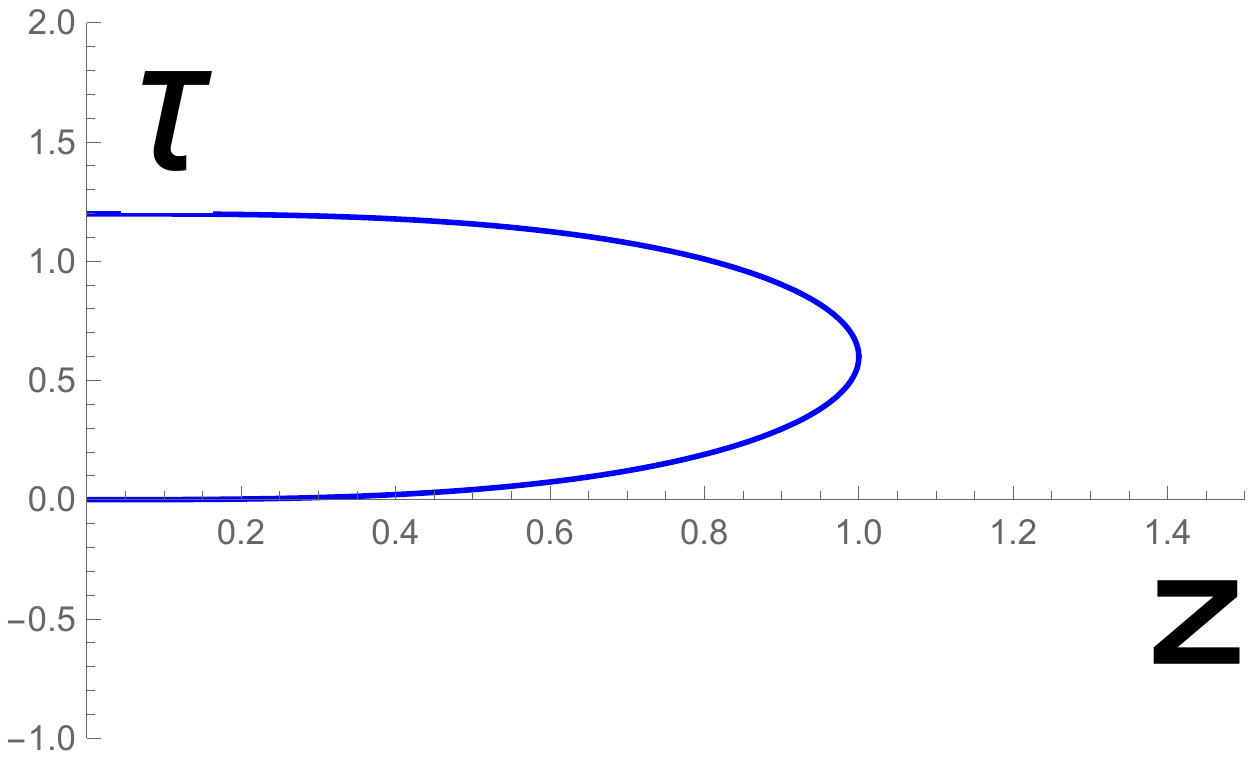}
    \caption{The profile of EOW brane with brane-localized scalar source, which may be called boundary Janus. The horizontal and vertical axis describe $z$ and $\tau$, respectively. In this plot we set $z_0=1.$}
    \label{fig:brane_profile_zerotension}
\end{figure}

Even if we set $V=T \neq 0$, we can get similar profile of $z(\tau)$. 
For example, in case of $T=-1/3,-1/30,-1/300$, $z(\tau)$ behaves like the figure below and their end point and beginning point on the boundary changes with respect to the value of  $T$.

\section{\texorpdfstring{AdS$_3$/BCFT$_2$}{AdS3/BCFT2} with round-shaped branes}\label{ch:polar}
\begin{figure}[H]
    \centering
    \includegraphics[width=.75\textwidth,page=1]{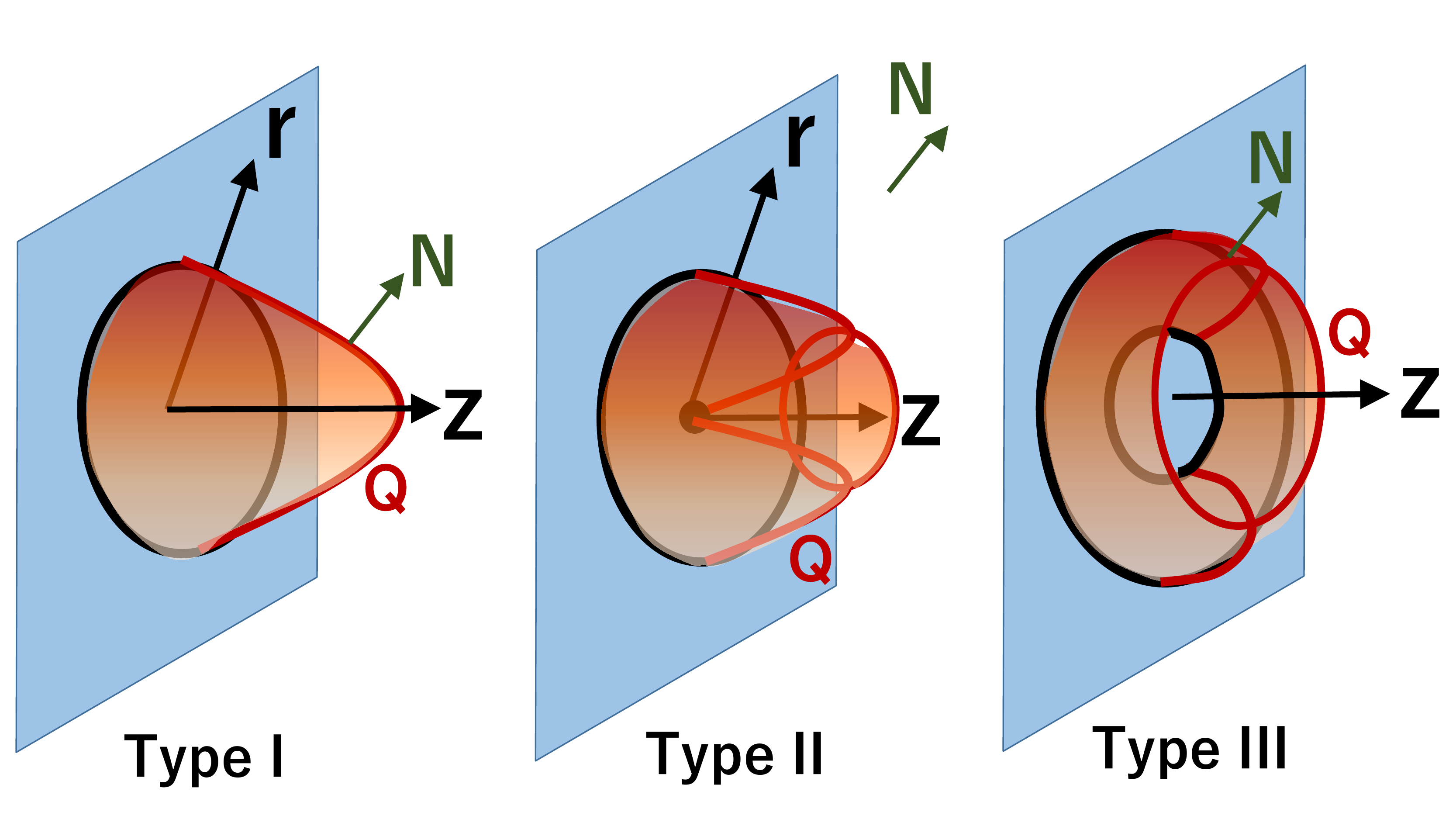}
    \caption{The setups of AdS$_3/$BCFT$_2$ with a round EOW brane: classified into Type I (left), II (middle) and III (right).}
    \label{fig:AdSBCFTP}
\end{figure}

Here we consider the AdS$_3/$BCFT$_2$ setup with round shaped branes as depicted in Fig.\ref{fig:AdSBCFTP}.
For this, we introduce
 the polar coordinate of Poincar\'{e} AdS$_3$:
\ba
ds^2=\frac{dz^2+dr^2+r^2d\theta^2}{z^2}.
\ea

We consider the EOW brane with the rotational invariant profile 
\ba
z=z(r),\ \ \ \phi=\phi(r).
\ea
The induced metric reads 
\ba\label{eq:inducemetriconpolarbrane}
ds^2=\frac{(1+\dot{z}(r)^2)dr^2+r^2d\theta^2}{z(r)^2},
\ea
where we set $\dot{z}=\de_r z$.
The normal vector of the EOW brane reads 
\ba
(N^z,N^r,N^\theta)=\frac{z}{\s{1+\dot{z}^2}}\left(1,-\dot{z},0\right).
\ea
This leads to
\ba
&& K_{rr}-h_{rr}K=\frac{(r+z\dot{z})\s{1+\dot{z}^2}}{rz^2},\no
&& K_{\theta\theta}-h_{\theta\theta}K=\frac{r^2(1+\dot{z}^2+z\ddot{z})}{z^2(1+\dot{z}^2)^{3/2}}.
\ea

The full equations of motion (\ref{Nbc}) are summarized into 
\ba\label{eq:polar ODE}
&& \dot{\phi}^2=-\frac{\ddot{z}}{2z\s{1+\dot{z}^2}}+\frac{\dot{z}\s{1+\dot{z}^2}}{2zr},\no
&& V(\phi)=\frac{z^2\dot{\phi}^2}{1+\dot{z}^2}-\frac{1}{\s{1+\dot{z}^2}}-\frac{z\dot{z}}{r\s{1+\dot{z}^2}}.
\ea

\subsection{Type I, Hemisphere-like shaped brane}\label{ch:TypeI}

The boundary conditions of Type I are
\ba\label{eq:cond of Type I}
z_0:=z(0)>0,\quad \dot{z}(0)=0,\quad \ddot{z}(0)<0.
\ea
The third condition demands the brane intersect with the asymptotic boundary. Solving $\dot{\p}^2\geq0$ will impose restrictions on $z(r)$:
\ba\label{eq:Type_sol}
z(r)=\int_0^r\frac{A(r)r}{\s{1-A^2(r)r^2}}dr+z_0,
\ea
where $A(r)$ is an arbitrary function satisfying $A(0)<0$ and $\dot{A}(r)\leq0$ on $I:=[0,r_1]$, and $r=r_1$ is an intersection point of the brane and the asymptotic boundary. In particular, when $A(r)=A<0$ is constant and $z_0=-(T+1)/A$, this solution reproduces the well-known hemispherical brane with constant tension $T$ and without matter fields. The solution \eqref{eq:Type_sol} is an improvement of this hemispherical brane. A more detailed derivation is summarized in the Appendix \ref{ap:round-sol}.

Try to stretch or compress the obtained solutions. In other words, when $z=z(r)$ is a solution of \eqref{eq:cond of Type I}, we find conditions in $\a>0$ for which $z=\a \cdot z(r)$ also becomes solution. Then there exists a $B(r)$ satisfying the following
\ba
\frac{B(r)r}{\s{1-B^2(r)r^2}}=\frac{\a A(r)r}{\s{1-A^2(r)r^2}},\quad B(0)<0,\quad \dot{B}(r)\leq 0,\quad \forall r\in I.
\ea
Therefore, $\a$ has an upper bound
\ba\label{eq:upper bound of alpha}
\a^2\leq \min_{r\in I}\left(1+\frac{\dot{A}(r)}{A^3(r)}\right).
\ea
For example, when $A(r)$ is constant, we cannot stretch spheres and only oblate spheres ($\a<1$) are allowed.

\begin{figure}[H]
    \centering
    \includegraphics[width=.4\textwidth,page=1]{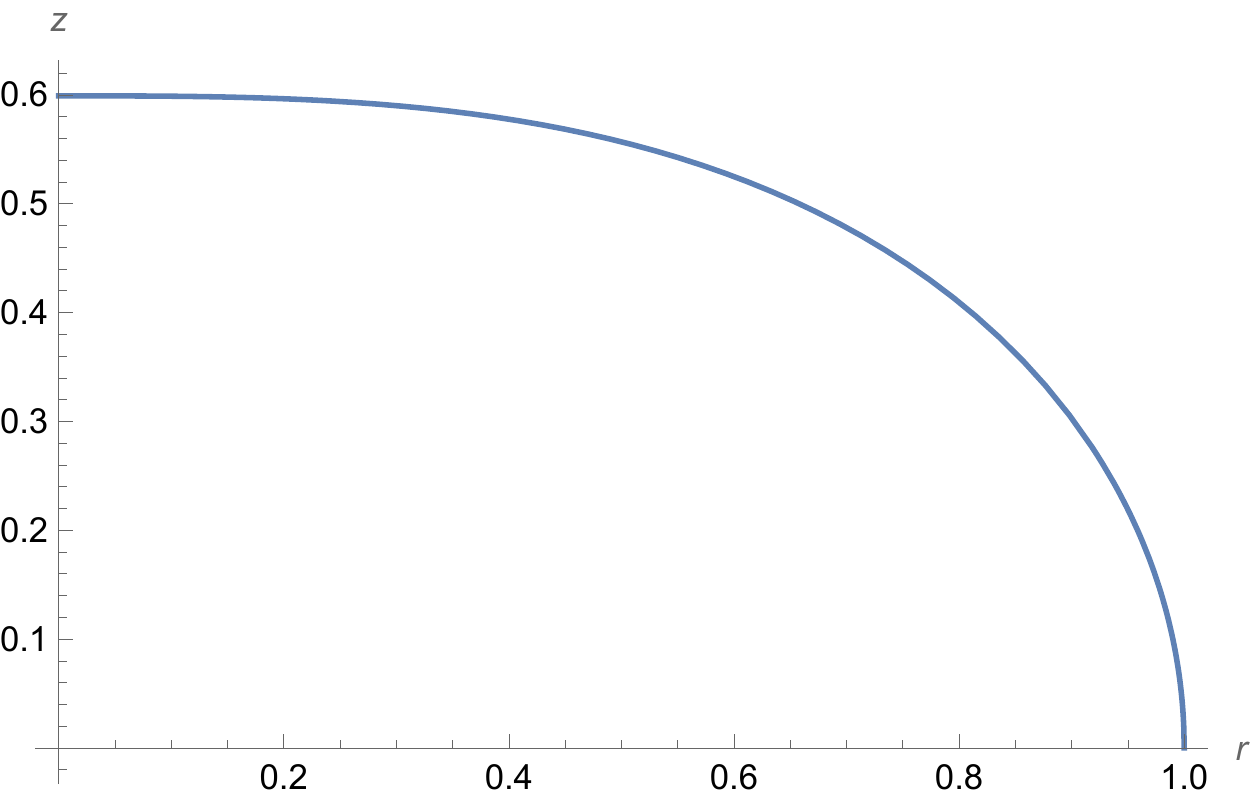}
 \includegraphics[width=.4\textwidth,page=1]{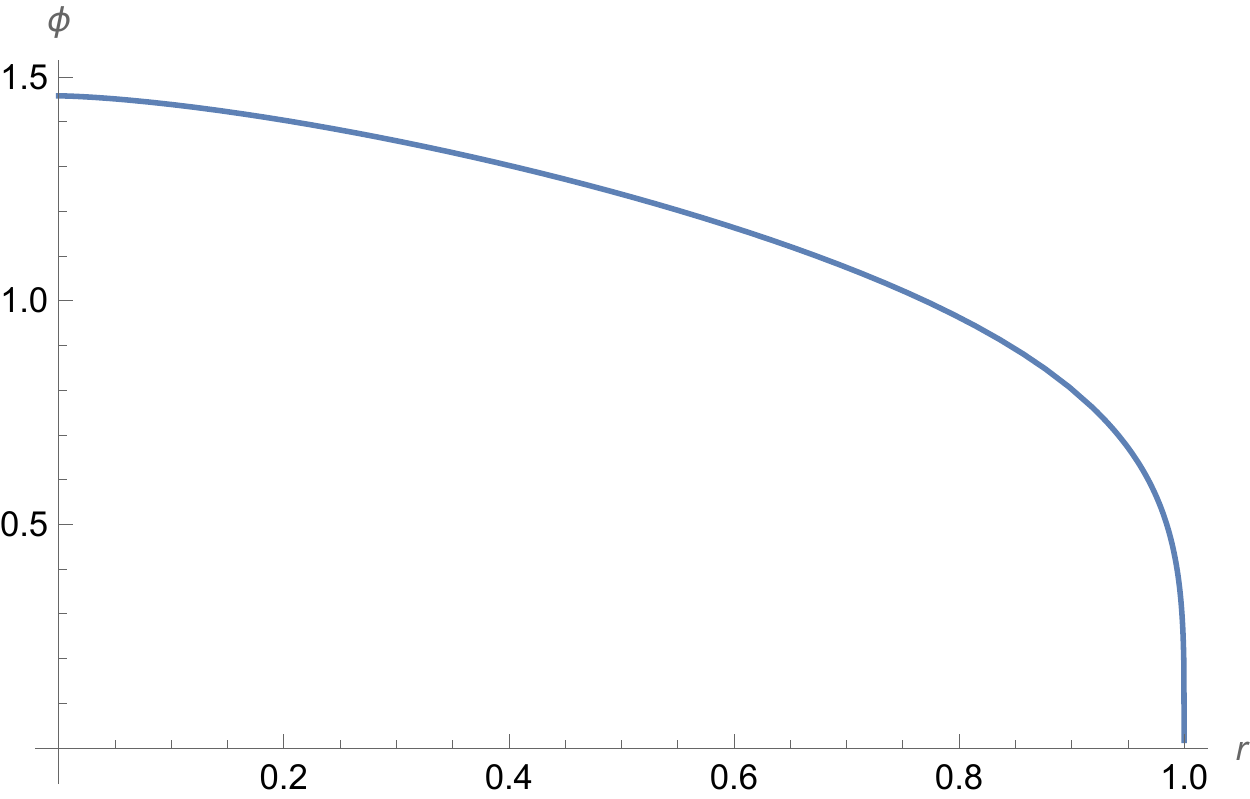}       \includegraphics[width=.4\textwidth,page=1]{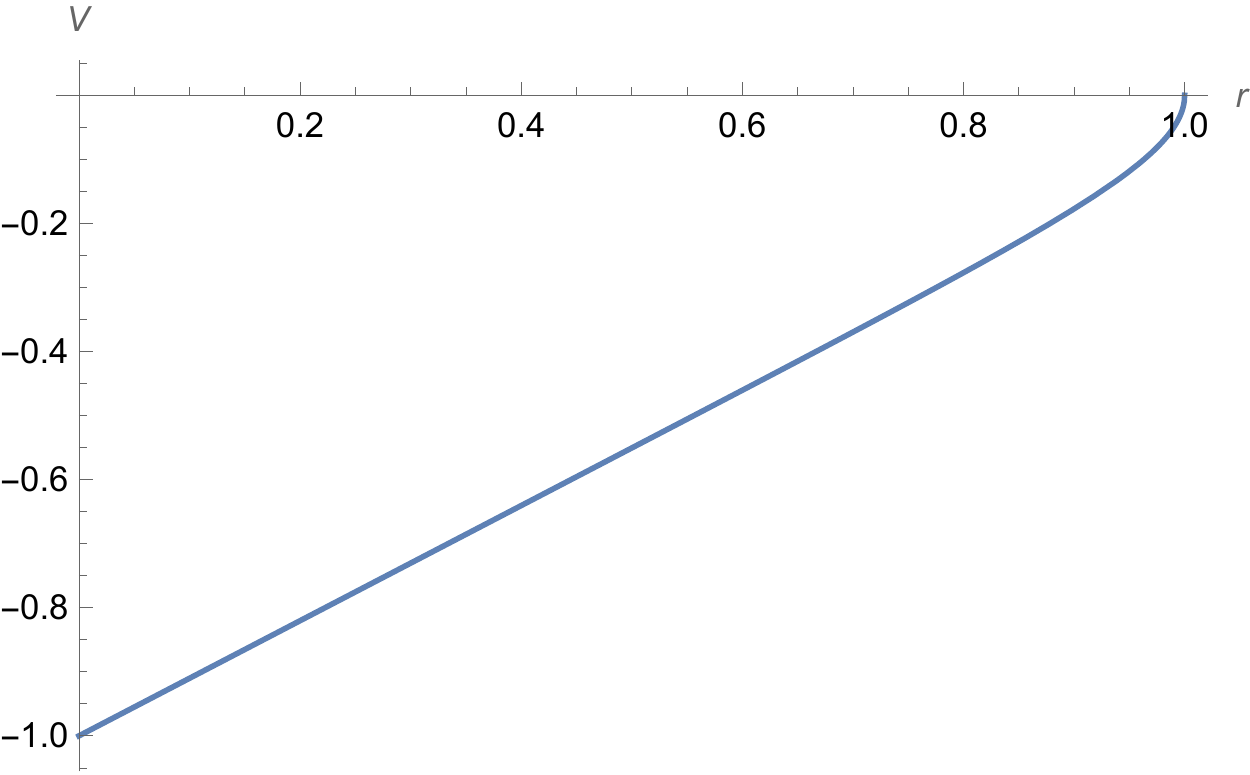}
    \includegraphics[width=.4\textwidth,page=1]{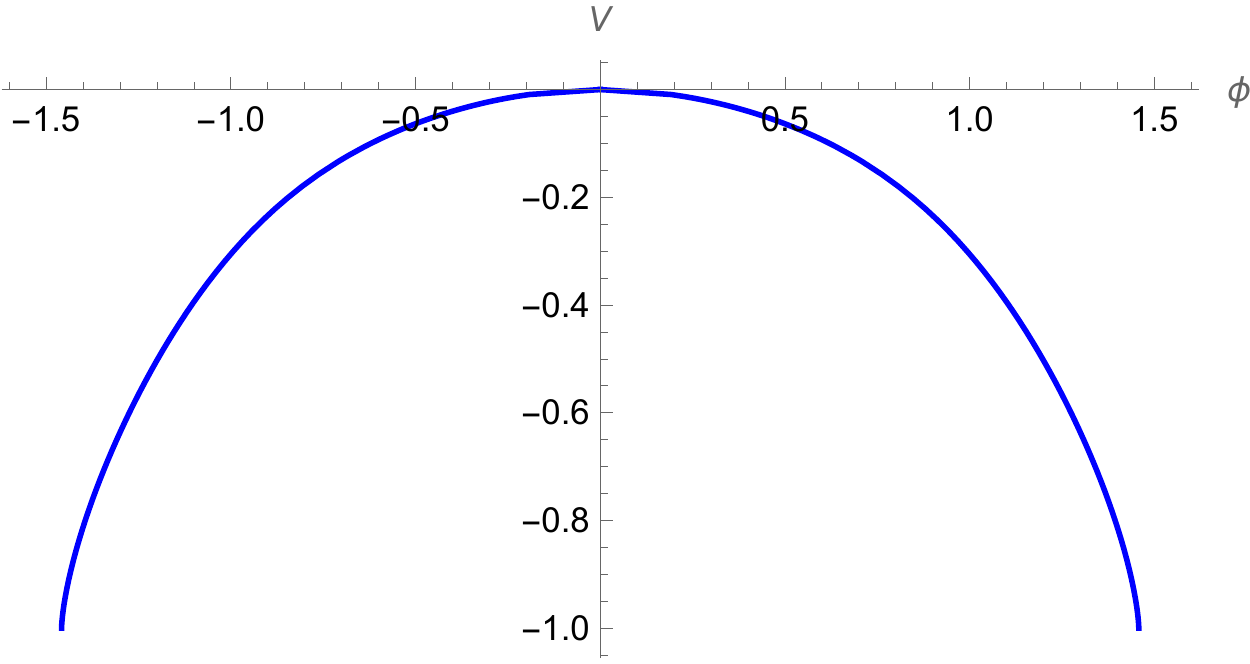}
 \caption{Configurations with $A(r) = -r$. The upper right, upper left, lower right, and lower left graphs are $z(r)$, $\p(r)$, $V(r)$, and $V(\p)$, respectively. In the fourth plot, because we can not determine the sign of $\phi$, we plot not only $(\p,V)$ but also $(-\p,V)$. The quadratic coefficient of $V(\p)$ is computed by $\lim_{r\to 1}V(r)/\p(r)^2$ and is $-0.25$.}
    \label{fig:TypeI_Ar_r}
\end{figure}

Calculate the mass of the scalar field $\phi$ on brane and the conformal dimension of corresponding boundary operator when $A(r) = -r$ and $r_1=1$. $z_0$ is determined so that $z(r_1) = 0$. As the result, $z(r)$ is complicated function expressed using a hypergeometric function. The numerical profile of this configuration is summarized in Fig.\ref{fig:TypeI_Ar_r}. 

Obviously, the induced metric \eqref{eq:inducemetriconpolarbrane} is not exact Poincar\'{e}, but coordinate transformation $\h=g(r):=z(r)/r$ tells that this is asymptotic AdS$_2$. Actually, since, near the asymptotic boundary $r=r_1$, $r$ is expanded by $\h$ as
\ba
r=g^{-1}( \eta ) = 1-\eta ^{2} +\frac{7}{6} \eta ^{4} -\frac{103}{144} \eta ^{6} +O\left( \eta ^{8}\right),
\ea
the induced metric can be read
\ba
h_{\h\h}=\frac{1}{\h^2}+O(\h),\quad h_{\th\th}=\frac{1}{\h^2}.
\ea
In this coordinate, the behavior of $\phi(\h)$ is
\ba
\phi ( \eta ) =-\sqrt{2} \eta ^{1/2} +\frac{\sqrt{2}}{5} \eta ^{5/2} +O\left( \eta ^{7/2}\right)
\ea
and conformal dimension of corresponding boundary operator reads as $\D=1/2<1$. This implies the mass $m$ of $\phi$ is $m^2=\D(\D-1)=-1/4<0$ and is consistent with the coefficient of potential $V(\phi)$. Therefore, we can conclude this deformation is relevant.

\begin{figure}[H]
    \centering
    \includegraphics[width=.4\textwidth,page=1]{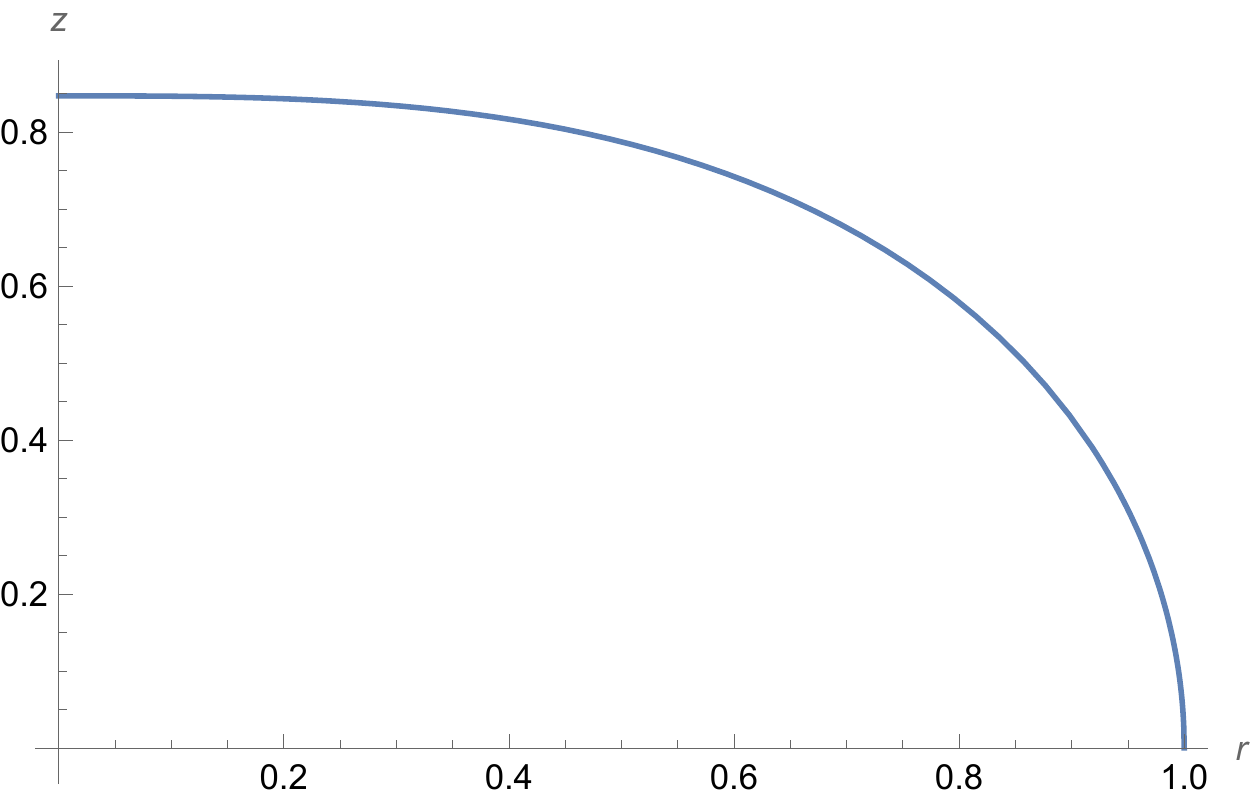}
 \includegraphics[width=.4\textwidth,page=1]{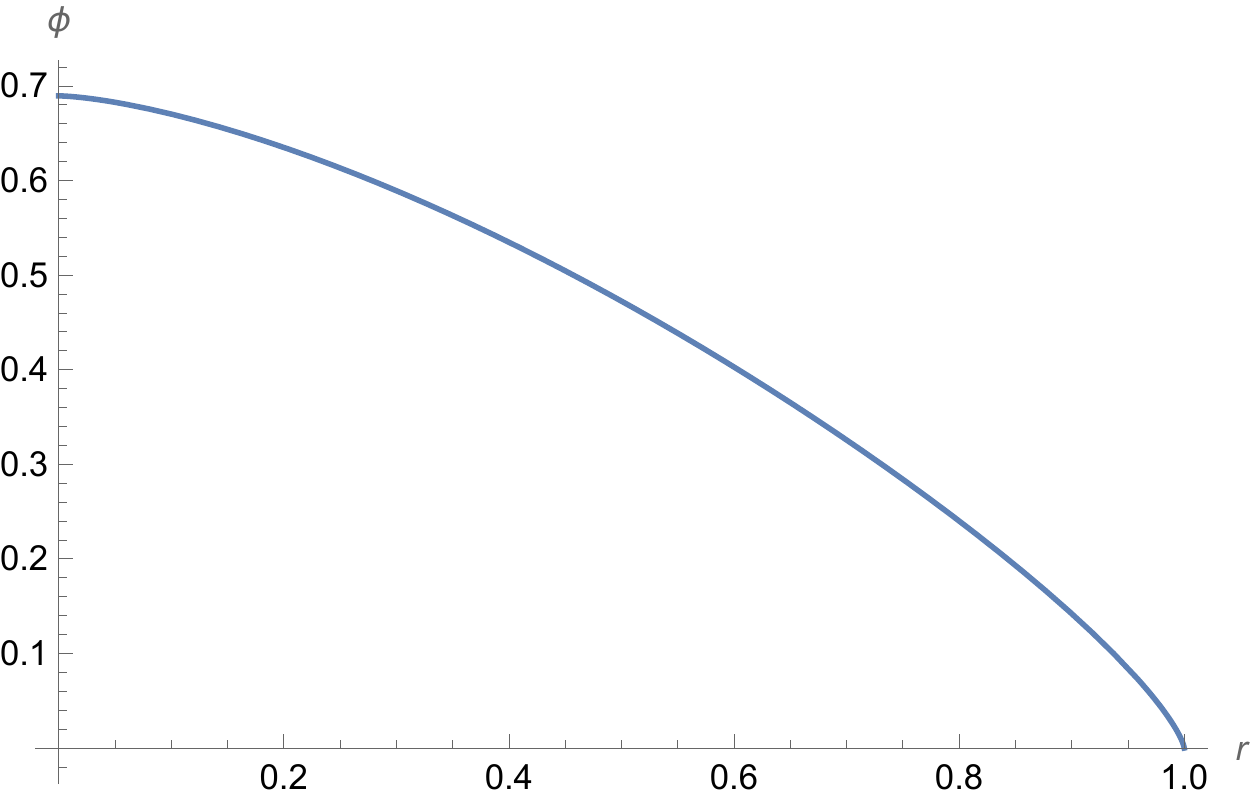}       \includegraphics[width=.4\textwidth,page=1]{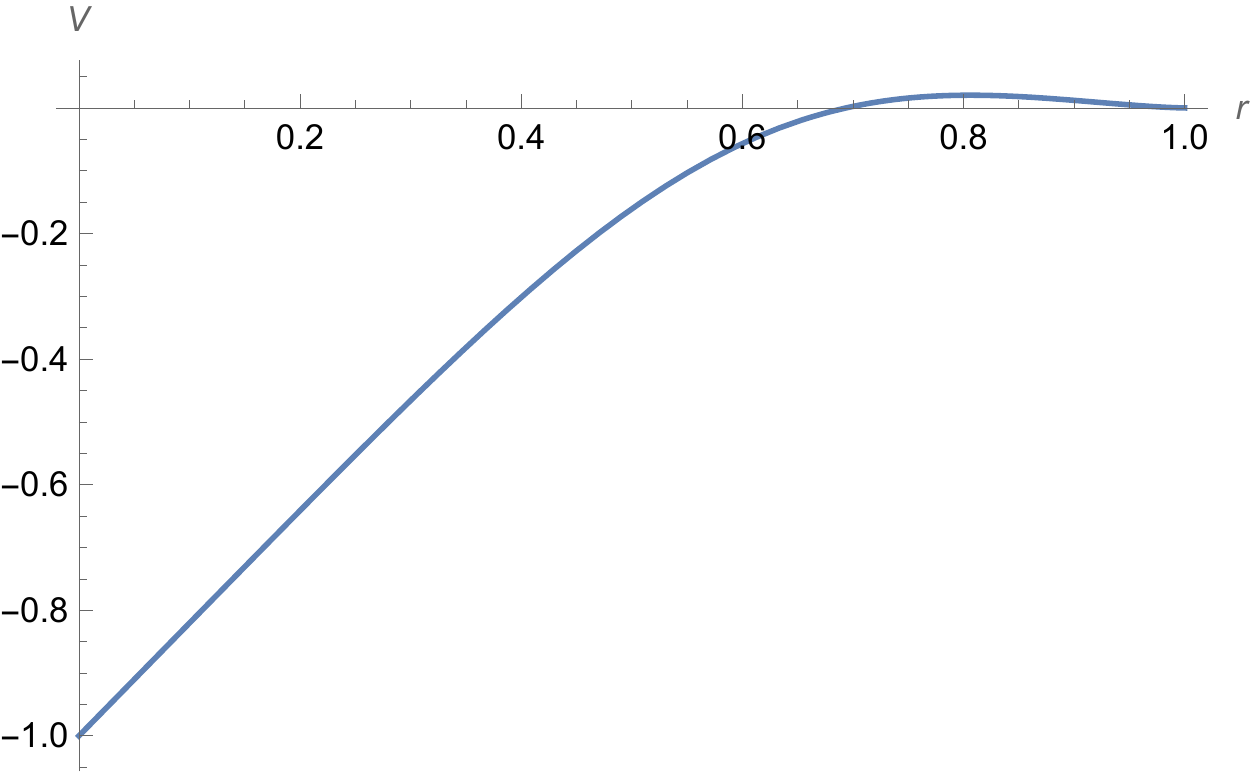}
    \includegraphics[width=.4\textwidth,page=1]{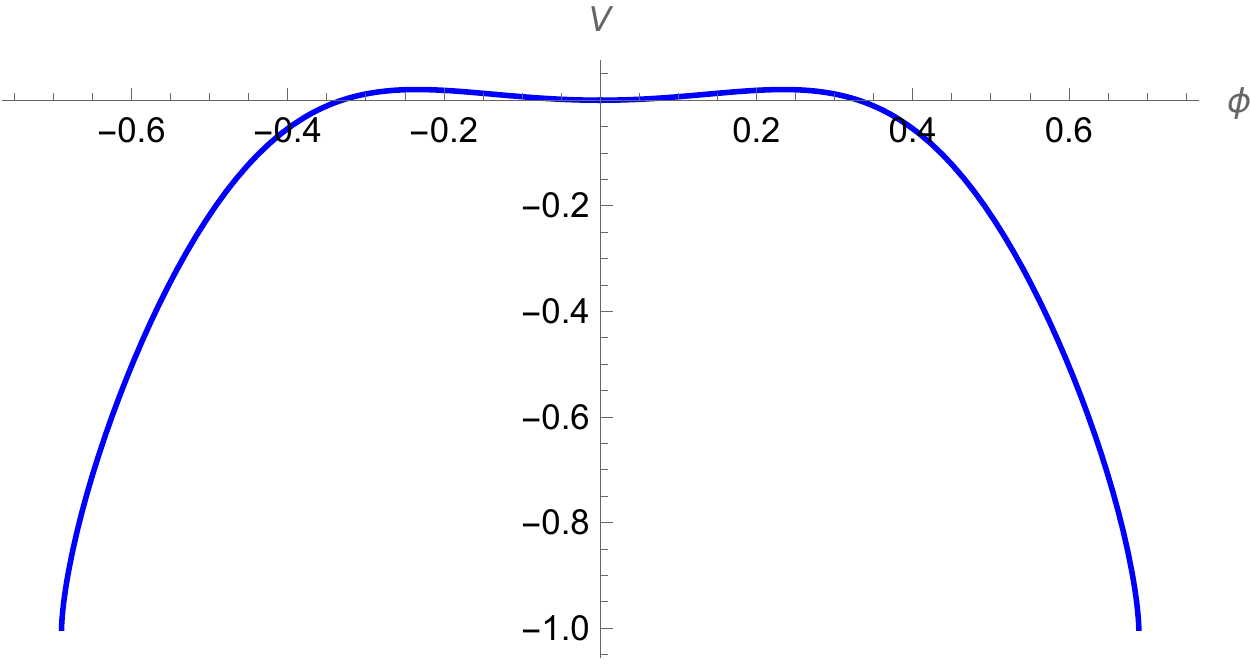}
 \caption{Configurations when $z(r)$ in fig.\ref{fig:TypeI_Ar_r} is multiplied by $\s{2}$. At this time, the $\p$-dependency near $r=1$ is $V(\p)\simeq 0.75\p^2$}
    \label{fig:TypeI_Ar_r_s2}
\end{figure}

Using \eqref{eq:upper bound of alpha}, however, we can stretch this configuration up to $\a = \sqrt{2}$ (Fig.\ref{fig:TypeI_Ar_r_s2}).  

As before, by introducing new coordinate $\h=\sqrt{2}g(r)$, we can get asymptotically AdS$_2$ induced metric. At this time
Near the asymptotic boundary, $\p$ behaves as follows:
\ba
\phi ( \eta ) = -\frac{\sqrt{2}}{3} \eta ^{3/2} +\frac{11\sqrt{2}}{84} \eta ^{7/2} +O\left( \eta ^{9/2}\right).
\ea
Hence, we can read $\D=3/2$ and $m^2=3/4$, and they match the result from potential, but this is irrelevant because $\D>1$. 

However, when $\a$ is slightly less than $\s{2}$, e.g., $\a = 1.414$, the result is 
\ba
\phi(\h)\simeq-0.012\h^{1/2}-20\h^{3/2},\quad V(\p)\simeq-1/4 \p^2,
\ea
and this is a relevant perturbation. Namely, $\a=\s{2}$ serves as the threshold between relevant and irrelevant deformation.

\subsection{Type II, Cone-like shaped brane}
In this and next subsections, analyses similar to those in the previous subsection will be performed for different configurations Type II and III.

The boundary conditions of Type II are
\ba
z(0)=0,\quad \dot{z}(0)>0.
\ea
This solution is
\ba
z(r)=-\int_0^r\frac{P(r)r-\k}{\s{1-(P(r)r-\k)^2}}dr
\ea
where $\k\in(0,1]$ controls the gradient $\dot{z}(0)=\k/\s{1-\k^2}$ and $P(x)$ is arbitrary function satisfying
\ba
\quad\dot{P}(r)\geq-\frac{\k}{r^2},\quad |P(0)|<\inf.
\ea

\begin{figure}[H]
    \centering
    \includegraphics[width=.4\textwidth,page=1]{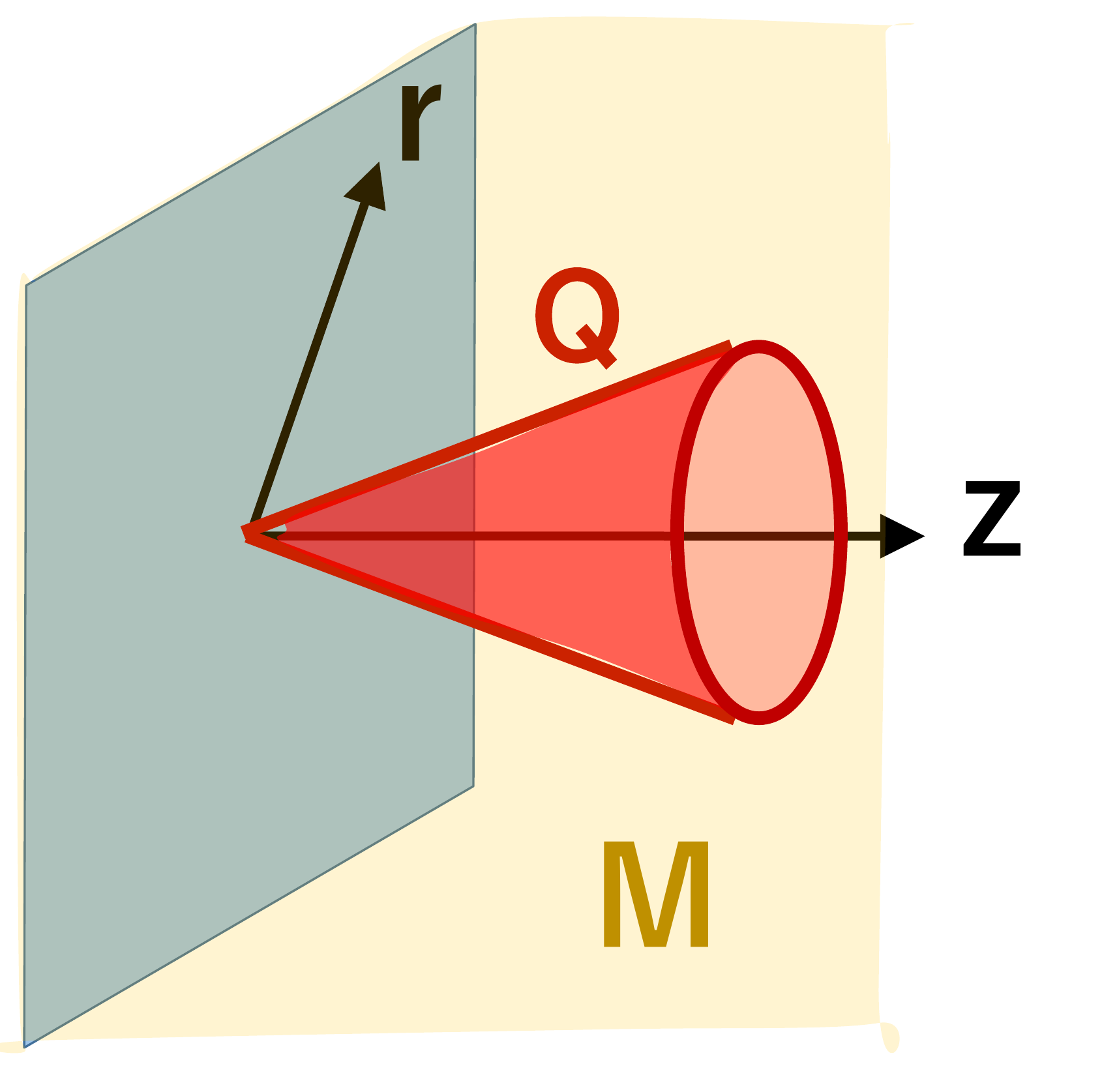}
    \caption{A sketch of a solution of AdS$_3/$BCFT$_2$ with a conical EOW brane. Note the gravity dual is the region outside of the cone.}
    \label{fig:AdSBCFTR}
\end{figure}

When $P(r)=0$, we can find a conical solution (Fig.\ref{fig:AdSBCFTR}):
\ba
z(r)=\frac{\k}{\s{1-\k^2}}r,\quad \phi ( r) =\frac{1}{\sqrt{2}\left( 1-\kappa ^{2}\right)^{1/4}}\log r,\quad V( \phi ) =\frac{-2+\kappa ^{2}}{2\sqrt{1-\kappa ^{2}}}.
\ea

In this case, the induced metric
\ba
ds^2=\frac{1}{\k^2}\frac{dr^2}{r^2}+\frac{1-\k^2}{\k^2}d\th^2
\ea
is not asymptotically AdS$_2$, but conformally flat because, by $r=e^{\s{1-\k^2}\r}$,
\ba
ds^2=\frac{1-\k^2}{\k^2}(d\r^2+d\th^2).
\ea
This means conical solution is not made by any boundary perturbations.

We see another example with $\k = 1$ and $P(r) = \l>0$. After some calculation, we get a configuration:
\ba\label{eq:semitorus}
&& z(r)=\frac{\s{\l r(2-\l r)}}{\l}, \no
&& \p(r) = \s{2}\left(\frac{2-\l r}{\l r}\right)^{1/4},\no
&& V(r)=-\frac{1}{2}\s{\frac{2-\l r}{\l r}}=-\frac{1}{4}\p^2.
\ea
This brane is half of a torus whose minor radius is 0, and $\p(r)$ and $V(r)$ diverge at $r=0$.

\begin{figure}[H]
    \centering
    \includegraphics[width=.5\textwidth,page=1]{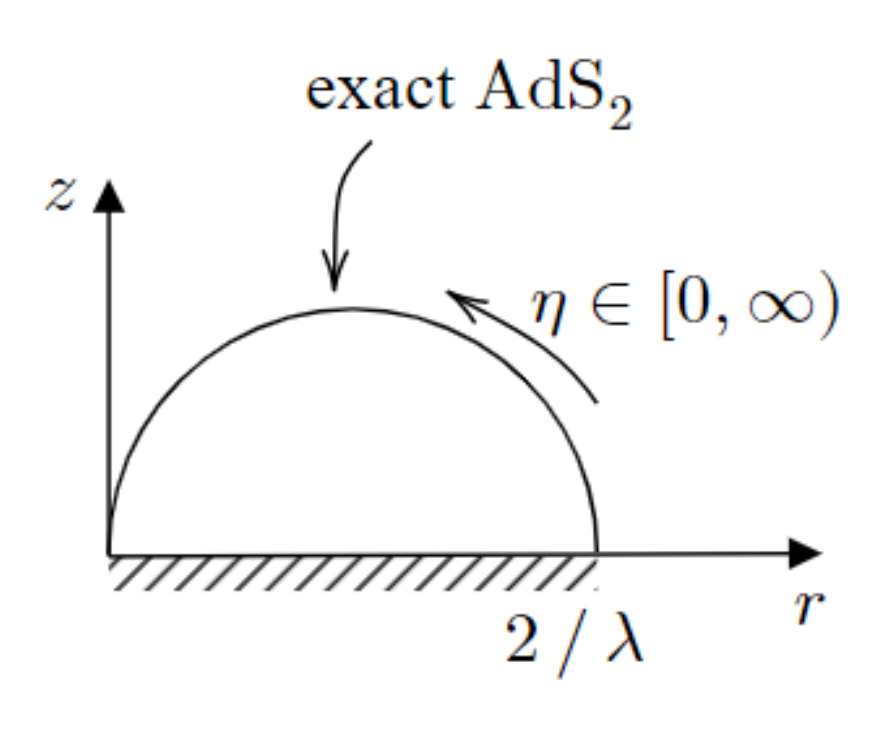}
    \caption{The configuration of \eqref{eq:semitorus}. The dual BCFT region is a disk with radius $2/\l$ removed its center and the induced metric on brane is exact AdS$_2$.}
    \label{fig:TypeII_circle_1}
\end{figure}

A geometry on this brane is exactly AdS$_2$ because the induced metric
\ba
ds^{2} =\frac{\l^{2} dr^{2} +( \l r)^{3}( 2-\l r) d\theta ^{2}}{( \l r)^{2}( 2-\l r)^{2}}
\ea
can be changed into
\ba
ds^2=\frac{d\h^2+d\theta^2}{\h^2}
\ea
by coordinates changing
\ba
r=\frac{2}{\l(1+\h^2)}.
\ea
This means $r=2/\l$, corresponding to $\eta=0$, is an asymptotic boundary of the brane but $r=0$ is deep interior.

Only for the boundary $r=2/\l$, we can apply calculation of conformal dimension of $\p$. Near the asymptotic boundary of the brane,
\ba
\phi ( r) =\sqrt{2}\left(\frac{2-\l r}{\l r}\right)^{1/4} \simeq \frac{2}{\l^{1/2}} \eta ^{1/2} -\frac{1}{\l^{1/2}} \eta ^{5/2}
\ea
This means that the conformal dimension of the dual scalar operator is $\D=1/2$ and the mass of the scalar field is $m^2=-1/4<0$.

\subsection{Type III, Annulus-like shaped brane}
The boundary conditions of the third case is
\ba\label{eq:third_condition}
r_0>0,\quad z(r_0)=0, \quad \dot{z}(r_0)>0
\ea
The Solution is
\ba
z(r)=\int_{r_0}^r\frac{A(r)r}{\s{1-A^2(r)r^2}}dr
\ea
and $A(r)$ satisfies
\ba\label{eq:third_constraint}
0<r_0 A(r_0)<1,\quad \dot{A}(r)\leq0.
\ea

For example, $A(r)=(2-r)/r$ and $r_0=1$ passes this conditions, the corresponding configuration is given by
\ba\label{eq:semi_torus}
z=\s{1-(r-2)^2},
\ea
or a half of torus whose minor radius is $r=1$ and major radius is $r=3$.

\begin{figure}[H]
    \centering
    \includegraphics[width=.4\textwidth,page=1]{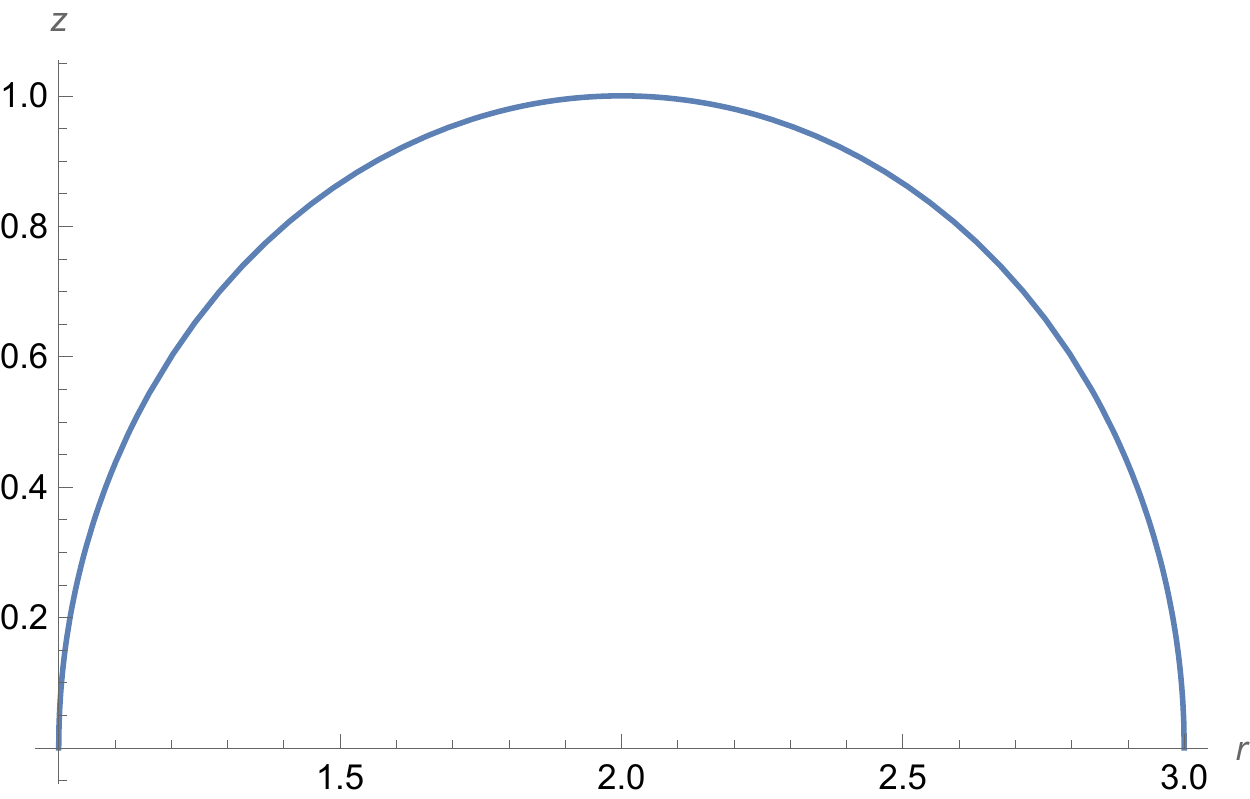}
 \includegraphics[width=.4\textwidth,page=1]{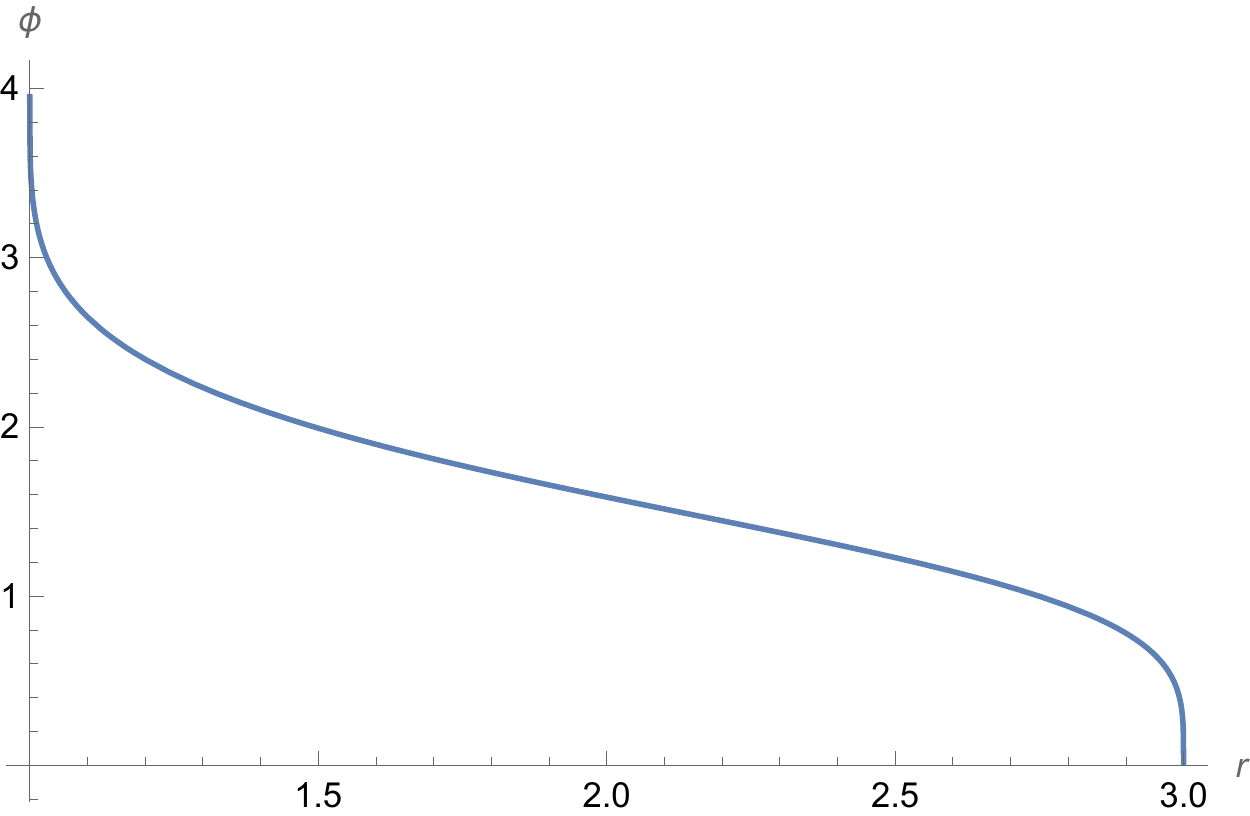}       \includegraphics[width=.4\textwidth,page=1]{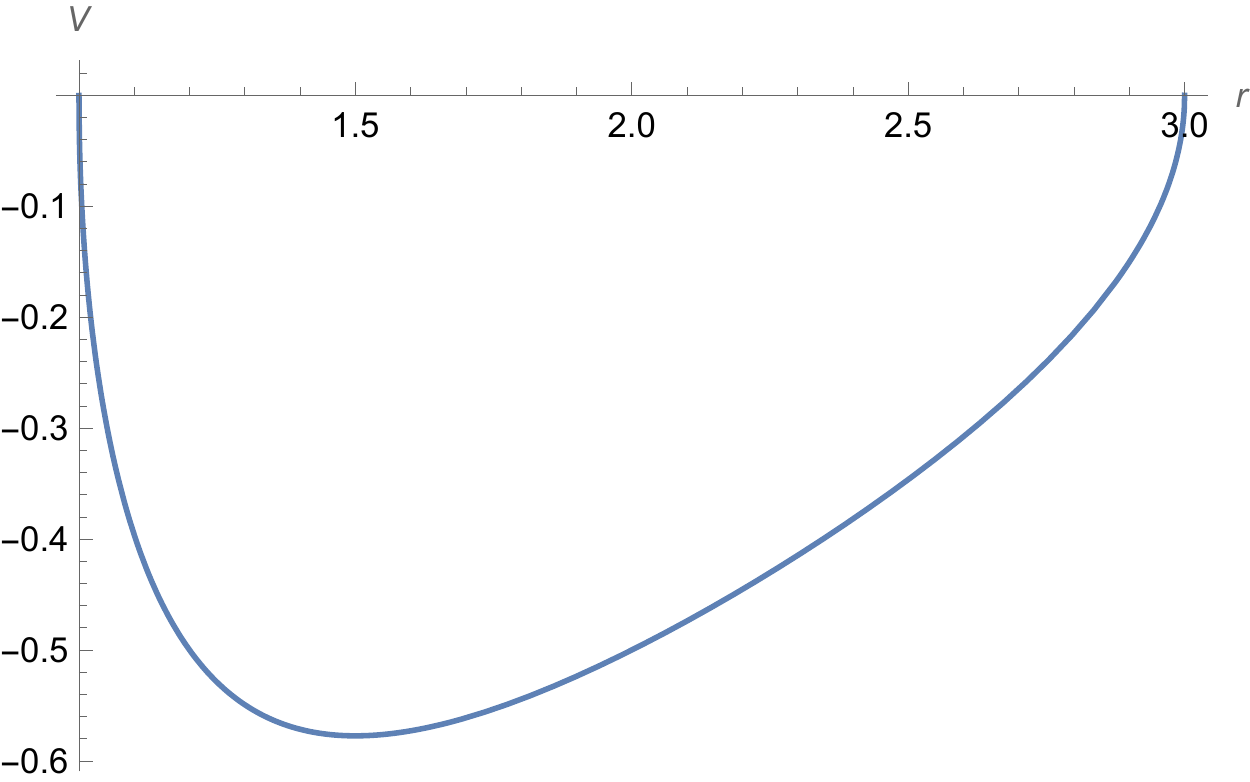}
    \includegraphics[width=.4\textwidth,page=1]{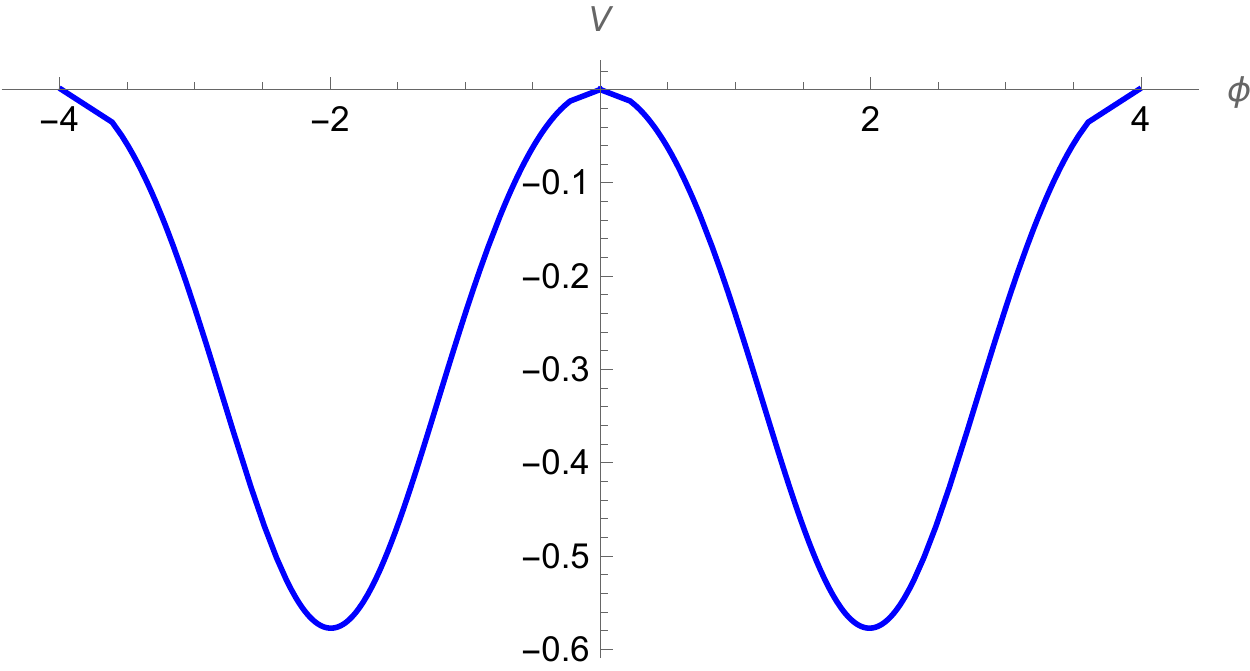}
 \caption{Configurations of half of a torus. The quadratic coefficient is -0.25 for both boundaries.}
    \label{fig:TypeIII_torus}
\end{figure}

This brane has two asymptotic boundaries. As in Type I, new coordinate $\h$ is introduced as follows:
\ba
r=g^{-1}(\h)=\frac{2\pm\s{1-3\h^2}}{1+\h^2}.
\ea
Near $r=1$, we choose negative sign, on the other hand, near $r=3$, positive sign.
Then induced metric for both boundaries becomes
\ba
ds^2=\frac{d\h^2}{\h^2-3\h^4}+\frac{d\th^2}{\h^2}.
\ea
$\p$ can also be expanded, for one boundary $r\simeq 1$,
\ba
\phi ( \eta ) =2 \eta ^{1/2} +\frac{3}{5} \eta ^{5/2} +O\left( \eta ^{9/2}\right),
\ea
and, for the other boundary $r\simeq 3$,
\ba
\phi ( \eta ) =-2 \eta ^{1/2} -\frac{3}{5} \eta ^{5/2} +O\left( \eta ^{9/2}\right),
\ea
where we ignore constant shift of $\p$.
Therefore dimension of dual operator is $\D=1/2<1$ and mass of scalar field is $m^2=-1/4<0$. This type of configuration is dual of Janus solution.

\begin{figure}[H]
    \centering
    \includegraphics[width=.4\textwidth,page=1]{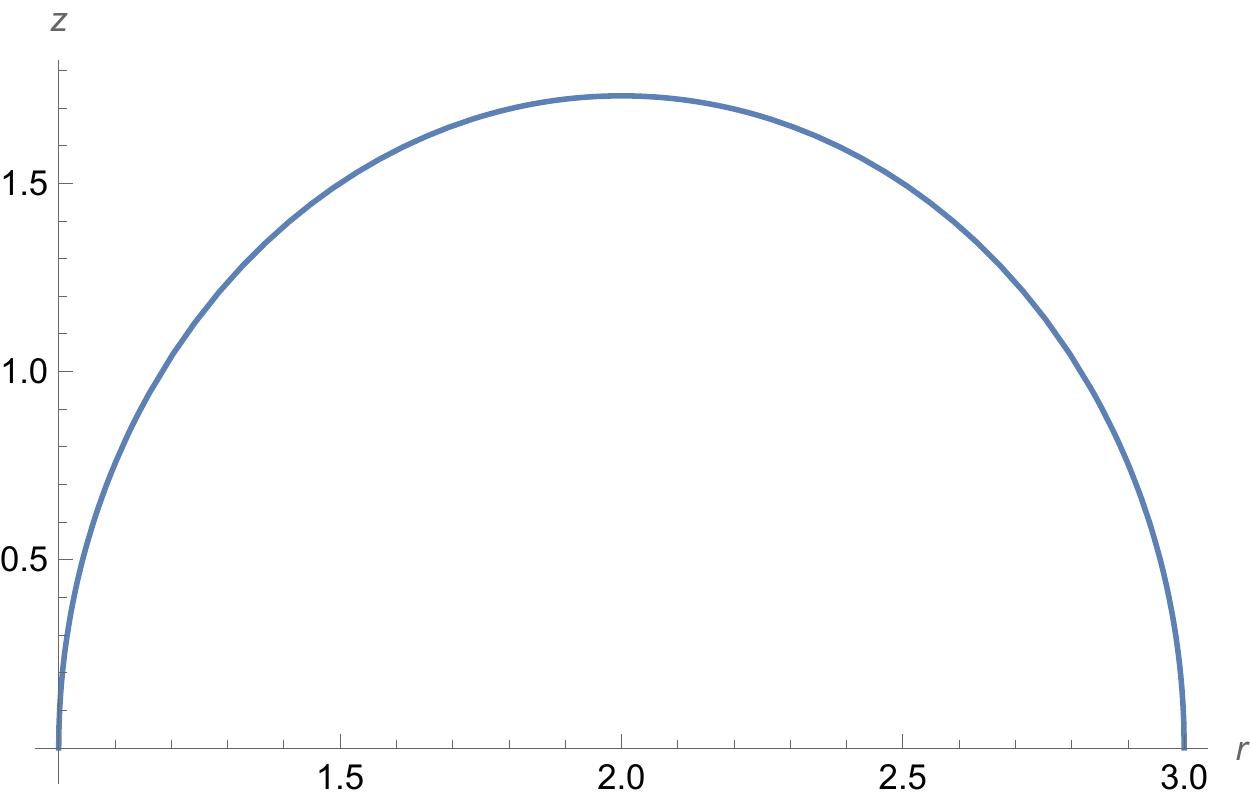}
 \includegraphics[width=.4\textwidth,page=1]{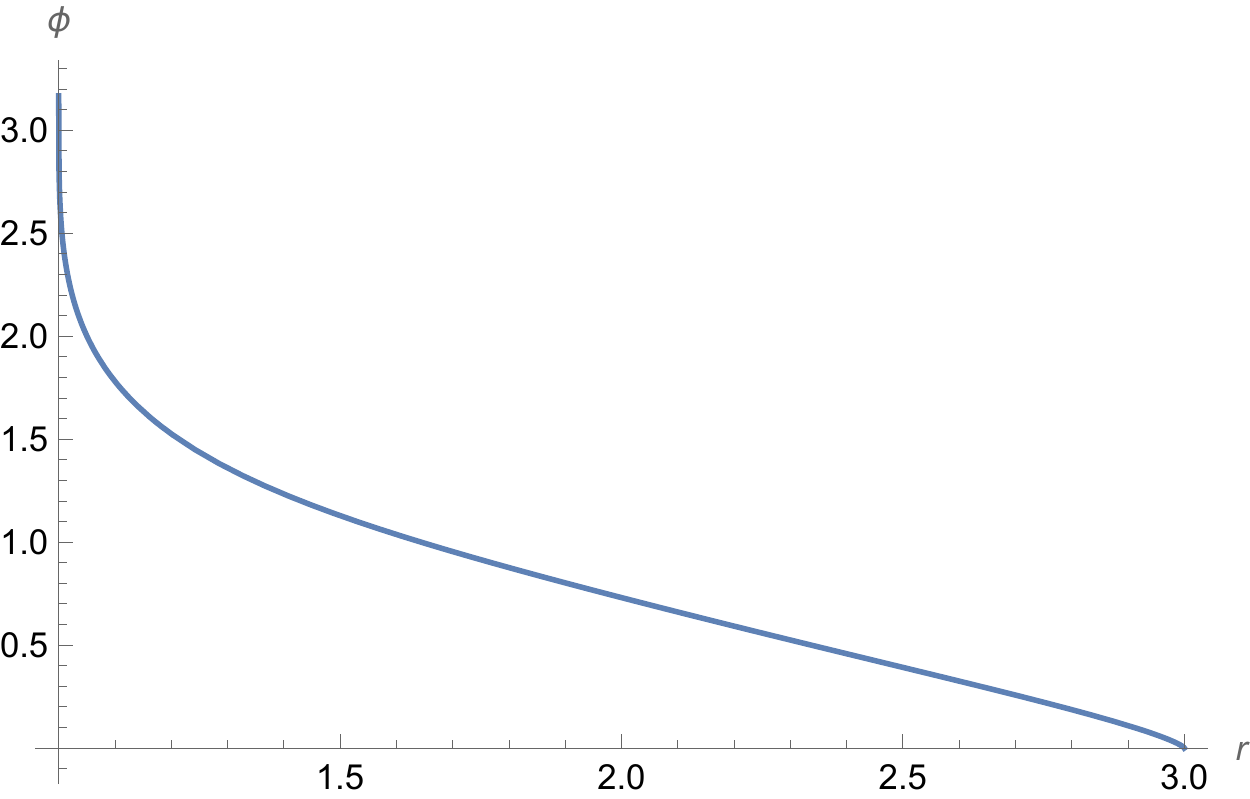}       \includegraphics[width=.4\textwidth,page=1]{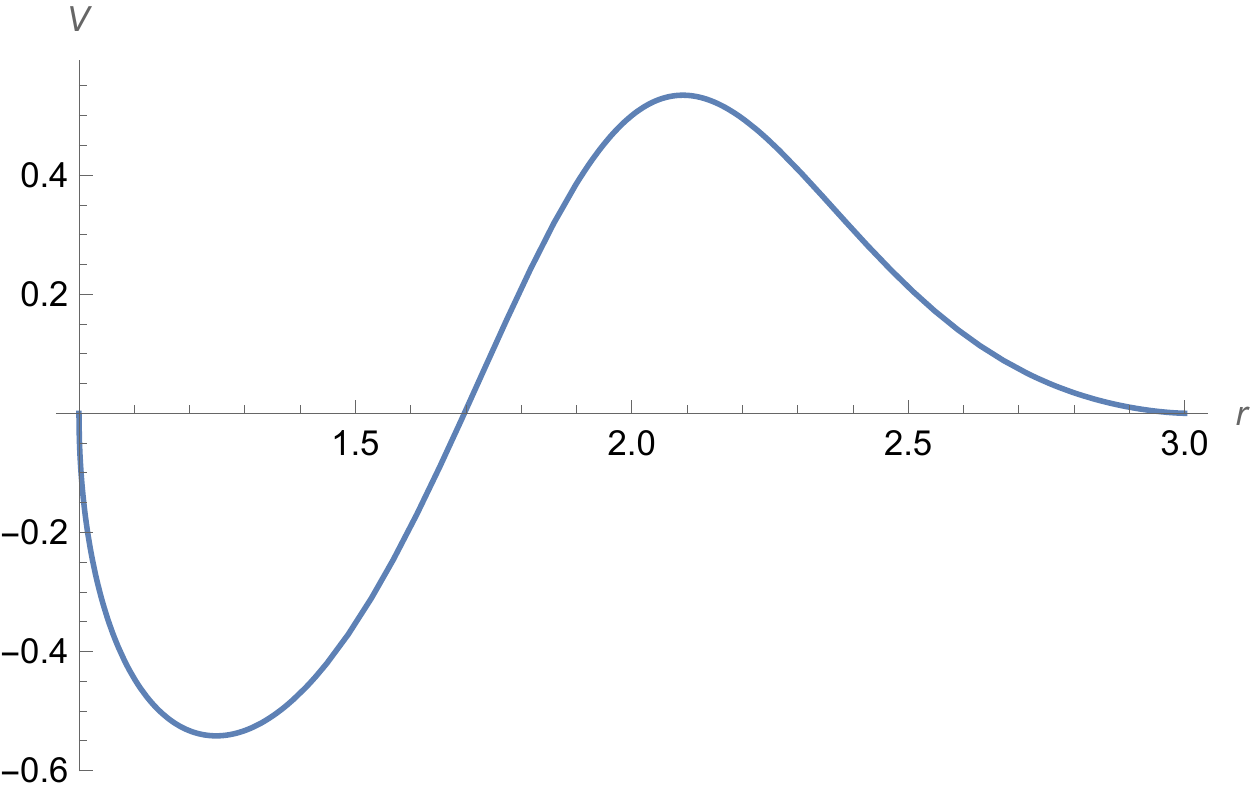}
    \includegraphics[width=.4\textwidth,page=1]{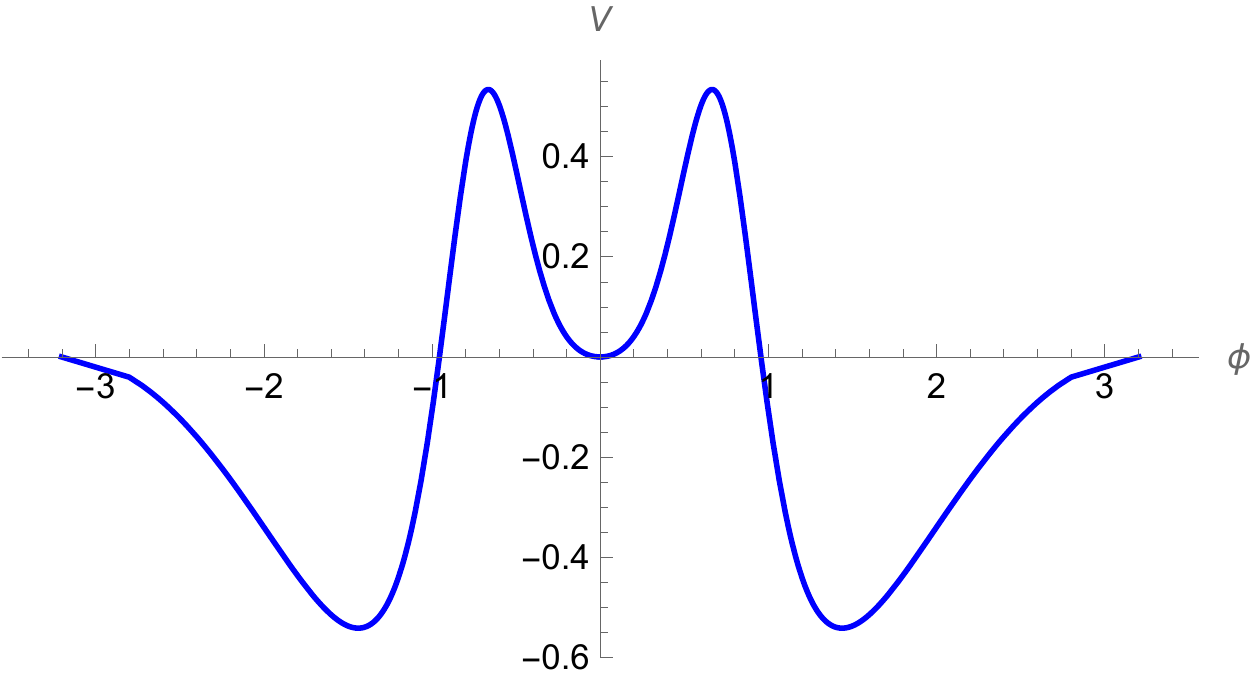}
 \caption{The $\s{3}$-stretched configuration of Fig.\ref{fig:TypeIII_torus}. The quadratic coefficients of $V(\p)$ are $-0.25$ at $r=1$ and $0.75$ at $r=3$.}
    \label{fig:TypeIII_torus2}
\end{figure}

By similar analysis in Type I, we can find that this configuration can be stretched by $\a\leq \s{3}$. When $\a=\s{3}$, at $r=1$, conformal weight is $\D=1/2$, but, at $r=3$, $\D=3/2$ (Fig.\ref{fig:TypeIII_torus2}). Hence, only the boundary at $r=3$ is irrelevant. Furthermore, for example, when $\a=1.732$, corresponding boundary RG flows are relevant for both boundary.

Next, we consider the case where brane does not intersect with the asymptotic boundary in $r>r_0$. We find that any function $z(r)$ satisfying $z(r_0)=0$, $\dot{z}(r)\geq0$ and $\ddot{z}(r)\leq0$ is a special solution of \eqref{eq:third_condition} because $A(r)=\dot{z}(r)/r\s{1+\dot{z}^2(r)}$ meets \eqref{eq:third_constraint}.

In the following we list the results for two example $z=r-1$ and $z=(r-1)^{2/3}$.

the former is an example where AdS$_2$ radius $L$ of asymptotic boundary of brane is $L\neq1$. The potential $V(\p)$ can be displayed analytically using $\p$ as follows:
\ba
&& \p(r) = 2^{3/4}\arctanh{\sqrt{\frac{r-1}{r}}},\no
&& V(r)=\frac{1-3r}{2\sqrt{2}r}=-\frac{2+\tanh^{2}\left( 2^{-3/4} \phi \right)}{2\sqrt{2}}=-\frac{1}{\s{2}}-\frac{1}{8}\p^2+O(\p^3).
\ea
Then the induced metric is
\ba
ds^2=\frac{2dr^{2} +r^{2} d\theta ^{2}}{( r-1)^{2}}=\frac{2d\eta ^{2}}{\eta ^{2}( \eta-1 )^{2}} +\frac{d\theta ^{2}}{\eta ^{2}},
\ea
where $\h=(r-1)/r$. Namely, the geometry of this brane is asymptotically AdS$_2$ with radius $L^2=2$ and
\ba
\phi ( \eta ) =2^{3/4} \eta ^{1/2} +\frac{2^{3/4}}{3} \eta ^{3/2} +O\left( \eta ^{5/2}\right).
\ea
Therefore conformal weight is consistent since
\ba
\D=\frac{1}{2}+\sqrt{\frac{1}{2^2}-\frac{1}{8}\cdot 2}=\frac{1}{2}<1.
\ea
In this way, the gradient of $z(r)$ at asymptotic boundary changes the boundary AdS radius $L$.

\begin{figure}[H]
    \centering
    \includegraphics[width=.4\textwidth,page=1]{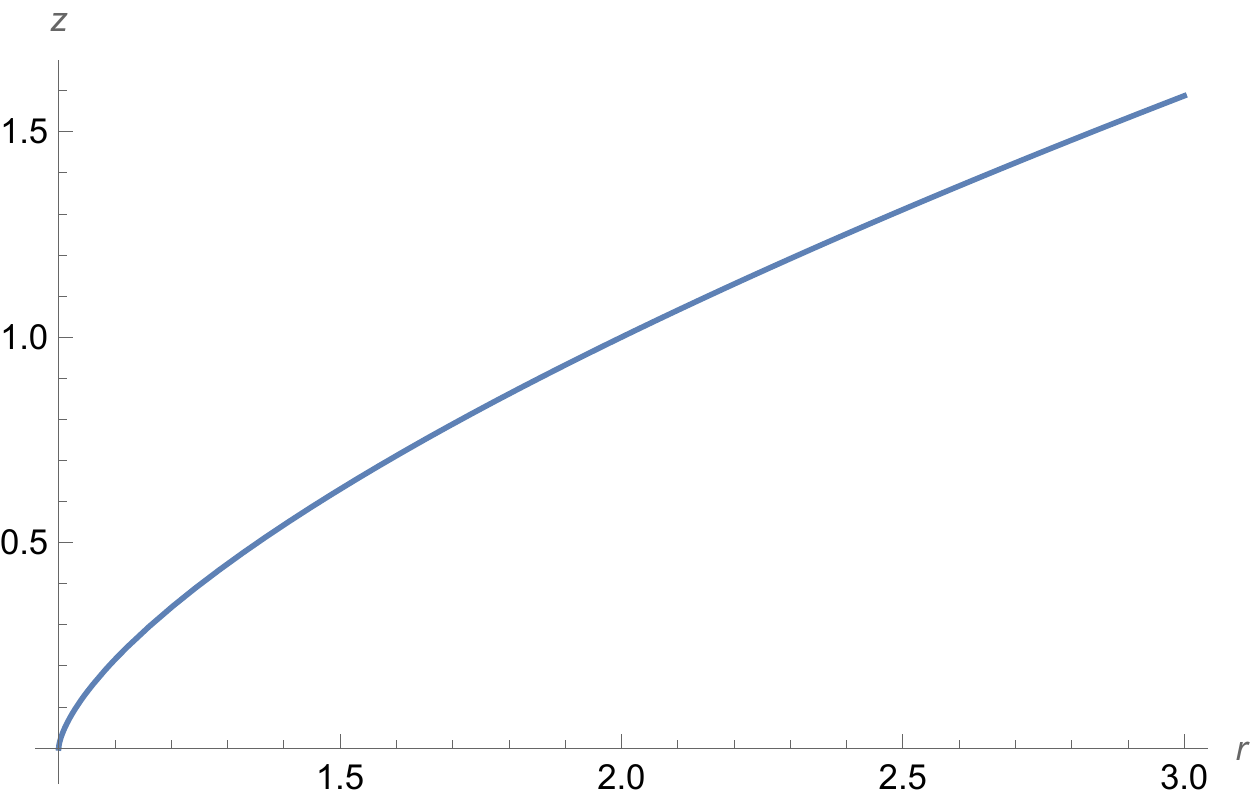}
 \includegraphics[width=.4\textwidth,page=1]{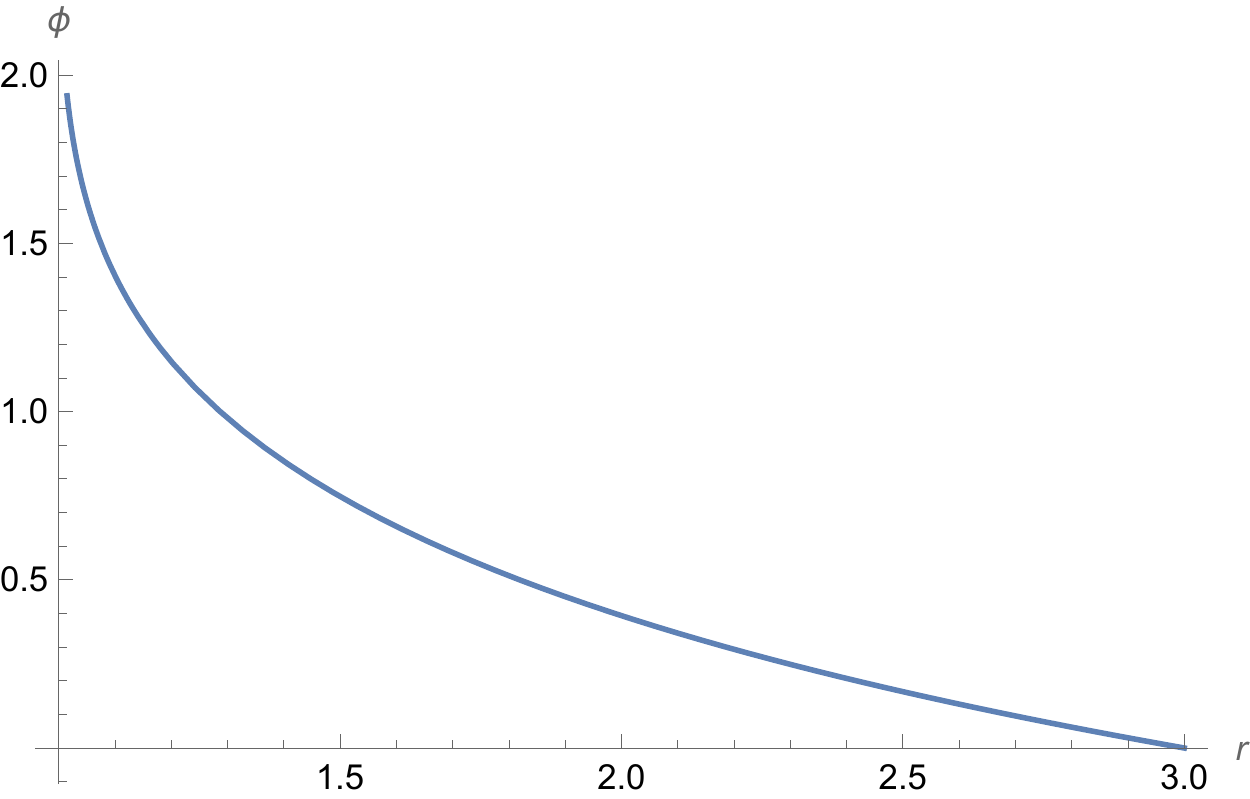}       \includegraphics[width=.4\textwidth,page=1]{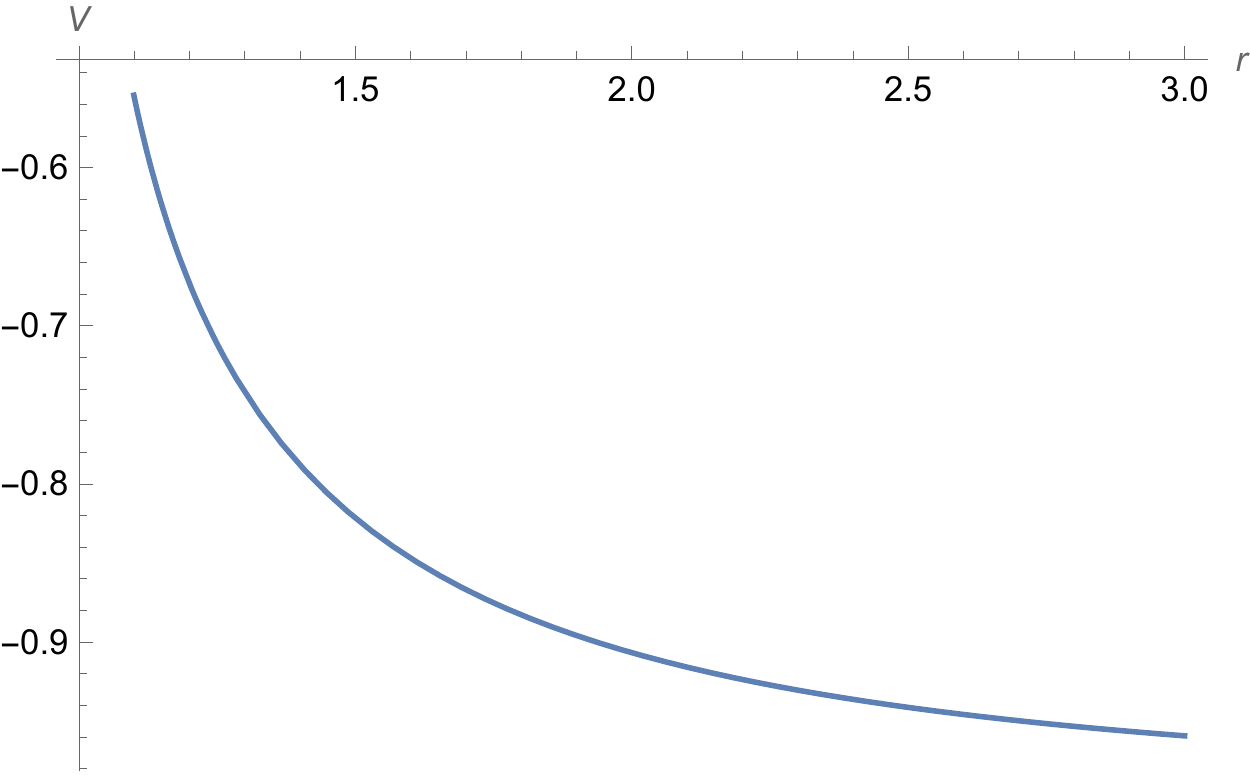}
    \includegraphics[width=.4\textwidth,page=1]{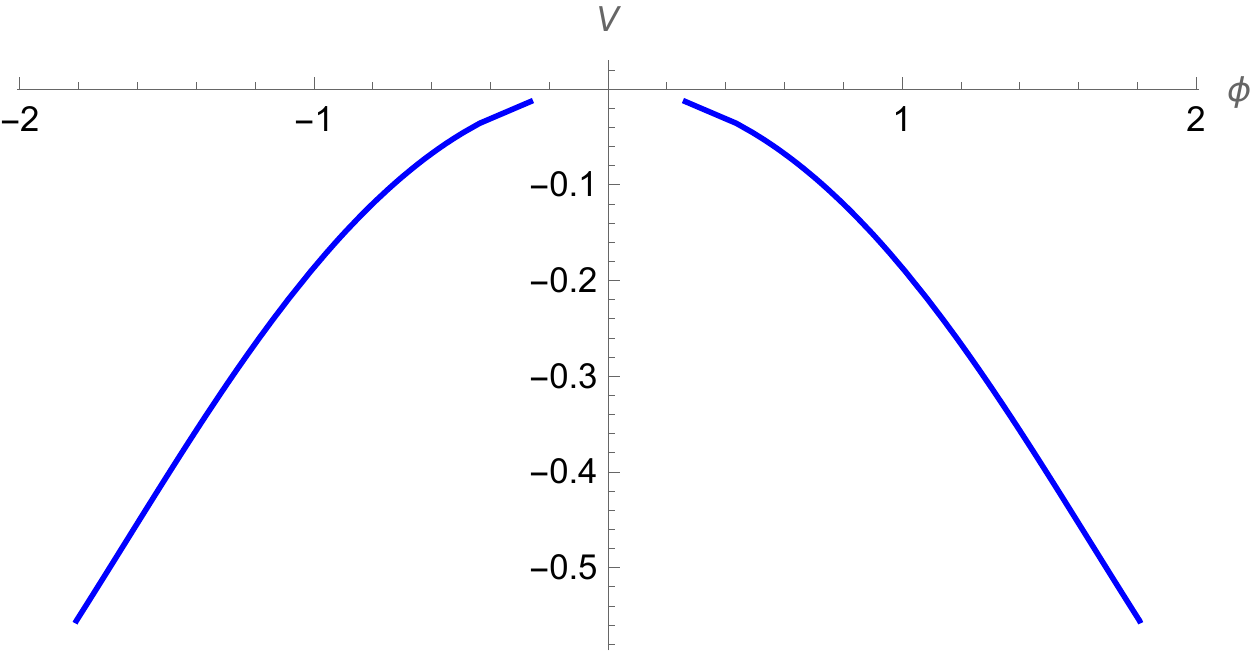}
 \caption{The profile of $z(r)=(r-1)^{2/3}$. In this way, when the series expansion of the inverse function $r(z)$ contains non-integer power term, the conformal dimension can be $\D\neq1/2$.}
    \label{fig:TypeIII_3}
\end{figure}

The latter is an example of $\D\neq 1/2$. Actually, the profile Fig.\ref{fig:TypeIII_3} are
\ba
&& h_{\h\h}=\frac{1}{\h^2}+O\left(\h^{-1}\right),\no
&& V(\p)=-\frac{3}{16}\p^2+O(\p^4),\no
&& \p(\h)=\sqrt{6} \eta^{1/4}+\frac{2}{3} \sqrt{\frac{2}{3}} \eta ^{3/4}+O\left(\h^{5/4}\right),
\ea
and we can read $m^2=-3/16$, $L^2=1$, and $\D=3/4$. Of course, these satisfy the relation $m^2L^2=\D(\D-1)$.

\subsection{Holographic entanglement entropy}
In this subsection, we calculate the entanglement entropy in this polar system in similar way as the previous chapter. 

The geodesic length \eqref{eq:geodesiclengthofAds} of AdS in polar coordinates is
\ba
D(P,P^\prime)=\cosh^{-1}\left(\frac{r^{2} +r^{\prime2} +z( r^\prime )^{2} -2r r^\prime \cos( \theta -\theta^\prime )}{2z( r^\prime ) \epsilon } +O( \epsilon )\right).
\ea
where $P=(\epsilon,r,\th)$ is a point in dual BCFT region and $P^\prime=(z(r^\prime),r^\prime,\th^\prime)$ is a point of the brane. Obviously, to minimize this length implies $\th=\th^\prime$. The derivative with respect to $r^\prime$ yields
\ba\label{eq:mininalconditionofzr}
2( r^\prime -r) z( r^\prime ) -\left(( r^\prime -r)^{2} -z( r^\prime )^{2}\right)\dot{z}( r^\prime ) =0.
\ea

\begin{figure}[H]
    \centering
    \includegraphics[width=.4\textwidth,page=1]{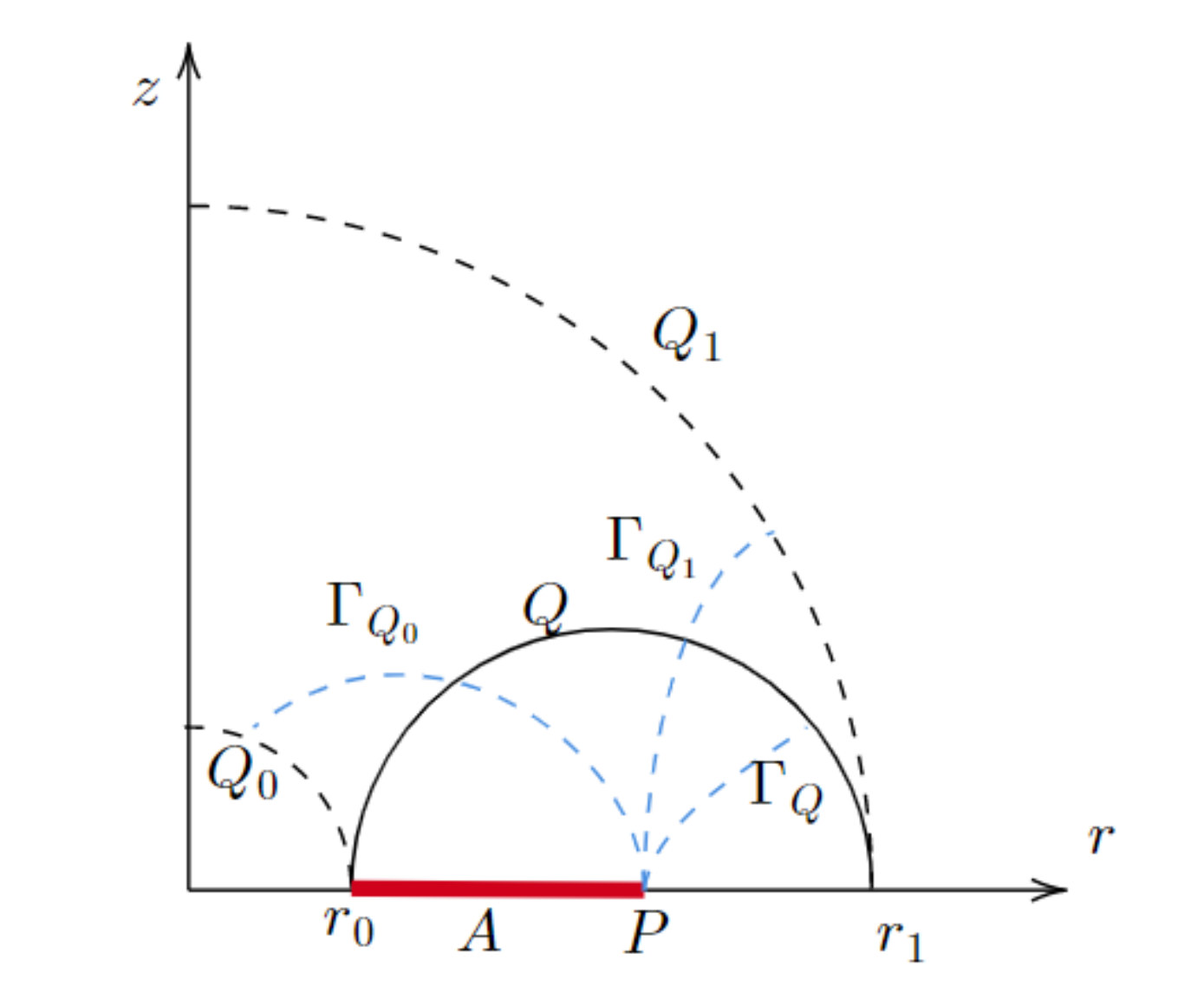}
 \caption{The branes $Q_0$ and $Q_1$ are the solution without perturbation and exact hemisphere. the blue dashed lines represent geodesic connecting a point $P$ and branes.}
    \label{fig:PolorEntropy}
\end{figure}

Specially, we calculate the entropy when the dual BCFT region is annulus $[r_0,r_1]$ and subregion $A$ is a interval $[r_0,\ell+r_0]$, i.e., $P=(\epsilon,\ell+r_0,\th)$ like Fig.\ref{fig:PolorEntropy}. Also, the geodesic connecting $P$ and brane $Q$ is denoted as $\G_Q$. If BCFT has no boundary perturbation, the shape of the brane is an exact hemisphere. Therefore, the entropy change $\D S$ of this RG flow is computed by
\ba
\D S=\frac{c}{6}\operatorname{Area}(\G_Q)-\frac{c}{6}\min\{\operatorname{Area}(\G_{Q_0}),\operatorname{Area}(\G_{Q_1})\}.
\ea
For the branes $Q_0$ and $Q_1$, solving \eqref{eq:mininalconditionofzr}, we obtain
\ba
r^\prime=\frac{(r_0+\ell)r_i^2}{(r_0+\ell)^2+r_i^2},
\quad\operatorname{Area}(\G_{Q_i})=\log\left| \frac{(r_0+\ell)^{2} -r_i^{2}}{r_i\epsilon }\right|,\quad i=0,1.
\ea
Hence, for example, when the brane $Q$ is simply $z=\sqrt{R^2-(r-r_c)^2}$, $R=(r_1-r_0)/2$, and $r_c=(r_1+r_0)/2$, we find
\ba
r^\prime =\frac{r_{c} (r_{c} -( r_{0} +\ell) )^{2} -R^{2} (r_{c} -2( r_{0} +\ell) )}{(r_{c} -( r_{0} +\ell) )^{2} +R^{2}},\quad \operatorname{Area}(\G_Q)=\log\frac{R^{2} -( r_{c} -( r_{0} +\ell ))^{2}}{R\epsilon }.
\ea

\begin{figure}[H]
    \centering
    \includegraphics[width=.5\textwidth,page=1]{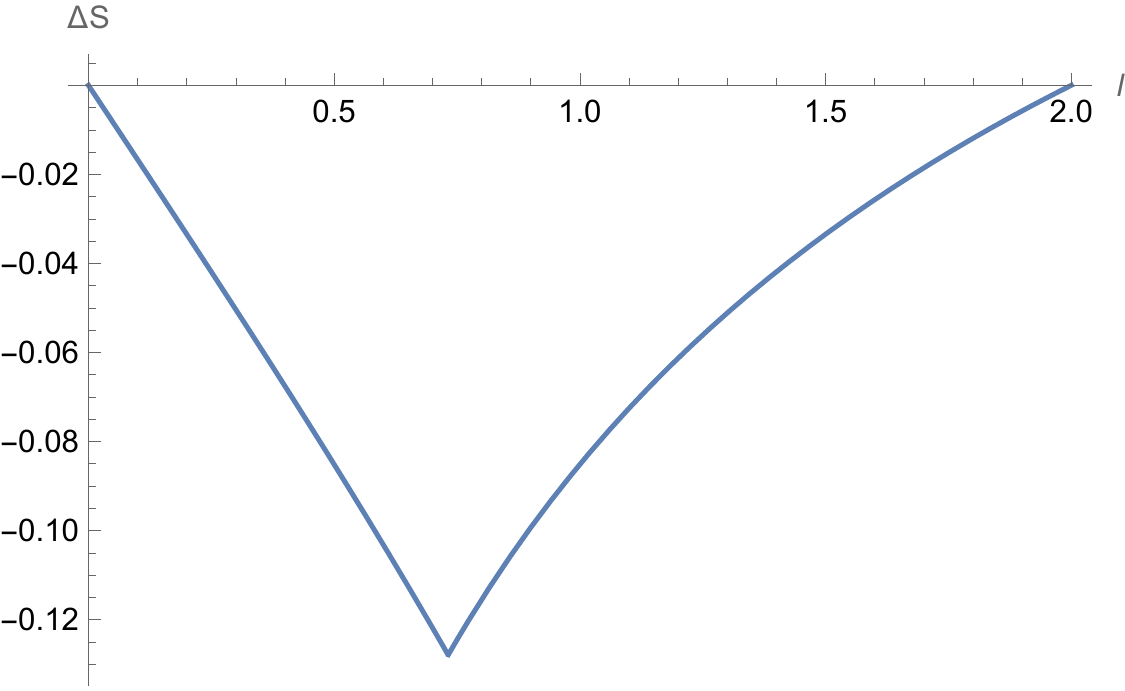}
 \caption{The horizontal axis is $\ell$ and the vertical axis is $\D S$ where $r_0=1$, $r_1=3$, and $c=1$. We can see a phase transition by comparison of $\G_{Q_0}$ and $\G_{Q_1}$.}
    \label{fig:BoundaryEntropyOfPolarBrane}
\end{figure}

As we can see in Fig.\ref{fig:BoundaryEntropyOfPolarBrane}, $\D S$ vanishes at $\ell = 0, 2R$ and decreases as the brane $Q$ flows to deep interior. These mean also that the corresponding RG flow is relevant.

\subsection{Boundary entropy in the presence of brane-localized scalar}
In this section, we calculate boundary entropy for simple case. Consider a case where BCFT region is $0\leq r\leq 1$ and brane $Q_\a$ is $z=\a \s{1-r^2},(0\leq\a\leq 1)$. If $\a=1$, brane $Q_1$ is a tension-less and has no matter part. Therefore, the boundary entropy $S_{\rm bdy}$ can be calculated by $S_{\rm bdy}=I_1-I_\a$, where
\ba
I_\a=-\frac{1}{16\pi G_N}\int_{M_\a}\s{g}(R+2)-\frac{1}{8\pi G_N}\int_{Q_\a}\s{h}\left(K-h^{ab}\partial_a\p\partial_b\p-V(\p)\right)
\ea
and $M_\a$ is gravitational dual region. After some calculation, we get
\ba
I_\a=\frac{1}{4\pi G_N}\int_{M_\a}\frac{r}{z^3}dzdrd\th+\frac{1}{8\pi G_N}\frac{1-\a^2}{\a^2}\int_{Q_\a}\frac{r}{1-r^2}drd\th.
\ea
Since this integral diverges, we introduce cutoff $z>\e$. Then the integral have finite value
\ba
I_\a=\frac{1}{4\pi G_{N}}\frac{\pi }{\epsilon ^{2}} -\frac{\alpha ^{2}}{8G_{N}} +\frac{1}{4G_{N}}\ln( \epsilon \alpha ).
\ea
Finally the boundary entropy is
\ba
S_{\rm bdy} =-\frac{1}{4G_{N}}\left(\ln \alpha +\frac{1}{2}\left( 1-\alpha ^{2}\right)\right) =-\frac{c}{6}\left(\ln \alpha +\frac{1}{2}\left( 1-\alpha ^{2}\right)\right) \leq 0.
\ea

\section{Analysis of higher dimensional AdS/BCFT}

Next we consider extensions of the results in previous sections to 
higher dimensional AdS/BCFT setups. 

\subsection{Planar branes}
We assume the bulk metric is Poincar\'{e} AdS$_{d+1}$:
\ba
ds^2=\frac{dz^2+d\tau^2+dx_1^2+\cdots+dx_{d-1}^2}{z^2}.
\ea
We specify the shape of the EOW brane $Q$ and the brane-localized scalar field on $Q$ by
\ba
z=z(\tau),\ \ \ \  \phi=\phi(\tau).
\ea
The induced metric on $Q$ reads
\ba
ds^2=\frac{(1+\dot{z}^2)d\tau^2+dx_1^2+\cdots+dx_{d-1}^2}{z^2}.
\ea
The normal vector of the EOW brane is
\ba
(N^z,N^\tau,N^{x_1},\cdots,N^{x_{d-1}})=-\frac{z}{\sqrt{1+\dot{z}^2}}(-1,\dot{z},0,\cdots,0).
\ea
Then, the extrinsic curvature satisfies
\ba
K_{\tau\tau}-h_{\tau\tau}K=\frac{(d-1) \sqrt{1+\dot{z}^2}}{z^2},\\
K_{x_i x_i}-h_{x_i x_i}K=\frac{(d-1)(1+\dot{z}^2)z\ddot{z}}{z^2 \left(\dot{z}^2+1\right)^{3/2}}.
\ea
By plugging them into\eqref{Nbc}, we get
\ba
\dot{\phi}^2=-\frac{\ddot{z}}{2z \sqrt{1+\dot{z}^2}},\\
V(\phi)=\frac{z^2 \dot{\phi}^2}{1+\dot{z}^2}-\frac{d-1}{\sqrt{1+\dot{z}^2}}.
\ea
By eliminating $\dot{\phi}$ from $V(\phi)$, we can express $V(\phi)$ as
\ba
-V(\phi)=\frac{2(d-1)(1+\dot{z}^2)+z\ddot{z}}{2(1+\dot{z}^2)^{3/2}}
\ea
We consider $V(\phi)=0$ case and then we get $2(d-1)(1+\dot{z}^2)+z\ddot{z}=0$.We can solve this equation by setting $z=y^{\frac{1}{2d-1}}$. Then $V(\phi)=0$ reads to
\ba
\ddot{y}=-2(d-1)(2d-1)y^{1-\frac{2}{2d-1}}.
\ea
By solving this equation, we finally find
\ba
1+\dot{z}^2=\frac{z_0^{4(d-1)}}{z^{4(d-1)}},\label{solda}
\ea
where $z_0$ is an integration constant i.e. the value of $z$ of turning point. According to this, we have
\ba
\ddot{z}=-\frac{2(d-1)z_0^{4(d-1)}}{z^{4d-3}},\\
\dot{\phi}^2=(d-1)\frac{z_0^{2(d-1)}}{z^{2d}} \label{soldb}
\ea
Around $z=0$, these behave as
\ba
&& z\approx\qty((2d-1)z_0^{2(d-1)}\tau)^{\frac{1}{2d-1}},\\
&& \phi\approx\phi_0+\frac{1}{\sqrt{d-1}}\qty(\frac{(2d-1)\tau}{z_0})^{\frac{d-1}{2d-1}}=\phi_0+\frac{1}{\sqrt{2d-1}}\qty(\frac{z}{z_0})^{d-1}.
\ea
Near $z=0$, the geometry is approximated by AdS $_d$
\ba
ds^2=\frac{dz^2+dx_1^2+\cdots+dx_{d-1}^2}{z^2}
\ea
and $z$ dependence of $\phi$ agrees with the marginal deformation $\phi=z^{d-1-\Delta}J+z^{\Delta}\expval{O},\ \Delta=d-1$ in the standard dictionary of AdS/CFT \cite{Gubser:1998bc,Witten:1998qj,Klebanov:1999tb}.

It is also useful to evaluate the on-shell action namely the free energy as an exercise of our later analysis of phase transition in section \ref{sec:phasetr}. The total action is given by $I$ in (\ref{totac}), namely,
\ba
I\!=\!-\!\frac{1}{16\pi G_N}\int_{M}\!\s{g}\left(R\!+\!d(d-1)\right)\!-\!\frac{1}{8\pi G_N}\int_\Sigma\!\s{h}K\!-\!\frac{1}{8\pi G_N}\int_Q\!\s{h}\left(K\!-\!h^{ab}\de_a\phi\de_b\phi\!-\!V(\phi)\right)\!+\!I_{c.t.},\nonumber
\ea
where we added the counter term $I_{c.t.}$.

We would like to evaluate the on-shell action for
the solution (\ref{solda}) and (\ref{soldb}) with the UV cut off 
$z\geq \ep$. We note that the extrinsic curvature takes the value $K|{\Sigma}=d$ on $\Sigma$ and $K|_Q=\frac{d-2}{d-1}h^{ab}\de_a\phi\de_b\phi$.
Then the action is evaluated as follows:
\ba
I&=&\frac{d}{8\pi G_N}-\frac{d}{8\pi G_N}\int_{\Sigma} \s{h}+\frac{1}{8\pi G_N(d-1)}\int \s{h}h^{ab}\de_a\phi\de_b\phi+I_{c.t.}\no
&=&\frac{d}{8\pi G_N}\int \frac{d^dxdz}{z^{d+1}}-\frac{d}{8\pi G_N\ep^{d}} 
+\frac{1}{8\pi G_N}\int \frac{d^dx}{z^d}\cdot \left(\frac{z_0}{z}\right)^{2(d-1)}\cdot \frac{1}{\s{1+\dot{z}^2}}+I_{c.t.}\no
&=&\frac{1}{8\pi G_N}\int d^dx\left(-\frac{1}{z^d}+\frac{1}{z^d}\right)=0,
\label{totovan}
\ea
where the counter term $I_{c.t.}$ removes all divergent terms proportional to 
$1/\ep^d$. In this way, we find that the total action or free energy vanishes. 
In section \ref{sec:phasetr}, we will extend this analysis at zero temperature to that at finite temperature to study phase transitions, focusing on the $d=2$ case.

\subsection{Round-shaped branes}
The polar AdS$_{d+1}$ metric is
\ba
ds^2=\frac{dz^{2} +dr^{2} +r^{2} d\O_{S^{d-1}}^{2}}{z^{2}}
\ea
where $d\O_{S^{d-1}}^{2}=\tilde{g}_{ij}d\th^id\th^j$ is a metric of $S^{d-1}$. We assume the shape of EOW brane and the scalar field depends on $r$
\ba
z=z(r),\quad \phi=\phi(r).
\ea
The induced metric reads
\ba
ds^2=\frac{(1+\dot{z}^2)dr^2+r^{2} d\O_{S^{d-1}}^{2}}{z^2}.
\ea
The normal vector of the EOW brane is
\ba
(N^z,N^r,N^{\th^1},\dots,N^{\th^{d-1}})=\frac{z}{\sqrt{1+\dot{z}^2}}(1,-\dot{z},0,\dots,0).
\ea
Then, we get
\ba
K_{rr}-h_{rr}K=( d-1)\frac{( r+z\dot{z})\sqrt{1+\dot{z}^{2}}}{rz^{2}},\no\\
K_{ij}-h_{ij}K=\frac{r^{2}\tilde{g}_{ij}}{z^{2}\sqrt{1+\dot{z}^{2}}}\left( d-1+\frac{z\ddot{z}}{1+\dot{z}^{2}} +( d-2)\frac{z\dot{z}}{r}\right).
\ea
By combining \eqref{Nbc} with them,we get
\ba
\dot{\phi }^{2} =\frac{\dot{z}}{2rz}\sqrt{1+\dot{z}^{2}} -\frac{\ddot{z}}{2z\sqrt{1+\dot{z}^{2}}},\no\\
V(\phi)=-\frac{d-1}{\sqrt{1+\dot{z}^{2}}} -\frac{( 2d-3) z\dot{z}}{2r\sqrt{1+\dot{z}^{2}}} -\frac{z\ddot{z}}{2\left( 1+\dot{z}^{2}\right)^{3/2}},
\ea
and the expression of $\dot{\phi}^2$ does not depend on the dimension $d$. Because the shape of brane is determined by the positivity condition of $\dot{\phi}^2$ in polar case, we can apply the same discussion as $d=3$ and only the mass and the conformal dimension are modified by potential.

\section{\texorpdfstring{AdS$_3$/BCFT$_2$}{AdS3/BCFT2} with brane-localized scalar at finite temperature and phase transition}\label{sec:phasetr}

In the standard AdS/CFT, we expect the Hawking-Page phase transition from the thermal AdS phase to AdS black hole phase as we increase the temperature \cite{Hawking:1982dh,Witten:1998qj,Witten:1998zw}, which is interpreted as the confinement/deconfinement phase transition.
In the AdS/BCFT, an analogous phase transition was found in \cite{Takayanagi:2011zk,Fujita:2011fp} for AdS$_3/$BCFT$_2$ without any matter fields. The purpose of this section is to extend this analysis to  AdS$_3/$BCFT$_2$ with a brane-localized scalar field.
We will set the potential to zero $V(\phi)=0$. Thus we have a massless scalar on the EOW brane. 

\begin{figure}[H]
    \centering
    \includegraphics[width=.3\textwidth,page=1]{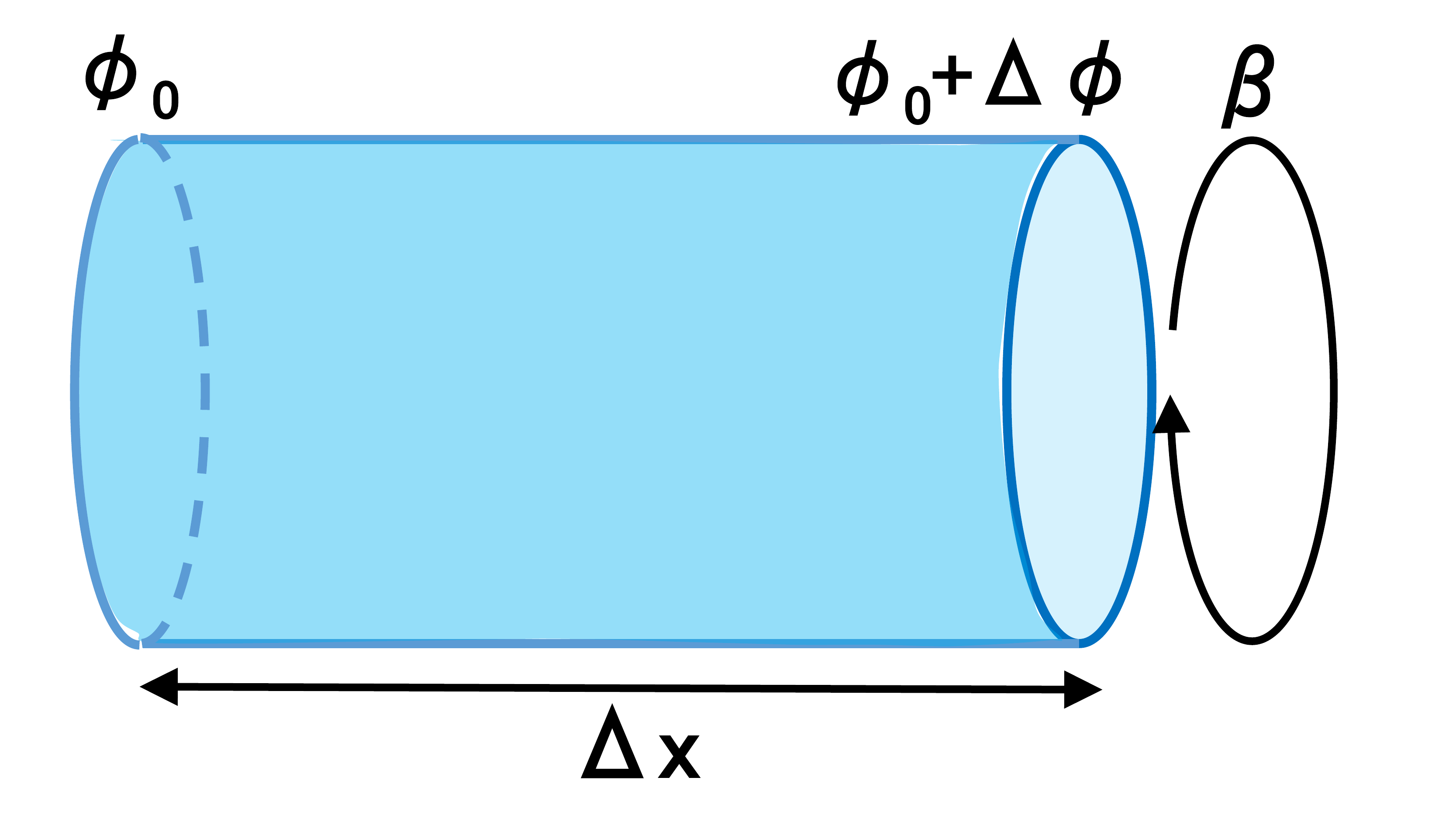}
    \includegraphics[width=.5\textwidth,page=1]{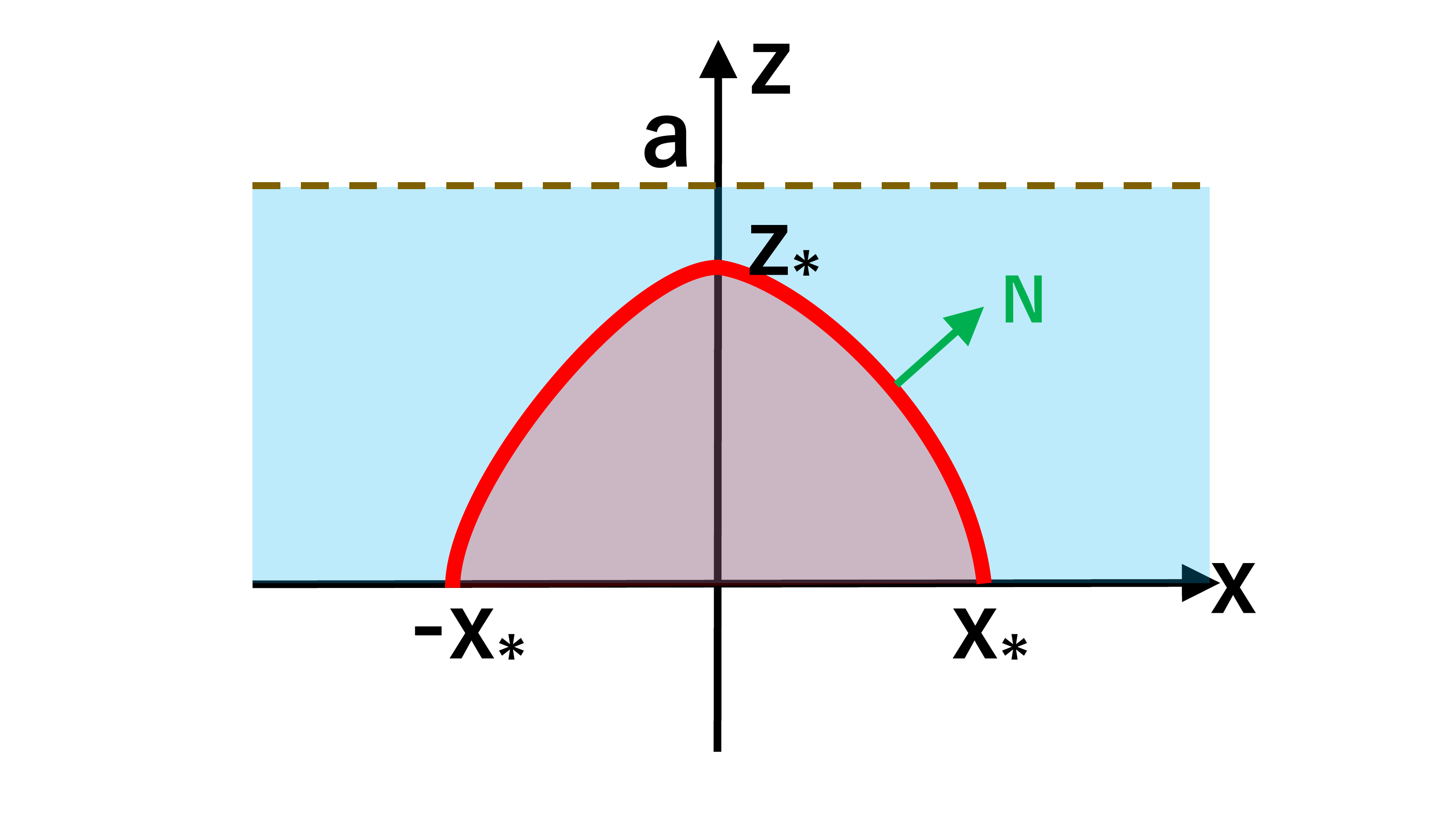}
    \caption{The cylinder geometry where the dual BCFT$_2$ is defined (left) and the profile of connected EOW brane in thermal AdS$_3$/BTZ black hole (right).}
    \label{fig:TAdSEOW}
\end{figure}

The BCFT$_2$ is defined on a cylinder, which is described by the coordinate $(\tau,x)$ such that
\ba
-x_*\leq x\leq x_*,\  \ \ \ \tau\sim\tau+\beta,
\ea
where $\Delta x=2x_*$ is the width of the cylinder and $\beta$ is the inverse temperature. See the left panel of Fig.\ref{fig:TAdSEOW} for a sketch. This is dual to a region surrounded by the EOW brane whose profile is specified by $z=z(x)$, assuming the translational invariance in the Euclidean time $\tau$ direction. There are two candidates of the bulk metric, namely the thermal AdS$_3$ and the BTZ metric.\footnote{
One may wonder if other bulks solutions give the brane solutions with the required boundary condition. However, this is not possible and we do not need to worry this. This is shown as follows. The general bulk solution to the vacuum Einstein equation with a negative cosmological constant is characterized by energy flux $T_{++}(x^+)$ and $T_{--}(x^-)$ as in the Banados's solution \cite{Banados:1998gg,Roberts:2012aq}. However, here we require the translation invariance in the time direction as we are interested in a static AdS/BCFT setup. Then the energy flux should take a constant value. Thus this leads to either thermal AdS or BTZ, depending on whether the energy density is positive or negative.} In addition we turn on the massless scalar field $\phi$ on the EOW brane, whose configuration is expressed as $\phi=\phi(x)$.
Below we write the derivative of $f$ w.r.t $x$ as $\dot{f}$.

In this model, the only input parameters are two dimensionless quantities:
the temperature times the width and the difference of the scalar field 
\ba
\frac{\Delta x}{\beta},\ \ \ \ \Delta\phi\equiv \phi(x_*)-\phi(-x_*),
\label{twopara}
\ea
owing to the conformal invariance.
 Once these two values are given, we can find a unique solution which satisfies the equation of motion and the boundary condition in the AdS/BCFT. In case we have multiple solutions we pick up the one with the smallest free energy. Below we will study the thermal AdS$_3$ solution and the BTZ one in detail. Then we will work out the phase structure for non-zero $\beta$ and $\Delta\phi$.

\subsection{Thermal \texorpdfstring{AdS$_3$}{AdS3} Solution}

Consider the thermal AdS$_3$ solution:
\ba
&& ds^2=\frac{d\tau^2}{z^2}+\frac{dz^2}{h(z)z^2}+\frac{h(z)dx^2}{z^2},\no
&& h(z)=1-\frac{z^2}{a^2}.
\ea
In the absence of EOW brane, we need to compactify the $x$ coordinate such that $x\sim x+2\pi a$ to avoid the conical singularity. However, in AdS/BCFT, we do not need to worry about the singularity if it is outside of 
the considered region surrounded by the EOW brane. Notice also that in the absence of the brane-localized scalar, the Neumann boundary condition fixes the width of the interval to be $\Delta x=2x_*=\pi a$ \cite{Takayanagi:2011zk,Fujita:2011fp}. However this is no longer true in our model when the scalar field $\phi$ is turned on.

The induced metric reads
\ba
ds^2=\frac{d\tau^2}{z^2}+\frac{h^2+\dot{z}^2}{hz^2}dx^2,
\ea
where we write $\frac{dz}{dx}$ as $\dot{z}$.
The profile of EOW brane is sketched in Fig.\ref{fig:TAdSEOW}.

The normal vector of EOW brane reads
\ba
(N^\tau,N^z,N^x)=\frac{z}{\s{h+\dot{z}^2/h}}(0,h,-\dot{z}/h).
\ea
The Neumann boundary condition (\ref{Nbc}) leads to
\ba
&& K_{\tau\tau}-Kh_{\tau\tau}=-\frac{V(\phi)}{z^2}
-\frac{h}{h^2+\dot{z}^2}\dot{\phi}^2,\no
&& K_{xx}-Kh_{xx}=\dot{\phi}^2-\frac{h^2+\dot{z}^2}{hz^2}V(\phi).
\ea
In our brane profile we can calculation the extrinsic curvature as follows:
\ba
&& K_{\tau\tau}-Kh_{\tau\tau}
=\frac{2h^3-zh^2h'-3zh'\dot{z}^2+2h\dot{z}^2+2hz\ddot{z}}{2z^2h\left(h+\frac{\dot{z}^2}{h}\right)^{\frac{3}{2}}},\no
&& K_{xx}-Kh_{xx}=\frac{h\s{h+\frac{\dot{z}^2}{h}}}{z^2},
\ea
where $h'=\frac{dh(z)}{dz}$.

\subsubsection{Analysis of Thermal \texorpdfstring{AdS$_3$}{AdS3} at \texorpdfstring{$V=0$}{V=0}}

By setting $V=0$, we find 
\ba
2hz\ddot{z}+(4h-3zh')\dot{z}^2+4h^3-h^2h'z=0.
\ea
We can rewrite this as follows
\ba
\frac{2}{z}h^{\frac{5}{2}}\de_x\left(z^2h^{-\frac{3}{2}}\dot{z}\right)
=h^2h'z-4h^3,
\ea
where notice again that $\dot{f}$ and $f'$ mean $\frac{df}{dx}$ and $\frac{df}{dz}$, respectively.

Now, we introduce $y=y(x)$ which is related to $z=z(x)$ via
\ba
\dot{y}=3z^2h^{-\frac{3}{2}}\dot{z}.
\ea
This is solved in terms of $z$ as
\ba
y=\frac{3a^2z}{\s{h(z)}}-3a^3\arcsin{\frac{z}{a}}.\label{yxre}
\ea
Note that $y$ behaves as follows
\ba
&& y\simeq z^3\ \ (z\to 0),\ \ \ \ 
y\simeq \frac{3a^{\frac{7}{2}}}{\s{2(a-z)}} \ \ (z\to a).
\ea

The function $y=y(x)$ satisfies the following equation of motion when we regard $x$ as the time:
\ba
\ddot{y}=-\frac{3(2a^2z-z^3)}{a^2\s{h(z)}}\equiv -\frac{d U(y)}{dy},
\ea
where we introduced the potential energy $U$. This potential $U$ is given in terms of $z$ as
\ba
U=9a^2\left(-\frac{z^2}{2}+\frac{a^4}{2(a^2-z^2)}\right),
\ea
which is a function of $y$ via (\ref{yxre}). It behaves like
\ba
&& U(y)\simeq \frac{9}{2}y^{\frac{4}{3}}\ \  (y\to 0),\ \ \ \  U(y)\simeq \frac{y^2}{2a^2}\ \ (y\to \infty).
\ea

The energy conservation leads to 
\ba
\frac{1}{2}\dot{y}^2+U(y)=U(y_*),
\ea
where $y_*$ is the turning point of the trajectory. The EOW brane is described by $y=y(x)$ such that at $x=0$ we have  $y=y_*$ and at 
$x=\pm x_*$ we have $y=0$.

Thus $z(x)$ and $\phi(x)$ are obtained by solving the first order differential equations 
\ba
&& 3z^2h^{-3/2}\frac{dz}{dx}\left(=\frac{dy}{dx}\right)=\s{2U(z_*)-2U(z)},\no
&& \frac{d\phi}{dx}=\frac{h^{3/4}}{z}\left(1+\frac{2h(z)}{9z^4}\left(U(z_*)-U(z)\right)\right)^{\frac{1}{4}}. \label{qeomtads}
\ea

\begin{figure}[H]
    \centering
    \includegraphics[width=.3\textwidth,page=1]{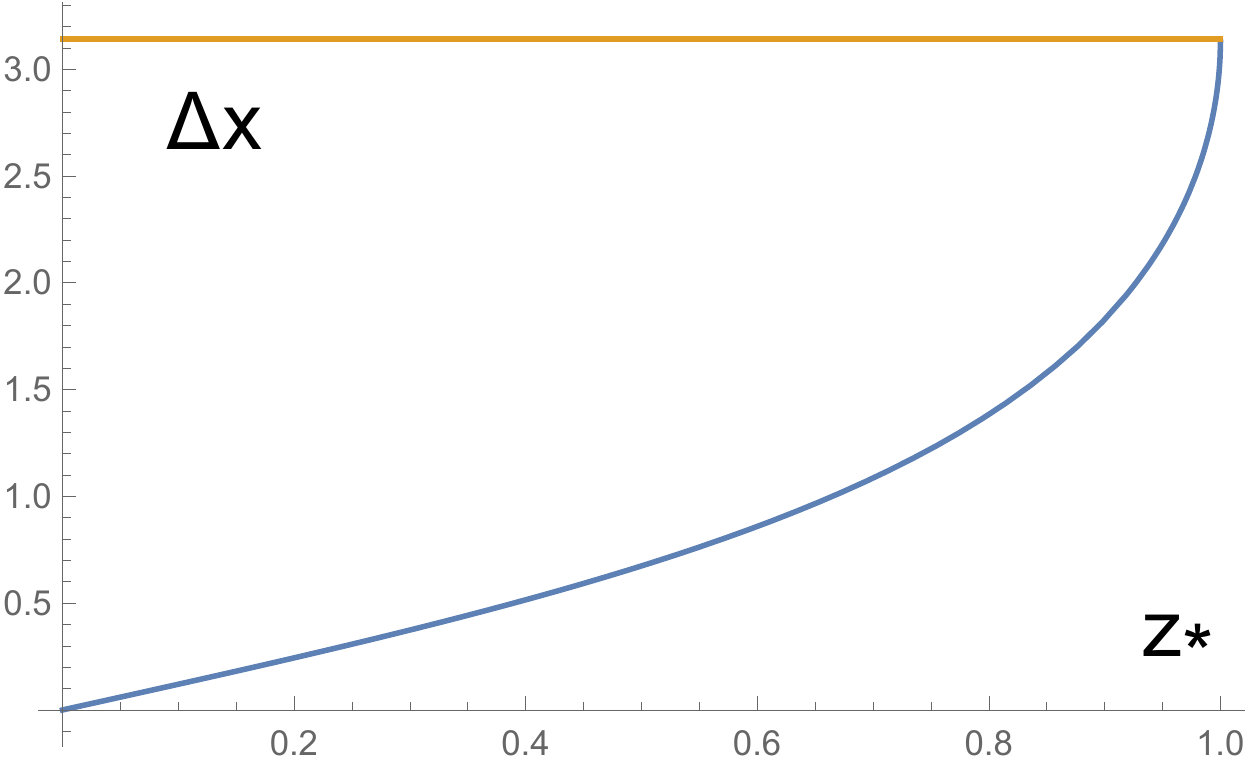}
     \includegraphics[width=.3\textwidth,page=1]{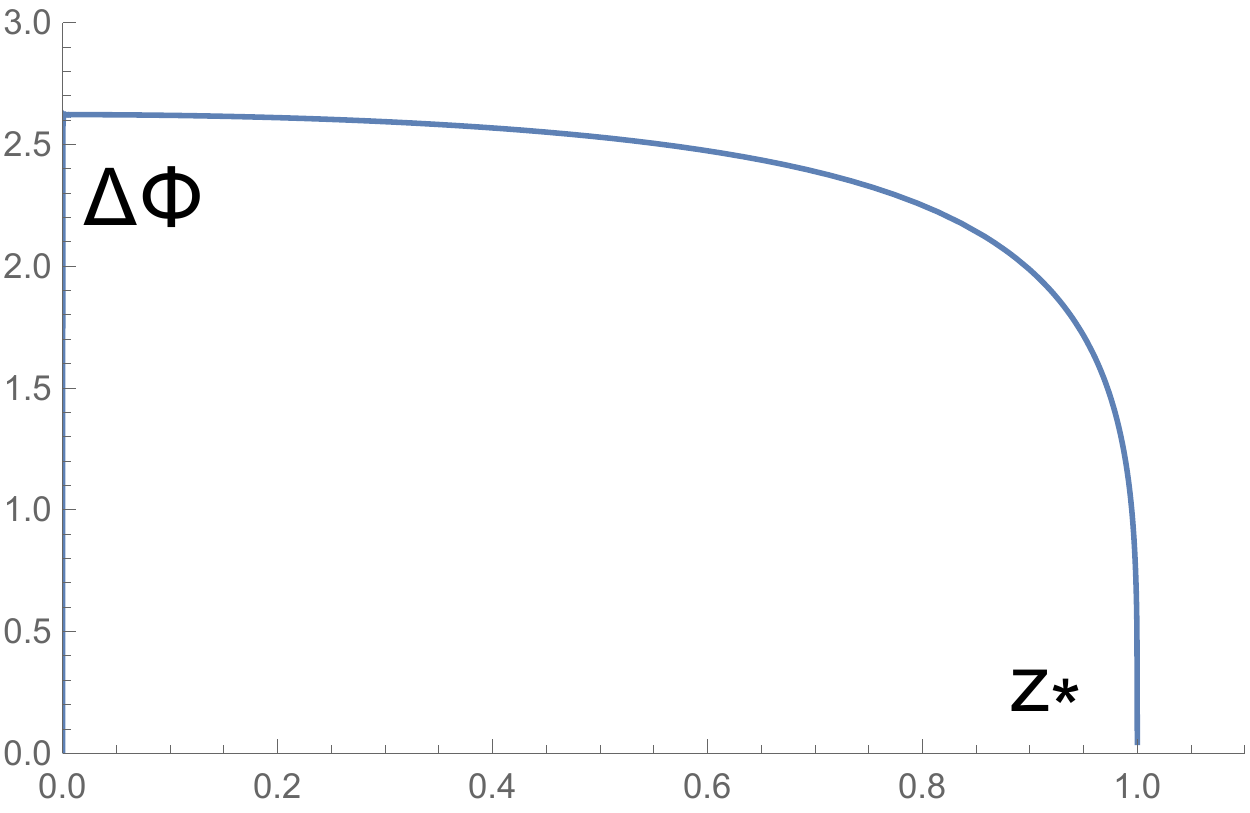}
    \caption{The profile of  the width $\Delta x=2x_*$ (left) and 
the shift of scalar field $\Delta\phi=\phi(x_*)-\phi(-x_*)$ (right) as function of $z_*$ on the EOW brane in thermal AdS$_3$. We set $a=1$. In the left graph, the orange horizontal line represents the value $\pi$.}
    \label{fig:TAdSXP}
\end{figure}

\begin{figure}[H]
    \centering
    \includegraphics[width=.3\textwidth,page=1]{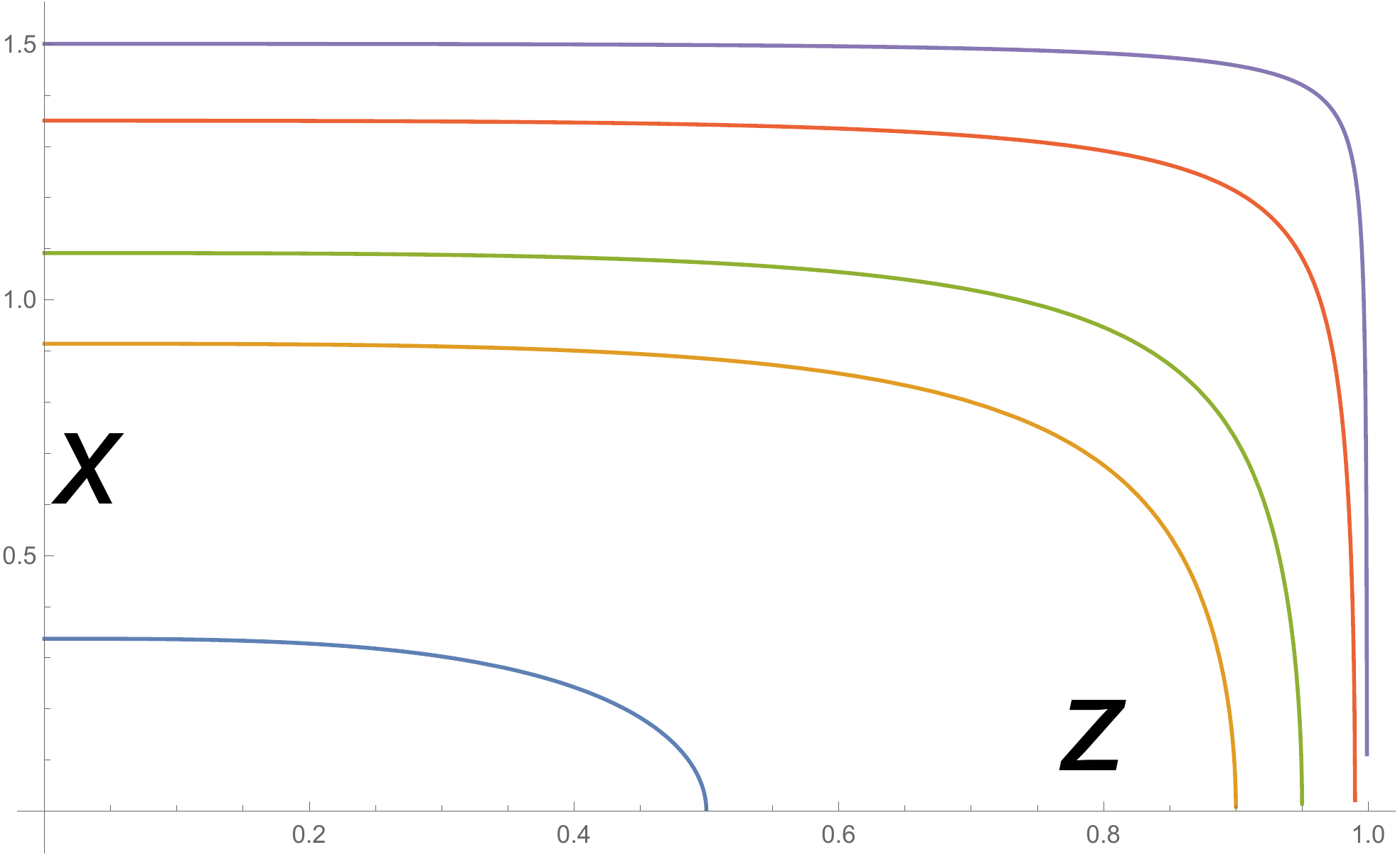}
    \caption{The profile of the EOW brane in thermal AdS$_3$. The blue, yellow, green and purple curve corresponds to $z_*=0.5, 0.9, 0.95, 0.99$ and $0.999$, respectively. We set $a=1$.}
    \label{fig:TAdSB}
\end{figure}

In Fig.\ref{fig:TAdSXP}, we plotted the value of the width $\Delta x=2x_*$ and 
the shift of scalar field $\Delta\phi=\phi(x_*)-\phi(-x_*)$ as function of $z_*$ at $a=1$. Note that $z_*$ takes the values in the range $0<z_*<a$ 
and that the limit $z_*\to 0$ (or equally $x_*\to 0$) reproduces the previous results for the Poincar\'{e} AdS$_3$ in section \ref{sec:bdyJanus}. Indeed, we can confirm (\ref{xxpola}) in the $z_*\to 0$ limit, which is rewritten as $x_*\sim 1.2z_*$ and $\Delta\phi\simeq 2.6$ in the present notation. Also notice that in the opposite limit $z_*\to a$, the solution is reduced to the known standard one \cite{Takayanagi:2011zk,Fujita:2011fp} without any scalar field, which means $\Delta x=\pi a$.   
Also refer to Fig.\ref{fig:TAdSB} for the plots of the brane profile $z=z(x)$.

Now, we consider how we can find solution when the width $\Delta x=2x_*$, the temperature $1/\beta$ and the strength of the external field $\Delta\phi=\phi(x_*)-\phi(-x_*)$ are given. 

In our thermal AdS brane solution, the temperature i.e. periodicity of $\tau$ is arbitrary. The integral of (\ref{qeomtads}) leads to the relation 
\ba
\frac{\Delta x}{a}=X_T\left(\frac{z_*}{a}\right),\ \ \ 
\Delta\phi=\Phi_T\left(\frac{z_*}{a}\right),
\ea
where the functions $X_T$ and $\Phi_T$ are given by Fig.\ref{fig:TAdSXP}.
Therefore, for a given $\Delta \phi$, we find $z_*/a$ and then for a given $\Delta x$, $a$ is determined. In this way, we can fully fix the solution. This implies that there is an upper bound of $\Delta \phi$.
This argument also shows that the independent input parameters are 
the two dimensionless quantities (\ref{twopara}),
as expected from the conformal symmetry of this system.
Note that we have 
\ba
X_T(0)=0,\ \ \ X_T(1)=\pi, \ \ \ \Phi_T(0)\simeq 2.6, \ \ \  \Phi_T(1)=0,
\ea
where $\Phi(0)$ agrees with (\ref{xxpola}).\footnote{It is intriguing to note that $X_T$ is always less than $\pi$ i.e. $\Delta x<a$. In the standard case without any matter field on the EOW brane, we have $\Delta x=a$, which means that the EOW brane covers the precisely semi-circle of the AdS boundary. 
In the presence of brane-localized scalar, the above results shows that it only covers less than a half of the boundary circle. This prohibits a construction of traversable wormhole in the present AdS/BCFT setup, providing a consistency check of our model.}

\subsubsection{Free Energy in Thermal \texorpdfstring{AdS$_3$}{AdS3}}

Now we would like to evaluate the free energy at a fixed temperature $1/\beta$, which is given by the on-shell action $I$ (\ref{totac}) itself, using our solution in thermal AdS$_3$. The total on-shell action, written by $I_T$, is given by 
\ba
I_T=\!-\frac{1}{16\pi G_N}\int \s{g}(R\!+\!2)\!-\!\frac{1}{8\pi G_N}\int_{\Sigma}\s{h}K\!-\!I_{c.t.}
\!-\frac{1}{8\pi G_N}\int_Q \s{h}\left(K\!-\!h^{ab}\de_a\phi\de_b\phi\!-\!V(\phi)\right),\no \label{actionbf}
\ea
where $I_{c.t.}$ is the local counter term which subtract $O(\ep^{-2})$ divergence for the standard geometric cut off $z\geq \ep$. Note that since we set $V=0$, the boundary condition (\ref{Nbc}) leads to $K=0$. We compactify the Euclidean time direction as $\tau\sim \tau+\beta$. we write the bulk 3d space by $M$ i.e. the region $z\leq z(x)$, which is surrounded by the AdS boundary $\Sigma$ and the EOW brane $Q$. We also note 
\ba
R=-6,\ \ K|_{\Sigma}=2+O(\ep^4).
\ea

We have to be careful when we subtract the counter term. The counter term corresponds to the contribution of Poincar\'{e} AdS$_3$ i.e. the metric with $h(z)=1$. However, we need to adjust the length of interval in $x$ direction \cite{Hawking:1982dh} such that 
\ba
\ti{x}_*=\s{h(\ep)}x_*\simeq \left(1-\frac{\ep^2}{2a^2}\right) x_*,
\ea
where $2\ti{x}_*$ is the length of the interval in the Poincar\'{e} Ad3$_3$, while $2x_*$ is the original one in thermal AdS$_3$. 

$I_{c.t}$ is explicitly given by the gravity action on the Poincar\'{e} AdS$_3$ with the two parallel EOW brane 
$x=\pm \ti{x}_*$:
\ba
I_{c.t}&=&-\frac{1}{16\pi G_N}\int \s{g^{(0)}}(R^{(0)}+2)-\frac{1}{8\pi G_N}\int_{\Sigma}\s{h^{(0)}}K^{(0)}\no
&=&\frac{2\beta\ti{x}_*}{4\pi G_N}\int^{\infty}_\ep \frac{dz}{z^3}-\frac{2\beta\ti{x}_*}{8\pi G_N}\cdot\frac{2}{\ep^2}.
\ea

Thus, the boundary term on $\Sigma$ cancels completely as usual in AdS:
\ba
&& -\frac{1}{8\pi G_N}\left[\int_{\Sigma}\s{h}K-\int_{\Sigma}\s{h^{(0)}}K^{(0)}\right] \no
&&=-\frac{\beta}{8\pi G_N}\left[(2x_*)\frac{2\s{h(\ep)}}{\ep^2}-(2\ti{x}_*)\frac{2}{\ep^2}\right]=0.
\ea

\begin{figure}[H]
    \centering
    \includegraphics[width=.3\textwidth,page=1]{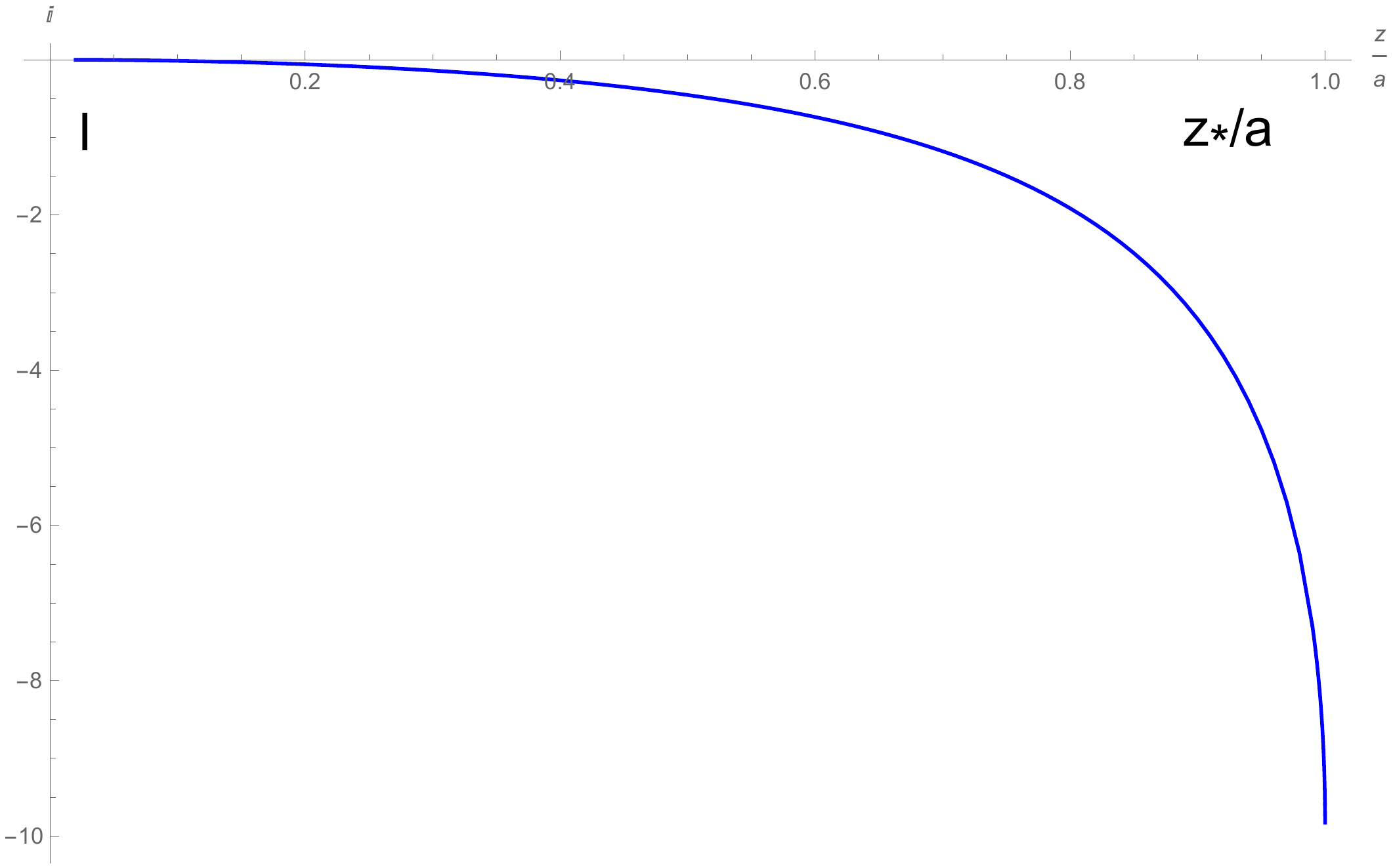}
    \hspace{5mm}
     \includegraphics[width=.3\textwidth,page=1]{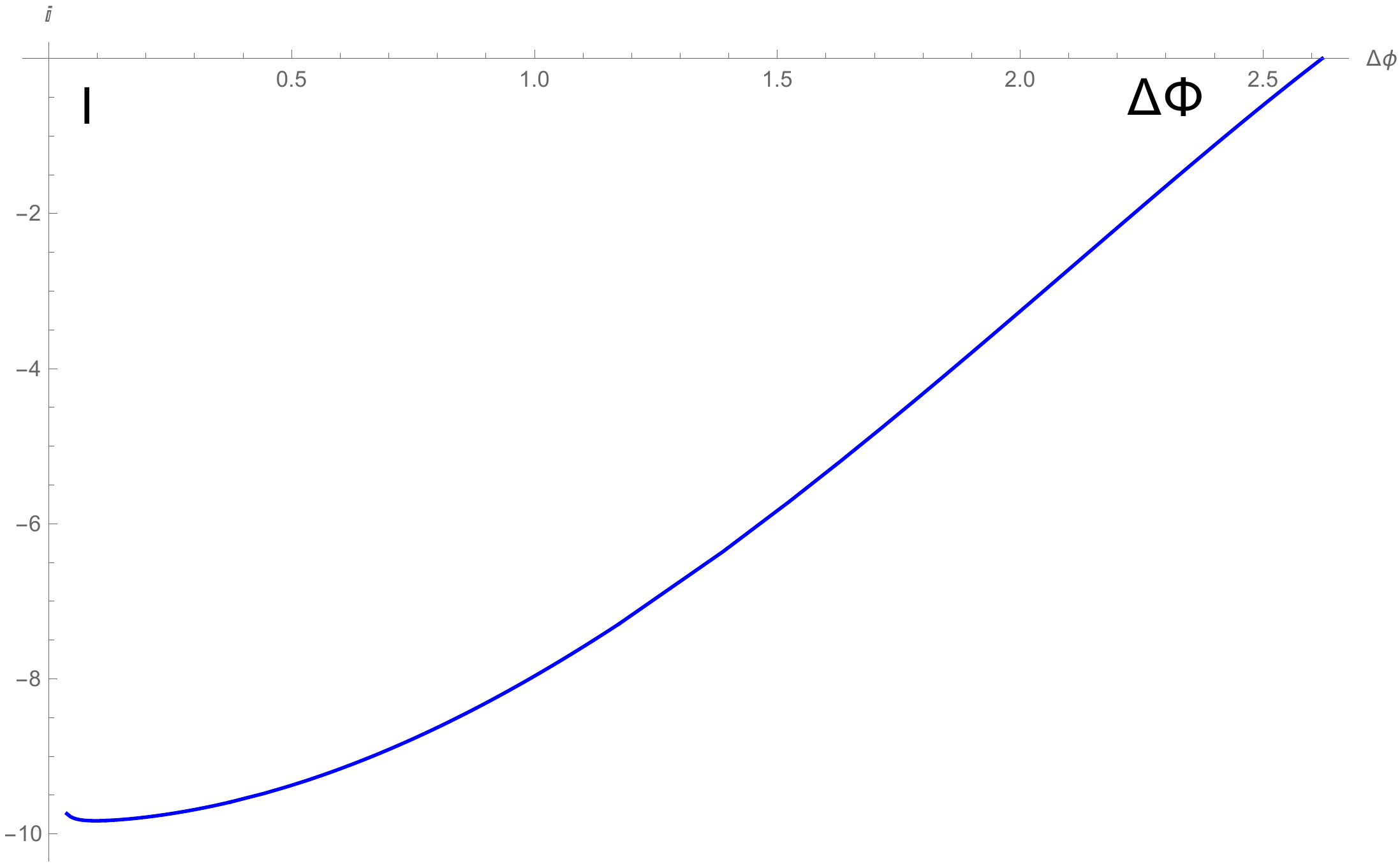}
    \caption{Plots of free energy $I_T$ as a function of $z_*/a$ (left) and as a function of $\Delta \phi$ (right). We set $16\pi G_N=1$ and $\frac{\beta}{\Delta x}=1$.}
    \label{fig:TAdSFE}
\end{figure}

Now we can evaluate the on-shell action in the thermal AdS$_3$ with the EOW brane as follows
\ba
&& I_T=\frac{\beta}{4\pi G_N}\int_M \frac{dzdx}{z^3}
-\frac{2\beta\ti{x}_*}{4\pi G_N}\int^{\infty}_\ep \frac{dz}{z^3}+\frac{\beta}{8\pi G_N}\int_Q dx
\s{\frac{h}{h^2+\dot{z}^2}}(\dot\phi)^2\no
&& =\frac{2\beta}{4\pi G_N}\int^{x_*}_0 dx\left[\frac{1}{2\ep^2}-\frac{1}{2z^2}\right]-\frac{2\beta}{4\pi G_N}\int^{\ti{x}_*}_0
\frac{dx}{2\ep^2}+\frac{\beta}{4\pi G_N}\int^{z_*}_0\frac{3dz}{\s{2h(z)(U(z_*)-U(z)}}\no
&&=\frac{\beta}{4\pi G_N}\left[\frac{x_*}{2a^2}-\int^{z_*}_0 dz\frac{3}{h(z)^{3/2}\s{2(U(z_*)-U(z))}}
+\int^{z_*}_0 dz\frac{3}{h(z)^{1/2}\s{2(U(z_*)-U(z))}}\right]\no
&&=-\frac{\beta \Delta x}{16\pi G_N a^2}=-\frac{c\beta\Delta x}{24\pi a^2},  \label{FTADS}  
\ea
where we employed the differential equation (\ref{qeomtads}) many times. 
It is useful to note that we can write the free energy (\ref{FTADS})  as follows
\ba
I_T=-\frac{1}{16\pi G_N}\cdot \frac{\beta}{\Delta x}\cdot\left[X_{T}(\Phi_T^{-1}(\Delta\phi))\right]^2,  \label{TADSfg}
\ea
such that it only depends on the two input parameters (\ref{twopara}). This is plotted in the left panel of Fig.\ref{fig:TAdSFE}.

We can show that $I_T\leq 0$. We have $I_T=0$ in the limit $z_*\to 0$ (or equally $x_*\to 0$) i.e. when the solution  approaches the one in the Poincar\'{e} AdS$_3$. This agrees with our previous result (\ref{totovan}).
In the opposite limit $z_*\to a$ we obtain 
\ba
I_T|_{z_*=a}=-\frac{\beta}{4\pi G_N}\cdot \frac{\pi}{4a}=-\frac{\pi c\beta}{24 \Delta x},
\ea
which agrees with the known result of the AdS/BCFT with no scalar field $\phi=0$ in \cite{Takayanagi:2011zk,Fujita:2011fp}.

\subsection{Connected EOW brane in BTZ}
Now, we consider a connected EOW brane in the BTZ black hole solution:
\ba
&& ds^2=\frac{h(z)d\tau^2}{z^2}+\frac{dz^2}{h(z)z^2}+\frac{dx^2}{z^2},\no
&& h(z)=1-\frac{z^2}{a^2}.
\ea
The induced metric on a EOW brane reads
\ba
ds^2=\frac{h(z)d\tau^2}{z^2}+\frac{h(z)+\dot{z}^2}{h(z)z^2}dx^2.
\ea

Using the normal vector of EOW brane
\ba
(N^\tau,N^z,N^x)=\frac{z}{\s{h+\dot{z}^2}}(0,h,-\dot{z}),
\ea
we find the Neumann boundary condition (\ref{Nbc}) explicitly 
\ba
&& K_{\tau\tau}-Kh_{\tau\tau}=-\frac{hV(\phi)}{z^2}
-\frac{h^2}{h+\dot{z}^2}\dot{\phi}^2,\no
&& K_{xx}-Kh_{xx}=\dot{\phi}^2-\frac{h+\dot{z}^2}{hz^2}V(\phi).
\ea
In our setup, we have
\ba
&& K_{\tau\tau}-Kh_{\tau\tau}
=\frac{h(2h^2-zh'\dot{z}^2+2h\dot{z}^2+2hz\ddot{z})}{2z^2\left(h+\dot{z}^2\right)^{\frac{3}{2}}},\no
&& K_{xx}-Kh_{xx}=\frac{(2h-zh')\s{h+\dot{z}^2}}{2z^2h},
\ea
where $h'=\frac{dh(z)}{dz}$.

\subsubsection{Analysis of BTZ at \texorpdfstring{$V=0$}{V=0}}

By setting $V=0$, we obtain
\ba
2hz\ddot{z}+(4h-2zh')\dot{z}^2+4h^2-zhh'=0.
\ea
We can rewrite this as follows
\ba
\frac{2h^2}{z}\de_x\left(z^2h^{-1}\dot{z}\right)
=hh'z-4h^2,
\ea
where notice again that $\dot{f}$ and $f'$ mean $\frac{df}{dx}$ and $\frac{df}{dz}$, respectively.

Introduce a new function $y=y(x)$, related to $z=z(x)$ via
\ba
\dot{y}=3\frac{z^2}{h(z)}\dot{z},
\ea
which is solved in terms of $z$ as
\ba
y=-3a^2z+3a^3\arctanh{\frac{z}{a}}.\label{btzyxre}
\ea
It is helpful to notice that $y$ behaves as
\ba
&& y\simeq z^3\ \ (z\to 0),\ \  \ \ y\simeq -3a^2+\frac{3}{2}\log\frac{2a}{a-z} \ \ (z\to a).
\ea

The function $y=y(x)$ satisfies 
\ba
\ddot{y}=-6z+\frac{3z^2h'}{2h}\equiv -\frac{d U(y)}{dy},
\ea
where we introduced the potential energy $U$. This potential $U$ is given in terms of $z$ as
\ba
U=9a^2\left(-\frac{z^2}{2}+\frac{a^4}{2(a^2-z^2)}\right),
\ea
which is the same as that in thermal AdS$_3$.
As a function of $y$ via (\ref{btzyxre}), this  behaves like
\ba
&& U(y)\simeq \frac{9}{2}y^{\frac{4}{3}}\ \  (y\to 0),
\ \ \ \ U(y)\simeq \frac{9a^4}{8}e^{y^{\frac{2}{3}}}\ \ (y\to \infty).
\ea

The energy conservation leads to 
\ba
\frac{1}{2}\dot{y}^2+U(y)=U(y_*),
\ea
where $y_*$ is the turning point of the trajectory. The EOW brane is described by $y=y(x)$ such that at $x=0$ we have  $y=y_*$ and at 
$x=\pm x_*$ we have $y=0$.

Thus $z(x)$ and $\phi(x)$ are obtained by solving the first order differential equations 
\ba
&& 3\frac{z^2}{h}\frac{dz}{dx}\left(=\frac{dy}{dx}\right)=\s{2U(z_*)-2U(z)},\no
&& \frac{d\phi}{dx}=\s{\frac{2h(z)-zh'(z)}{2z^2h(z)}}\left(h(z)+\frac{2h(z)^2}{9z^4}\left(U(z_*)-U(z)\right)\right)^{\frac{1}{4}}. \label{qeombtz}
\ea

\begin{figure}[H]
    \centering
    \includegraphics[width=.4\textwidth,page=1]{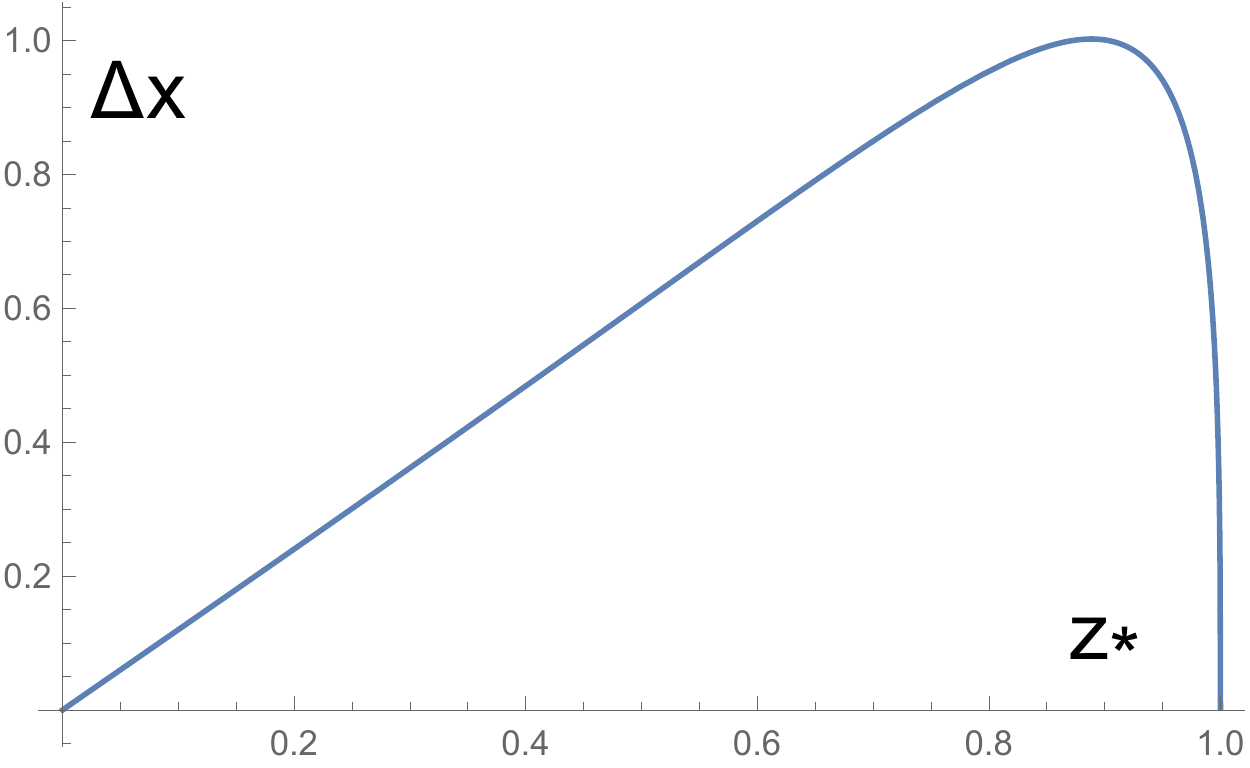}
     \includegraphics[width=.4\textwidth,page=1]{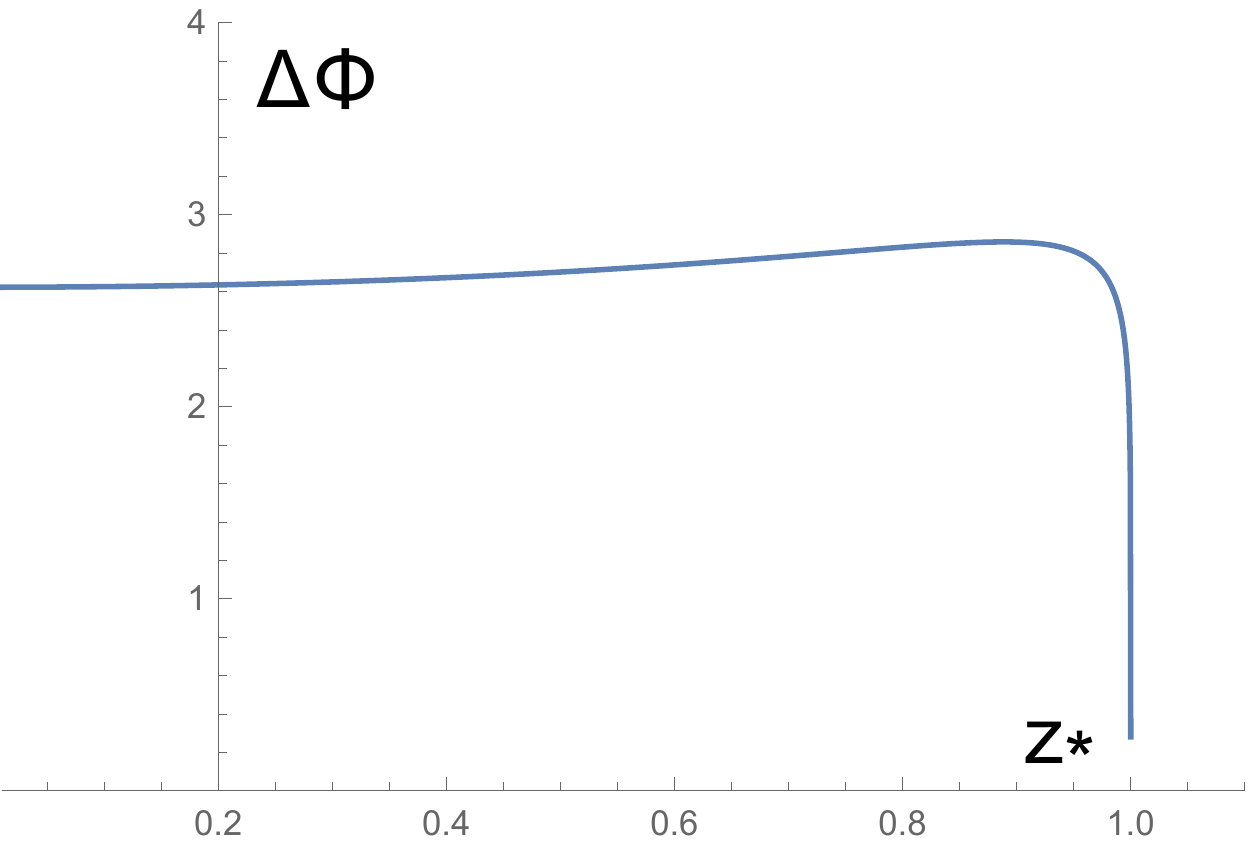}
    \caption{The profile of  the width $\Delta x=2x_*$ (left) and 
the shift of scalar field $\Delta\phi=\phi(x_*)-\phi(-x_*)$ (right) as function of $z_*$ on the EOW brane in BTZ. We set $a=1$.}
    \label{fig:BTZXP}
\end{figure}

\begin{figure}[H]
    \centering
    \includegraphics[width=.4\textwidth,page=1]{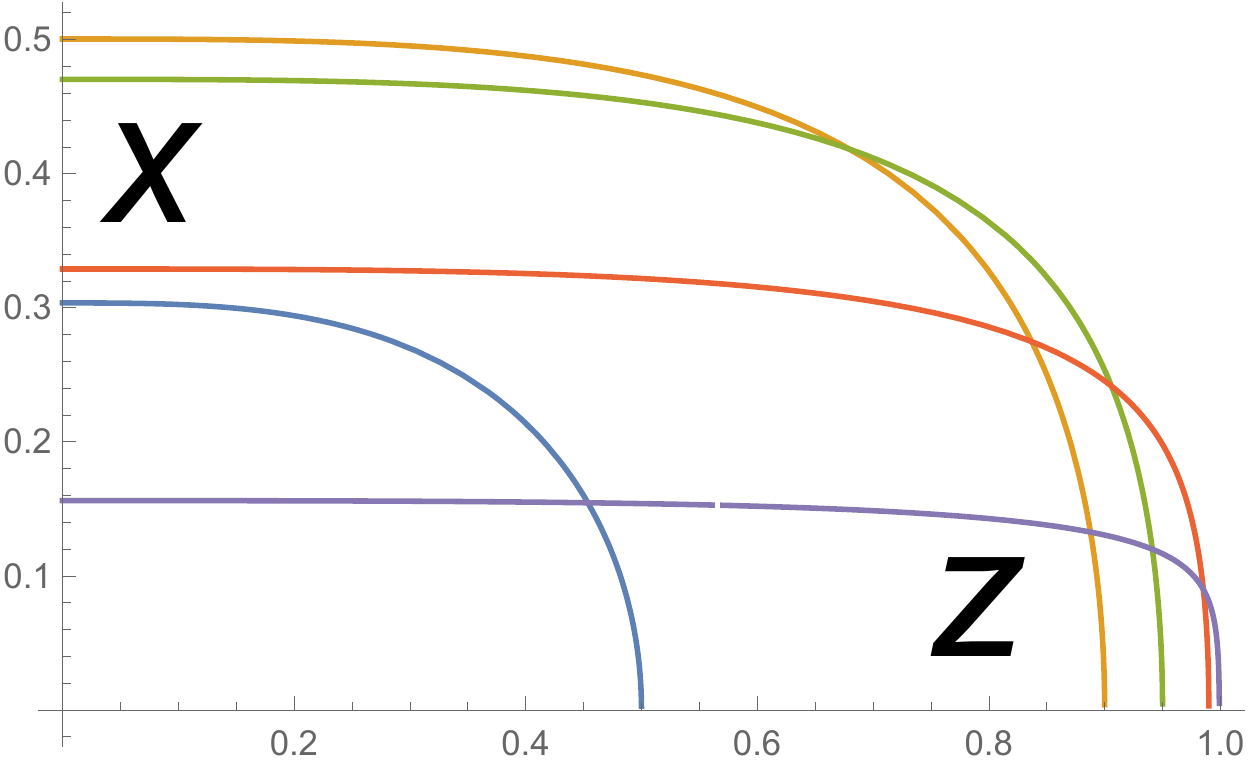}
    \caption{The profile of the EOW brane in BTZ. The blue, yellow, green, red and purple curve corresponds to $z_*=0.5, 0.9, 0.95, 0.99$ and $0.999$, respectively. We set $a=1$.}
    \label{fig:BTZB}
\end{figure}

In Fig.\ref{fig:BTZXP}, we showed the value of the width $\Delta x=2x_*$ and 
the shift of scalar field $\Delta\phi=\phi(x_*)-\phi(-x_*)$ as function of $z_*$ at $a=1$. Note that $z_*$ takes the values in the range $0<z_*<a$ 
and that the limit $z_*\to 0$ reproduces the previous results for the Poincar\'{e} AdS$_3$ in section \ref{sec:bdyJanus}. Also notice that in the limit $z_*\to a$, the solution is reduced to the know standard one without any scalar field i.e. the disconnected vertical solution $x=x_*$ and $x=-x_*$. Refer to Fig.\ref{fig:BTZB} for the plots of the brane profile.

In our BTZ solution, the temperature i.e. periodicity of $\tau$ is arbitrary, assuming $z_*<a$. If $z_*=a$, then we consider the disconnected solution where the temperature is fixed as $1/\beta=1/(2\pi a)$ as usual.
The integral of (\ref{qeombtz}) leads to the relation 
\ba
\frac{\Delta x}{a}=X_B\left(\frac{z_*}{a}\right),\ \ \ 
\Delta\phi=\Phi_B\left(\frac{z_*}{a}\right),
\ea
where the functions $X_B$ and $\Phi_B$ are given by Fig.\ref{fig:BTZXP}.
Therefore, for a given $\Delta \phi$, we find $z_*/a$ and then the given $\Delta x$ fixes the value of $a$. In this way we can fully fixes the solution once the two input parameters (\ref{twopara}) are given. However, we will see shortly below, the connected EOW brane solution in BTZ is never favored as it has a larger free energy.

\subsubsection{Free Energy for a connected EOW brane in BTZ}
Now we would like to evaluate the gravity action to calculate the free energy (\ref{actionbf}) for our on-shell solution in BTZ with the EOW brane, which is written as $I_B$. When we subtract the Poincar\'{e} AdS$_3$ solution as the counter term, we need to rescale the inverse temperature as follows
\ba
\ti{\beta}=\s{h(\ep)}\beta\simeq \left(1-\frac{\ep^2}{2a^2}\right) \beta,
\ea
where $\ti{\beta}$ is the periodicity of $\tau$ in the Poincar\'{e} AdS$_3$, while $\beta$ is the original one in BTZ. 

The counter term $I_{c.t.}$ is explicitly given by the gravity action on the Poincar\'{e} AdS$_3$ with the two parallel EOW branes
$x=\pm x_*$:
\ba
I_{c.t.}&=&-\frac{1}{16\pi G_N}\int \s{g^{(0)}}(R^{(0)}+2)-\frac{1}{8\pi G_N}\int_{\Sigma}\s{h^{(0)}}K^{(0)}\no
&=&\frac{2\ti{\beta}x_*}{4\pi G_N}\int^{\infty}_\ep \frac{dz}{z^3}-\frac{2\ti{\beta}x_*}{8\pi G_N}\cdot\frac{2}{\ep^2}.
\ea

Thus, the boundary term on $\Sigma$ cancels completely as usual in AdS:
\ba
&& -\frac{1}{8\pi G_N}\left[\int_{\Sigma}\s{h}K-\int_{\Sigma}\s{h^{(0)}}K^{(0)}\right] \no
&&=-\frac{2x_*}{8\pi G_N}\left[\beta\frac{2\s{h(\ep)}}{\ep^2}-\ti{\beta}\frac{2}{\ep^2}\right]=0.
\ea

Now we can evaluate the on-shell action in the BTZ with the EOW brane as follows
\ba
&& I_B=\frac{\beta}{4\pi G_N}\int_M \frac{dzdx}{z^3}
-\frac{2\ti{\beta}x_*}{4\pi G_N}\int^{\infty}_\ep \frac{dz}{z^3}+\frac{\beta}{8\pi G_N}\int_Q dx
\frac{h}{\s{h^2+\dot{z}^2}}\dot\phi^2\no
&& =\frac{2\beta}{4\pi G_N}\int^{x_*}_0 dx\left[\frac{1}{2\ep^2}-\frac{1}{2z^2}\right]-\frac{2\ti{\beta}}{4\pi G_N}\int^{\ti{x}_*}_0
\frac{dx}{2\ep^2}+\frac{2\beta}{8\pi G_N}\int^{x_*}_0 \frac{dx}{z^2}
\no
&&=\frac{\beta\Delta x}{16\pi G_N a^2}=\frac{c\beta\Delta x}{24\pi a^2}.
\ea
This shows that $I_B\geq 0$. We have $I_B=0$ in the limit $z_*\to 0$ i.e. when the solution approaches the one in the Poincar\'{e} AdS$_3$. This agrees with our previous result (\ref{totovan}). Also in the opposite limit  $z_*\to a$ we find $I_B=0$. However, since $I_B\geq 0$, this connected solution in BTZ will not be favored thermodynamically.

\subsection{Disconnected EOW Solution in BTZ}

\begin{figure}[H]
    \centering
    \includegraphics[width=.5\textwidth,page=1]{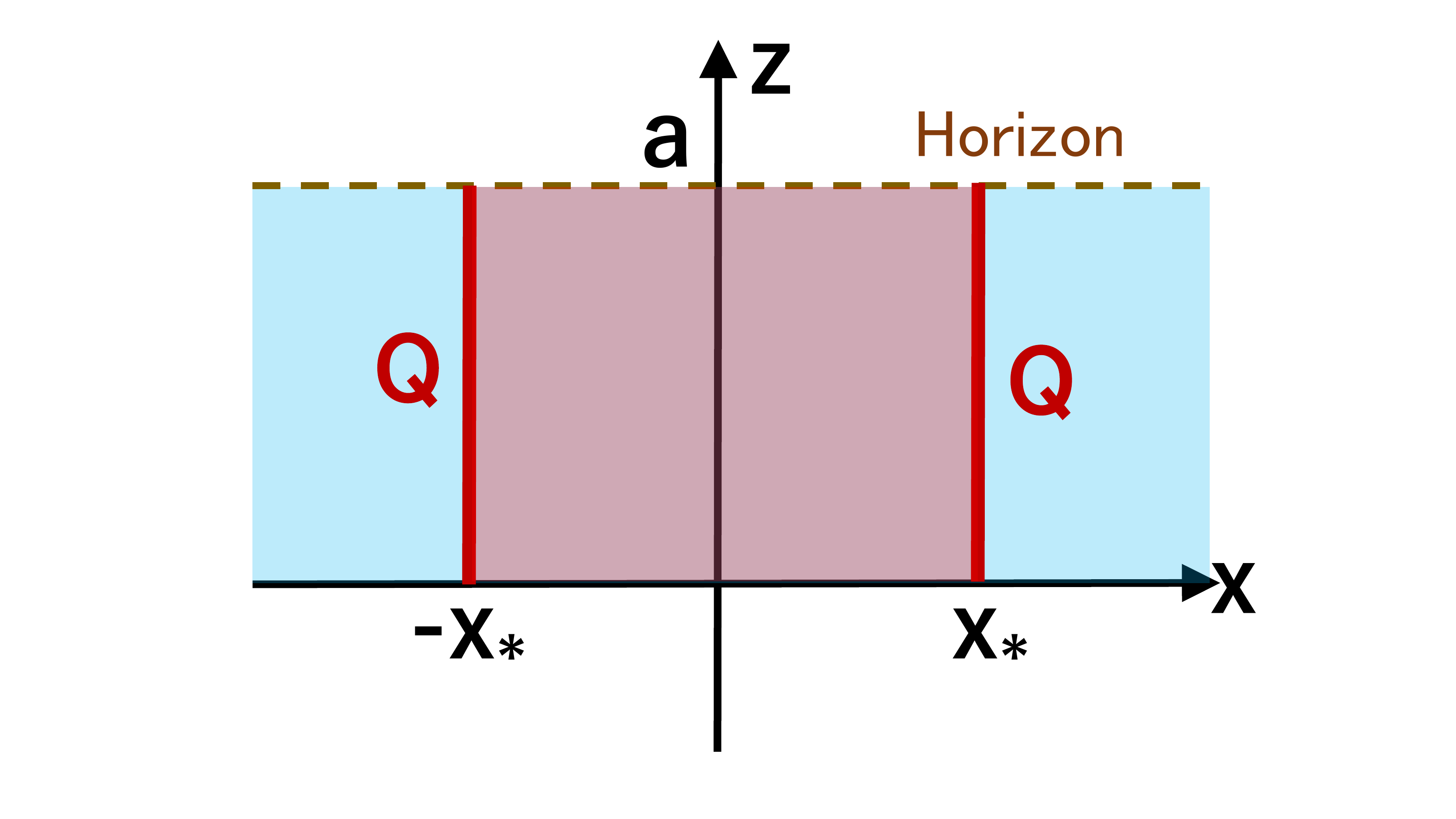}
    \caption{A sketch of disconnected EOW branes in BTZ.}
    \label{fig:BTZD}
\end{figure}

It is important to note that there is another solution of EOW branes in BTZ even when $\Delta \phi\neq 0$. This is the solution with two disconnected EOW branes as sketched in Fig.\ref{fig:BTZD}. This is given by  setting the scalar field to be a constant value $\phi=\phi_1$ and $\phi=\phi_2$ in each of the two disconnected EOW branes. Thus the solution is the same as the standard AdS/BCFT without any scalar field with vanishing tension \cite{Takayanagi:2011zk,Fujita:2011fp}. Thus the free energy reads
\ba
I_D=-\frac{\pi \Delta x}{4G_N\beta}=-\frac{\pi c \Delta x}{6\beta}. \label{discbtz}
\ea
Note that such a disconnected EOW brane solution is not possible in thermal AdS$_3$ due to an obvious topological reason. 

\subsection{Comparing Free Energy and Phase Transition}

When we have multiple classical solutions in gravity for input values of $\frac{\Delta x}{\beta}$ and $\Delta \phi$ from the BCFT, we need to select the one with the minimal values of action (or the free energy). 
Since the connected EOW brane in BTZ has the non-negative free energy, we only have two candidates of dominant solutions: one is a connected brane in thermal AdS$_3$ and the other is two disconnected branes in BTZ, whose 
 free energies $I_T$ and $I_D$ are given by  (\ref{FTADS}) and 
(\ref{discbtz}), respectively.

\begin{figure}[H]
    \centering
     \includegraphics[width=.5\textwidth,page=1]{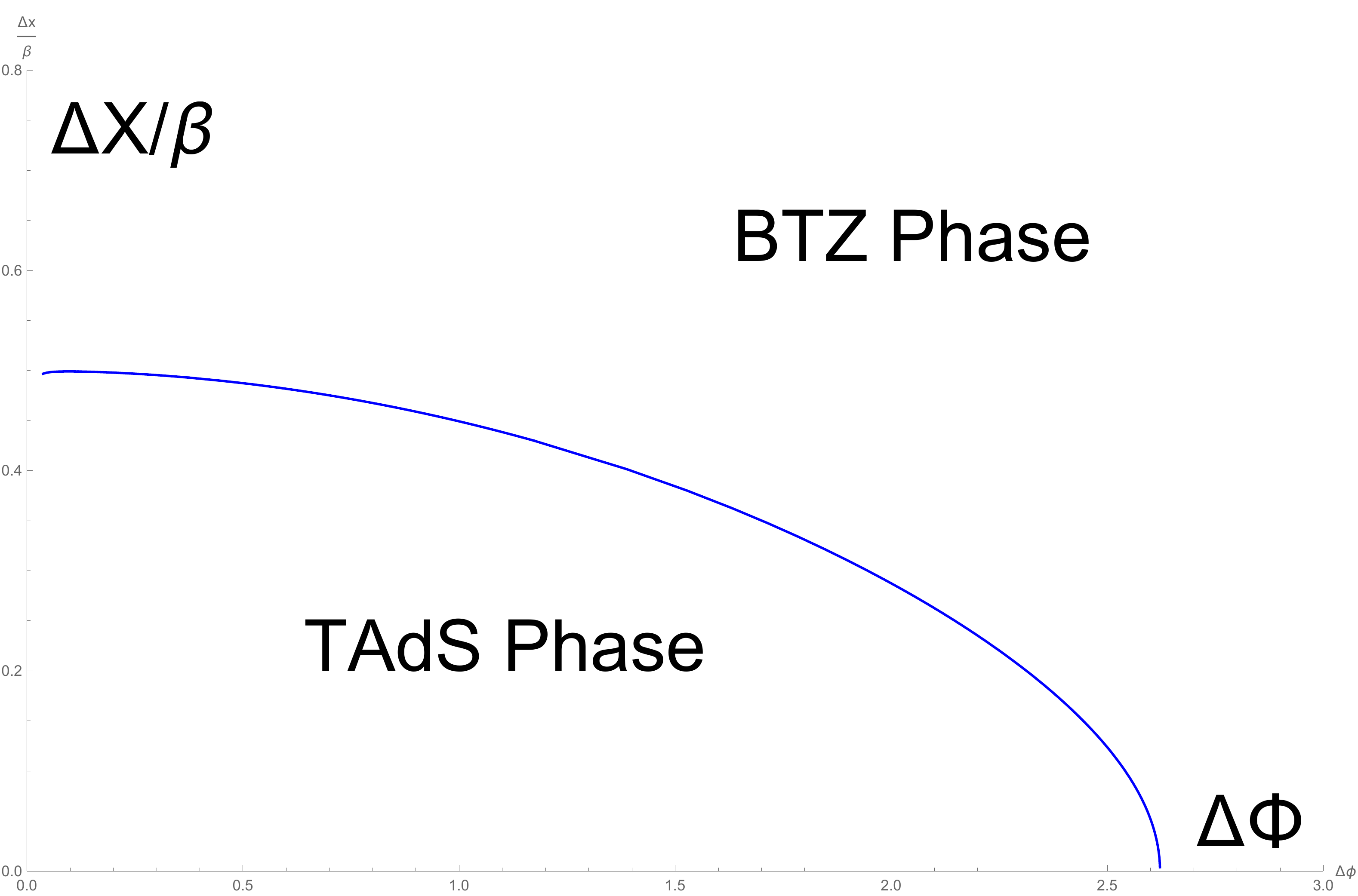}
    \caption{We plotted the phase diagram, where the horizontal and vertical axis correspond to $\Delta \phi$ and $\frac{\Delta x}{\beta}$, respectively. }
    \label{fig:Compare}
\end{figure}

The connected EOW brane in thermal AdS$_3$ is favored when
$I_T\leq I_D$, which leads to the condition
$\frac{\Delta x}{\beta}\leq \left(\frac{\Delta x}{\beta}\right)_c$, where this the temperature at the phase transition point, as a function of $\Delta \phi$, is given by
\ba
\left(\frac{\Delta x}{\beta}\right)_c =\frac{1}{2\pi}X_T\left(\Phi^{-1}_T(\Delta \phi)\right). \label{crtemp}
\ea
Thus we can conclude that in the low temperature phase $\frac{\Delta x}{\beta}\leq \left(\frac{\Delta x}{\beta}\right)_c$ and high temperature phase $\frac{\Delta x}{\beta}\geq \left(\frac{\Delta x}{\beta}\right)_c$,
the connected EOW brane in thermal AdS$_3$ and the disconnected EOW branes in BTZ are realized, respectively. This phase structure is depicted in Fig.\ref{fig:Compare}.

\section{Transition matrix from AdS/BCFT and phase transition}
In previous sections, we have introduced a brane-localized scalar field $\phi$ into the minimal AdS/BCFT construction \cite{Takayanagi:2011zk,Fujita:2011fp} and realized various novel configurations by designing the scalar. One of the most simple setup in this framework is realized by setting the potential $V(\phi)$ to zero, i.e. considering a massless free scalar. In this case, the nontriviality can be introduced by considering a brane which has two asymptotic infinities and impose different boundary values of $\phi$ on each. Such setups are discussed in section \ref{sec:bdyJanus} and section \ref{sec:phasetr}. 

In this section, we construct a setup in this framework and argue that there exists a novel phase transition in its time evolution behavior induced by the parameter $\Delta \phi$. More precisely, we consider a 1D quantum many-body system and a class of time-dependent transition matrices which is characterized by a single parameter. After seeing how this single parameter can be related to $\Delta \phi$ in an AdS$_3$ gravity via the AdS/BCFT correspondence, we will argue that the time growth of the pseudo entropy \cite{Nakata:2020luh,Mollabashi:2020yie,Mollabashi:2021xsd} in this setup is $\mathcal{O}(t)$ when $\Delta \phi$ is sufficiently small, $\mathcal{O}(t^0)$ when $\Delta \phi$ is sufficiently large, and $\mathcal{O}(\log(t))$ at the phase transition point. This behavior is similar to the measurement-induced phase transition of the entanglement entropy in 1D quantum many-body systems \cite{Skinner:2018tjl,Li:2018mcv,Li:2019zju} when changing the measurement rate. 

In the following, we will first explain the basic concepts of transition matrices and pseudo entropy. Then we will introduce our setup in the BCFT language and explain how it can be related to setups studied in section \ref{sec:bdyJanus} and section \ref{sec:phasetr} via the AdS/BCFT correspondence. After that, we will explain the details of the phase transition mentioned above and why we expect it exists. 

\subsection{Pseudo Entropy and The Setup}
Pseudo entropy is a generalization of entanglement entropy, and it was firstly introduced in \cite{Nakata:2020luh}. Let us consider a bipartite system composited by $A$ and $B$. The total Hilbert space is given by 
\begin{align}
    \mathcal{H}_{\rm tot} = \mathcal{H}_A \otimes \mathcal{H}_B. 
\end{align}
For two pure states $\ket{\psi},\ket{\varphi} \in \mathcal{H}_{\rm tot}$ which are not orthogonal to each other, the transition matrix is defined as 
\begin{align}
    \mathcal{T}^{\psi|\varphi} \equiv \frac{|\psi\rangle\langle \varphi|}{\langle \varphi|\psi\rangle}.
\end{align} 
Such a transition matrix describes a post-selection or weak measurement \cite{Aharonov:1988xu,RevModPhys.86.307} procedure where one prepares an initial state $\ket{\psi}$, performs operations on it, and then post-selects it to $\ket{\varphi}$. For example, the weak value \cite{Aharonov:1988xu,RevModPhys.86.307} of an operator $O$ can be obtained by computing ${\rm Tr} \left(\mathcal{T}^{\psi|\varphi}O\right)$.

The reduced transition matrix on $A$ is defined as 
\begin{align}
    \mathcal{T}_A^{\psi|\varphi} = 
    {\rm Tr}_B 
    \mathcal{T}^{\psi|\varphi}.
\end{align}
Note that the (reduced) transition matrix reduces to a standard (reduced) density matrix when $\ket{\varphi} = \ket{\psi}$. The pseudo entropy is defined by formally computing the von Neumann entropy 
\begin{align}
    S(\mathcal{T}_A^{\psi|\varphi}) = -{\rm Tr} \left[ \mathcal{T}_A^{\psi|\varphi} \log \mathcal{T}_A^{\psi|\varphi} \right]. 
\end{align}
Note that the pseudo entropy is not an entropy in the sense that it does not satisfy all the axioms of entropy such as reality or convexity. However, it reduces to entanglement entropy when $\ket{\varphi} = \ket{\psi}$ and hence is a generalization of it. Moreover, in holographic setups, the leading order of the pseudo entropy can be computed using the same recipe for holographic entanglement entropy \cite{Nakata:2020luh}. 

While the entanglement entropy of a pure state counts the number of EPR pairs one can distill by performing local operations and classical communications (LOCC) on the state, the operational meaning of pseudo entropy for general transition matrices is unclear. For a special class of transition matrices, the pseudo entropy can be interpreted as the number of EPR pairs one can distill in the post-selection procedure described by $\mathcal{T^{\psi|\varphi}}$ \cite{Nakata:2020luh}. However, despite the fact that its physical meaning is not completely clear, it has been found that pseudo entropy is useful for studying various physical systems such as phase transitions in quantum many-body systems \cite{Mollabashi:2020yie,Mollabashi:2021xsd}. 

In the following, let us consider a 1D system defined on an infinite line. Starting from an area law state $\ket{\psi_0}$ and time evolving it with a gapless Hamiltonian $H$, the entanglement will grow linearly and the state will become a volume law state. More precisely, the entanglement entropy of a single interval with length $l$ in the time-evolved state $e^{-iHt}\ket{\psi_0}$ will grow as $O(t)$ and saturate to $\mathcal{O}(l)$. This is well-known as the global quench setup \cite{Calabrese:2006rx,Calabrese:2016xau}. 

In CFT, such an initial state $\ket{\psi_0}$ can be realized by a regularized boundary state \cite{Calabrese:2006rx,Calabrese:2016xau}
\begin{align}
    \ket{\psi_0} \propto e^{-\alpha H} \ket{B}.
\end{align}
In the path integral language, one can consider a Euclidean strip parameterized by $(\tau,x)$, where $\tau\in[-\alpha,\alpha]$ is the Euclidean time and $x\in(-\infty,\infty)$ is the spatial coordinate. Here, the boundary conditions imposed on $\tau = -\alpha$ and $\tau = \alpha$ are the same. In this case, when focusing on the time slice $\tau = \tau'$, the state realized by the path integral over $-\alpha \leq \tau \leq \tau'$ and that over $\tau' \leq \tau \leq \alpha$ are $e^{-\tau'-\alpha H }\ket{B}$ and $\bra{B} e^{\tau'-\alpha H }$ respectively. By performing the analytic continuation $\tau' \rightarrow it$, one can realize the time evolved state $e^{-it-\alpha H }\ket{B}$. It is well-known that the effective temperature of such a state is given by $T_{\rm eff} = 1/4\alpha$. 

Let us consider a generalization of the setup described above. Instead of imposing the same boundary condition on the two boundaries, let us impose different ones. Denoting the boundary conditions imposed on $\tau=-\alpha$ and $\tau = \alpha$ as $\ket{B_1}$ and $\bra{B_2}$, respectively, the current path integral describes the following transition matrix 
\begin{align}\label{eq:TMreal}
    \mathcal{T} = \frac{e^{-iHt} e^{-\alpha H} |B_1\rangle \langle B_2 | e^{-\alpha H} e^{iHt} }{\langle B_2| e^{-\alpha H} e^{iHt} e^{-iHt} e^{-\alpha H} |B_1 \rangle}
\end{align}
after performing the analytical continuation $\tau \rightarrow it$. In the following, let us divide the subsystem $A$ given by $x\geq 0$ and consider the pseudo entropy associated with it.  

\begin{figure}
    \centering
    \includegraphics[width=.4\textwidth,page=1]{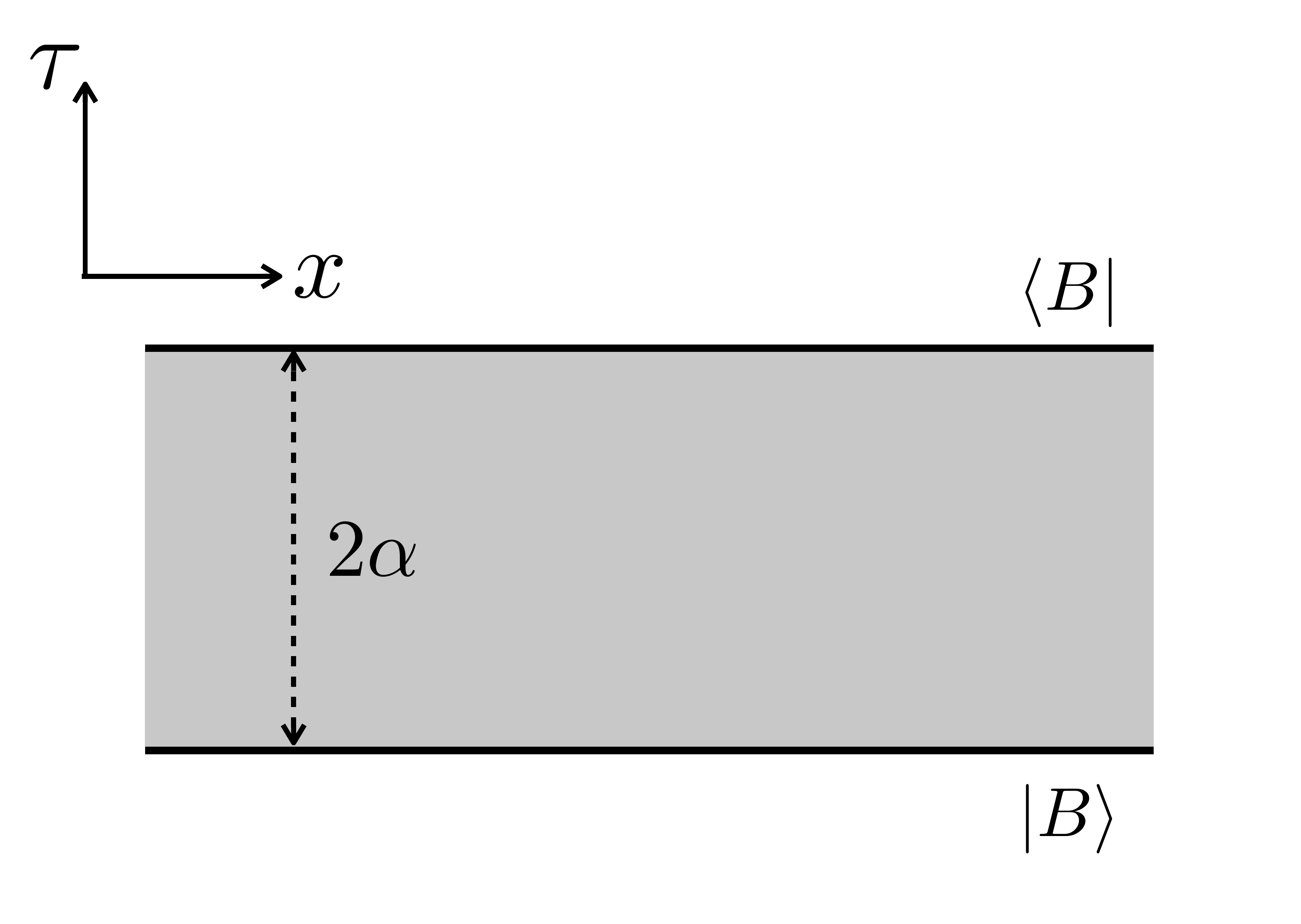}
    \includegraphics[width=.4\textwidth,page=1]{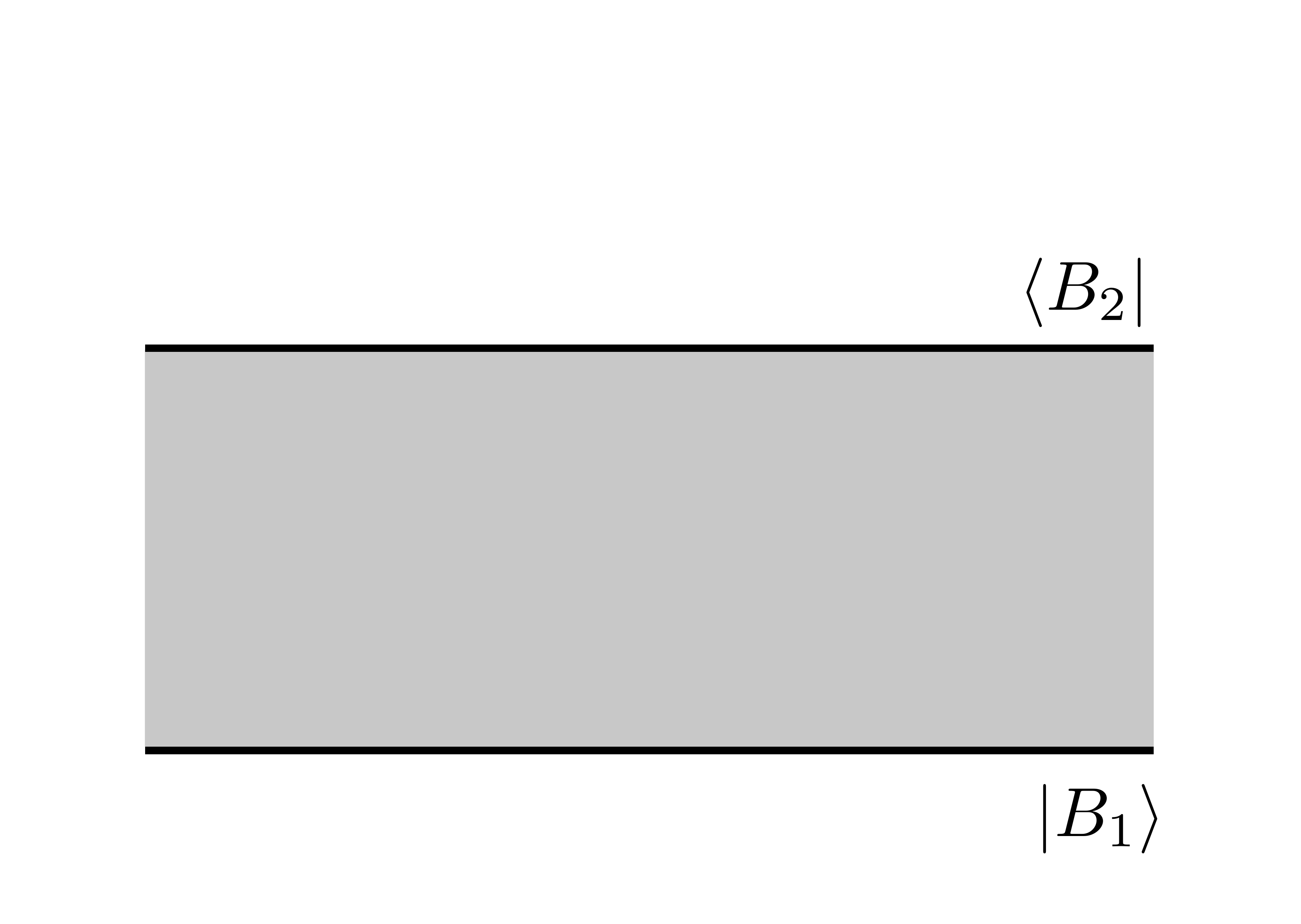}
    \caption{The Euclidean path integral setup corresponding to a density matrix after a global quench (left) and that corresponding to the transition matrix \eqref{eq:TMreal} (right), respectively.}
    \label{fig:GQpathint}
\end{figure}

\subsection{Holographic Duals and Phase Transition} 

The setups with brane-localized free massless scalar naturally realize a class of the transition matrices \eqref{eq:TMreal} in the AdS/BCFT correspondence. The boundary values of the brane-localized scalar field $\phi$ on the two asymptotic infinities correspond to the boundary states imposed on the two boundaries $\tau = \pm \alpha$ on the BCFT side. In this sense, the boundary value of $\phi$ parameterizes the boundary states. Accordingly, we can write 
\begin{align}
    \ket{B_1} = \ket{B(\phi_1)},~~\ket{B_2} = \ket{B(\phi_2)},
\end{align}
where 
\begin{align}
    \phi_1= \phi|_{z=0,\tau=-\alpha},~~\phi_2= \phi|_{z=0,\tau=\alpha}. 
\end{align}
The gravity dual of the Euclidean setup can been straightforward obtained from the results in section \ref{sec:phasetr}. In section \ref{sec:phasetr}, we have discussed the gravity dual of a BCFT defined on a manifold with topology $S^1\times I$, where we regard $S^1$ as the Euclidean time direction parameterized by $\tau$ and $I$ as the spatial direction parameterized by $x$. Now we are considering a BCFT defined on $I \times \mathbb{R}$, where we regard $I=[-\alpha,\alpha]$ as the Euclidean time direction parameterized by $\tau$ and $\mathbb{R}$ as the spatial direction parameterized by $x$. While not precise, it is useful to consider the infinite system $\mathbb{R}$ as the infinite radius limit of an $S^1$. In this sense, the definitions of $x$ and $\tau$ used here are opposite to that used in section \ref{sec:phasetr}. 

The $\Delta x$, $\beta$ and $\Delta \phi$ in section \ref{sec:phasetr} correspond to $2\alpha$, $\infty$ and $\Delta \phi$ in the current section, respectively. Looking at the $\Delta x /\beta = 0 $ part of figure \ref{fig:Compare}, one can see that there is a phase transition when changing $\Delta \phi$ and the phase transition point is given by $\Delta \phi = \Delta \phi_c \equiv 2K[-1] \simeq 2.6$. Since the coordinates of space and time are different between section \ref{sec:phasetr} and this section, let us call the BTZ phase (thermal AdS phase) in section \ref{sec:phasetr} the large $\Delta \phi$ phase (small $\Delta \phi$ phase) in this section to avoid confusion. Let us consider the Lorentzian counterparts for all these cases. 

\subsubsection{Small \texorpdfstring{$\Delta \phi$}{Delta Phi} Phase}
\begin{figure}
    \centering
     \includegraphics[width=.4\textwidth,page=1]{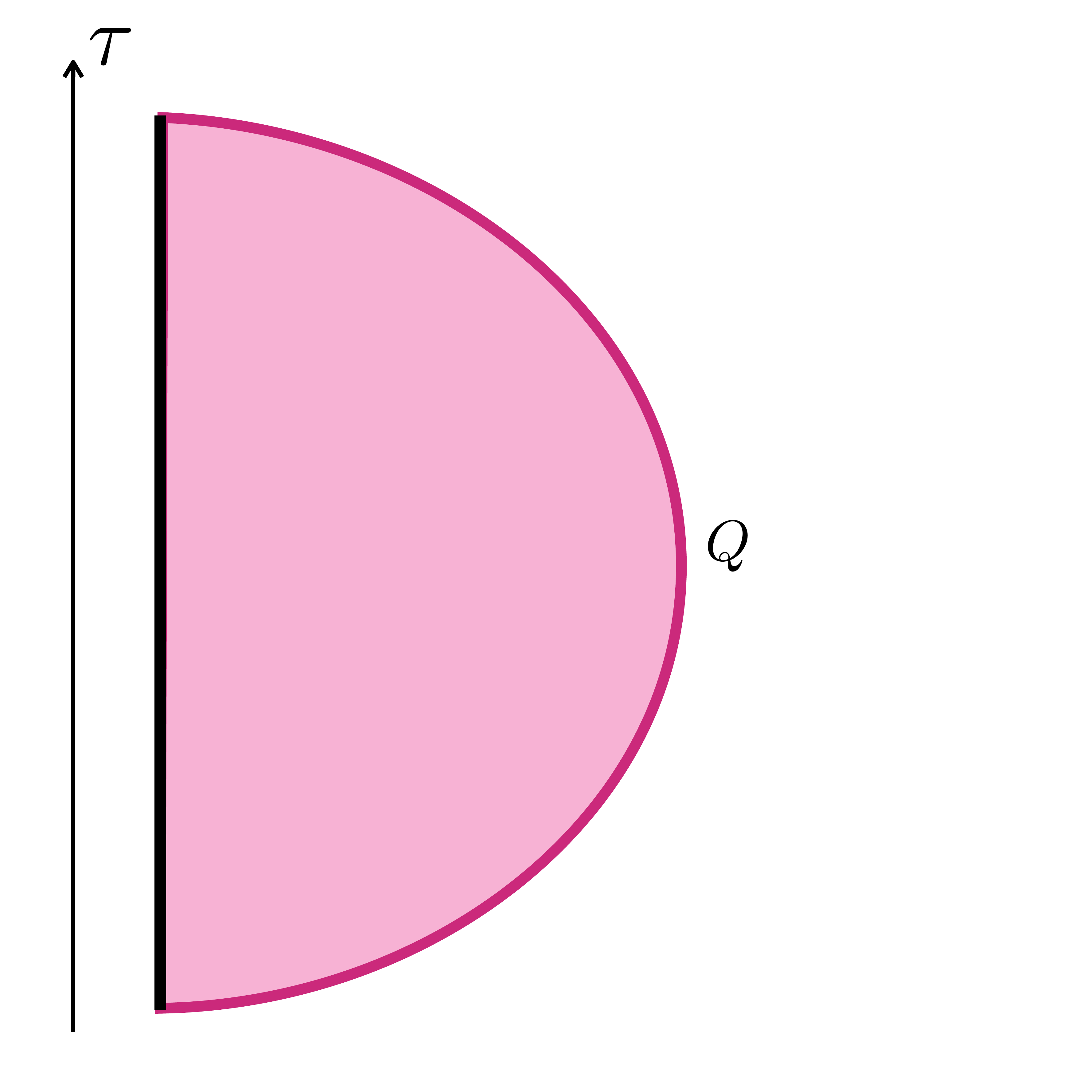}
     \includegraphics[width=.4\textwidth,page=1]{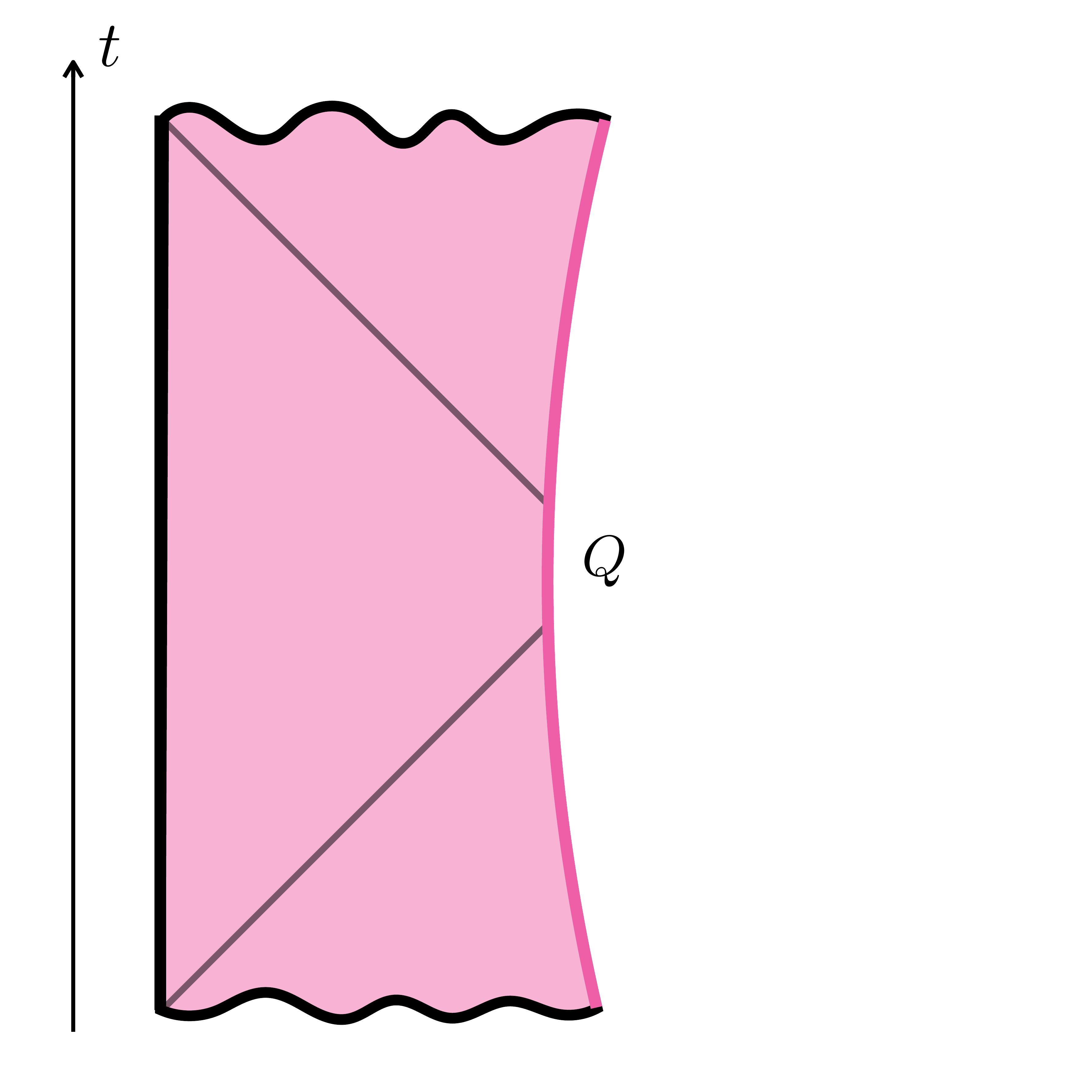}
    \caption{The $x={\rm const.}$ slice of the Euclidean setup and the Lorentzian counterpart of the small $\Delta$ phase. The Lorentzian turns out to be a one-sided BTZ black hole with temperature $1/(2\pi a)$, and an end-of-the-world brane falling behind the horizon.}
    \label{fig:smallphase}
\end{figure}

The small $\Delta \phi$ phase corresponds to the thermal AdS phase in section \ref{sec:phasetr}. The Euclidean gravity dual is given by a BTZ metric\footnote{Again, note that the time and space are opposite in section \ref{sec:phasetr} and this setion.} 
\begin{align}
& ds^2=\frac{h(z)d\tau^2}{z^2}+\frac{dx^2}{z^2}+\frac{dz^2}{h(z)z^2},\no
& h(z)=1-\frac{z^2}{a^2}.
\end{align}
with an end-of-the-world brane connecting the boundaries at the Euclidean past and future. Here, $a$ is determined by solving 
\begin{align}
    \frac{2\alpha}{a} = X_T\left(\frac{z_*}{a}\right),~~\Delta \phi = \Phi_T\left(\frac{z_*}{a}\right),
\end{align}
where the functions $X_T$ and $\Phi_T$ are given by figure \ref{fig:TAdSXP}. 
The effective temperature of the BTZ metric is given by 
\begin{align}
    T_{\rm eff} = \frac{1}{2\pi a }.
\end{align}
Accordingly, the corresponding Lorentzian setup is a one-sided BTZ black hole with temperature $1/(2\pi a)$ and an end-of-the-world brane falling behind the horizon. See figure \ref{fig:smallphase} for a sketch. A crucial feature of the current setup is that, while the Euclidean path integral is not time reflection symmetric on the BCFT side, the geometry of the gravity dual is. All the non-reflection symmetric elements in the gravity dual is the scalar field $\phi$. This feature allows us to get a real geometry corresponding to the Lorentzian setup by simply perform the analytical continuation. This is also true in other phases. 

The pseudo entropy can be computed from the HRT surface extending from $(z,x,t) = (0,0,t)$ and ends on the brane. To grasp an intuition, let us firstly consider the $\Delta \phi = 0$ case. This is nothing but the standard global quench setup \cite{Calabrese:2006rx,Calabrese:2016xau} or the Hartman-Maldacena setup \cite{Hartman:2013qma}. In this case, the computation can be analytically performed \cite{Hartman:2013qma} and the pseudo entropy reduces to the standard entanglement entropy after a global quench and turns out to be \cite{Hartman:2013qma}
\begin{align}
    S(\mathcal{\rho}_A) = \frac{c}{6}\frac{\pi}{2\alpha} t +\cdots = \frac{c}{6}  \frac{1}{a} t + \cdots. 
\end{align}
at late time. The linear behavior reflects the fact that the end-of-the-world brane goes deeper behind the horizon as the time evolves, and the coefficient reflects the temperature of the background geometry. 

Although we are not going to explicitly compute the time evolution of the pseudo entropy for more general $\Delta \phi \neq 0$, it is natural to expect the late time behaviors of the brane dynamics are similar for different $\Delta \phi$. Therefore, we conjecture that the leading order at late time is also 
\begin{align}
    S(\mathcal{T}_A) = \frac{c}{6}  \frac{1}{a} t + \cdots. 
\end{align}
even in this case. We leave testing this conjecture as a future problem. See figure for a sketch of what we have found in the small $\Delta \phi$ phase. 

\begin{figure}[H]
    \centering
     \includegraphics[width=.5\textwidth,page=1]{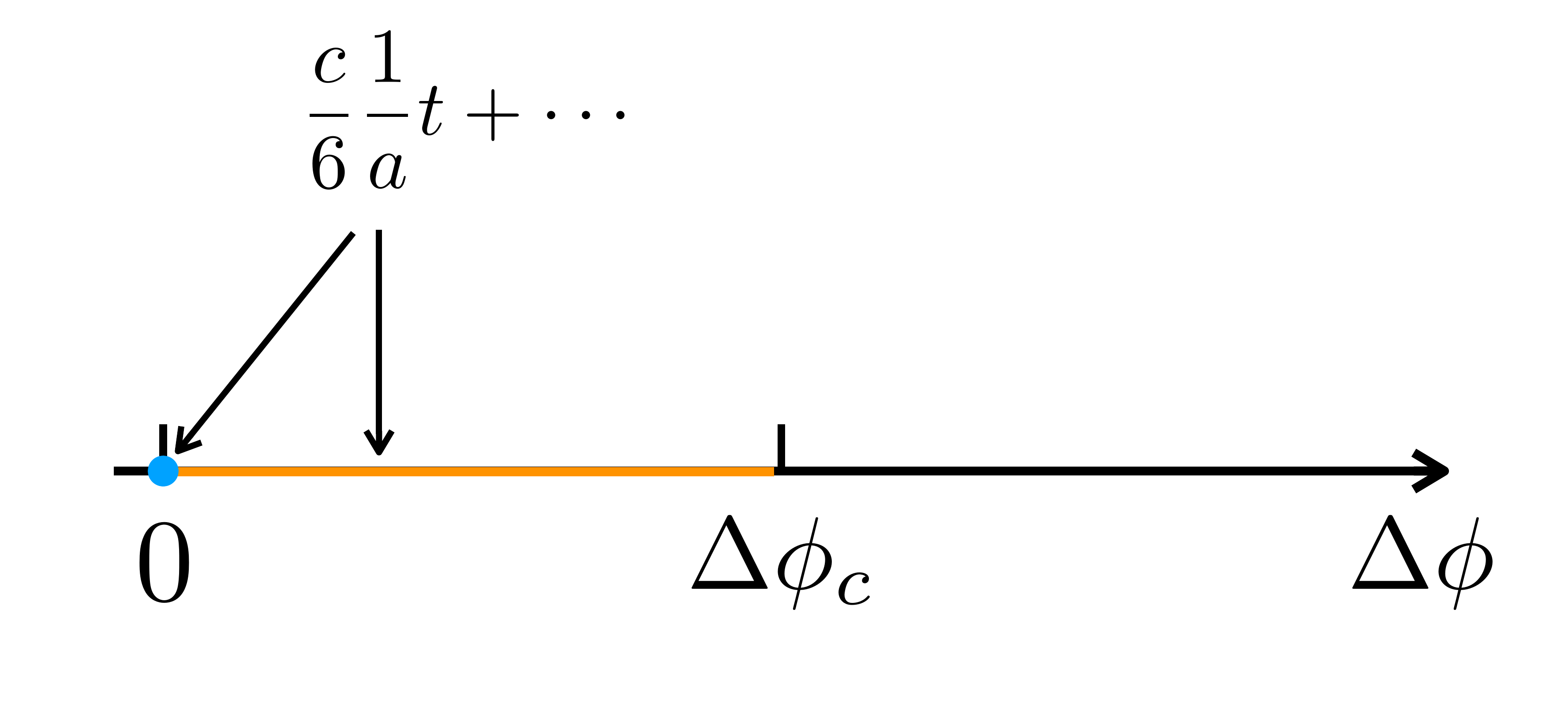}
    \caption{The leading time evolution of the pseudo entropy in the small $\Delta \phi$ phase. The blue point $\Delta \phi = 0$ is a case that we know the precise answer, while the behavior at the orange region remains to be a conjecture.}
    \label{fig:smallphasediagram}
\end{figure}

\subsubsection{Large \texorpdfstring{$\Delta \phi$}{Delta Phi} Phase}
\begin{figure}[H]
    \centering
     \includegraphics[width=.4\textwidth,page=1]{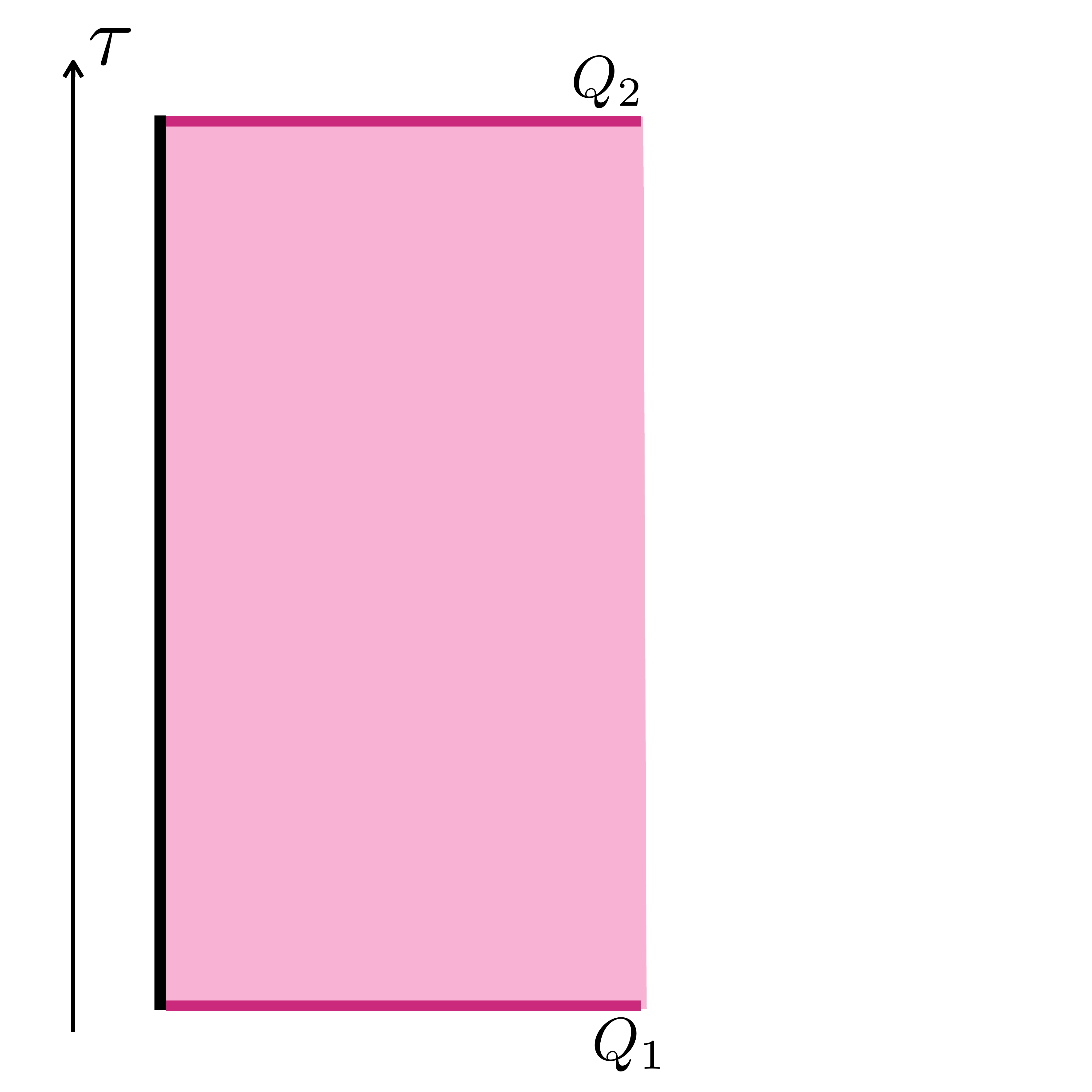}
     \includegraphics[width=.4\textwidth,page=1]{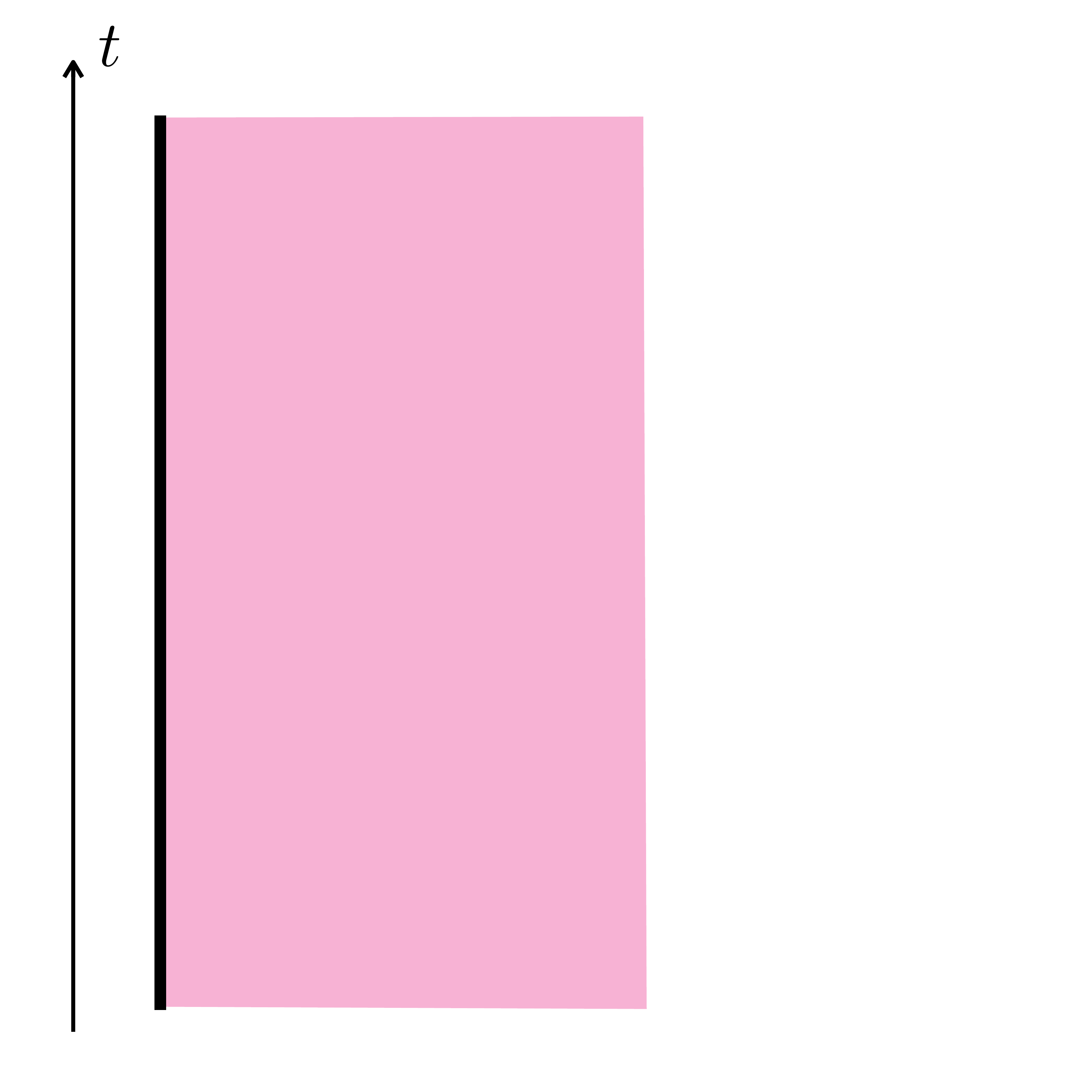}
    \caption{The $x={\rm const.}$ slice of the Euclidean setup and the Lorentzian counterpart of the large $\Delta$ phase. The Lorentzian turns out to be a static thermal AdS with temperature $1/(2\pi a)$, and there is non end-of-the-world branes involved in the Lorentzian configuration.}
    \label{fig:largephase}
\end{figure}

\begin{figure}[H]
    \centering
     \includegraphics[width=.5\textwidth,page=1]{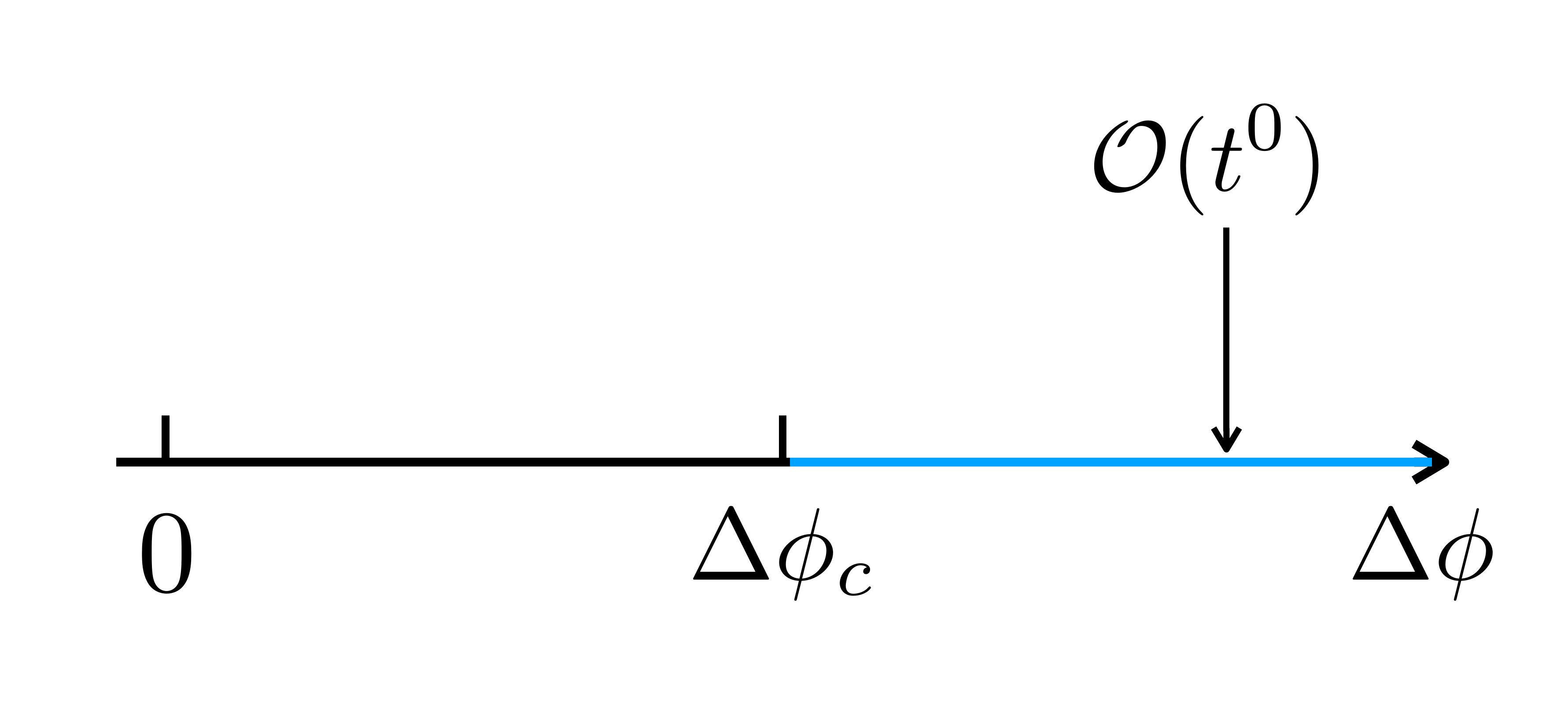}
    \caption{The leading time evolution of the pseudo entropy in the large $\Delta \phi$ phase.}
    \label{fig:largephasediagram}
\end{figure}

The large $\Delta \phi$ phase corresponds to the BTZ phase in section \ref{sec:phasetr}. The Euclidean gravity dual is given by a thermal AdS 
\begin{align}
    & ds^2=\frac{h(z)d\tau^2}{z^2}+\frac{dz^2}{h(z)z^2}+\frac{dx^2}{z^2},\no
    & h(z)=1-\frac{z^2}{a^2}.
\end{align}
with two end-of-the-world branes extending from the two boundaries at the Euclidean past and future. 
Here, $a$ is determined by solving 
\begin{align}
    \frac{2\alpha}{a} = X_B\left(\frac{z_*}{a}\right),~~\Delta \phi = \Phi_B\left(\frac{z_*}{a}\right), 
\end{align}
where the functions $X_B$ and $\Phi_B$ are given by figure \ref{fig:BTZXP}. The effective temperature of the thermal AdS metric is given by 
\begin{align}
    T_{\rm eff} = \frac{1}{2\pi a }.
\end{align}
It is straightforward to find that the corresponding Lorentzian setup is a static thermal AdS with temperature $1/(2\pi a)$ and no time evolution. Moreover, there is no end-of-the-world branes involved in the Lorentzian configuration. See figure \ref{fig:largephase} for a sketch. 
Accordingly, the holographic pseudo entropy has no time evolution either, i.e. 
\begin{align}
    S(\mathcal{T}_A) = \mathcal{O}(t^0). 
\end{align}
See figure \ref{fig:largephasediagram} for a sketch of what we have found in the large $\Delta \phi$ phase.

\subsubsection{The Phase Transition Point}

\begin{figure}[H]
    \centering
     \includegraphics[width=.4\textwidth,page=1]{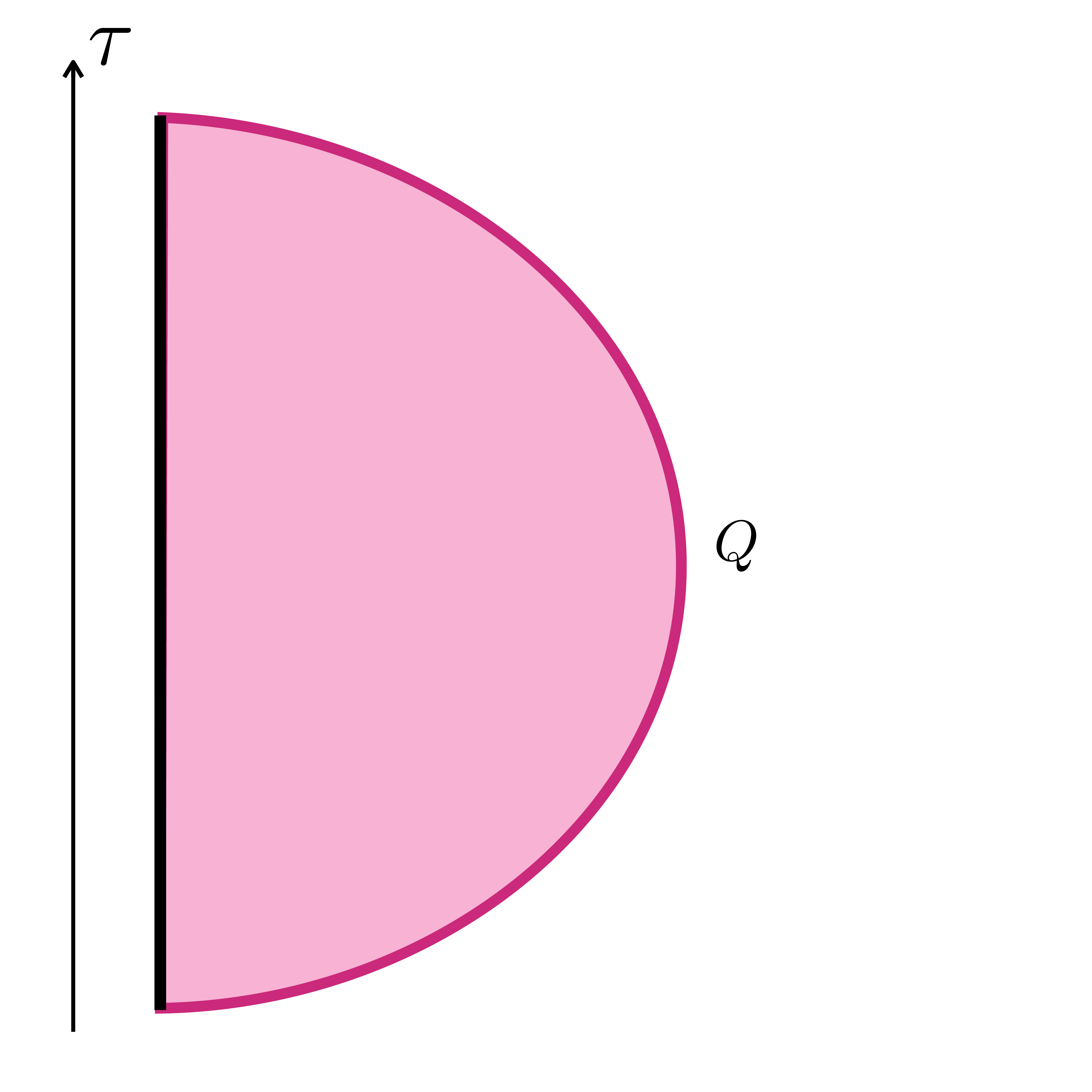}
     \includegraphics[width=.4\textwidth,page=1]{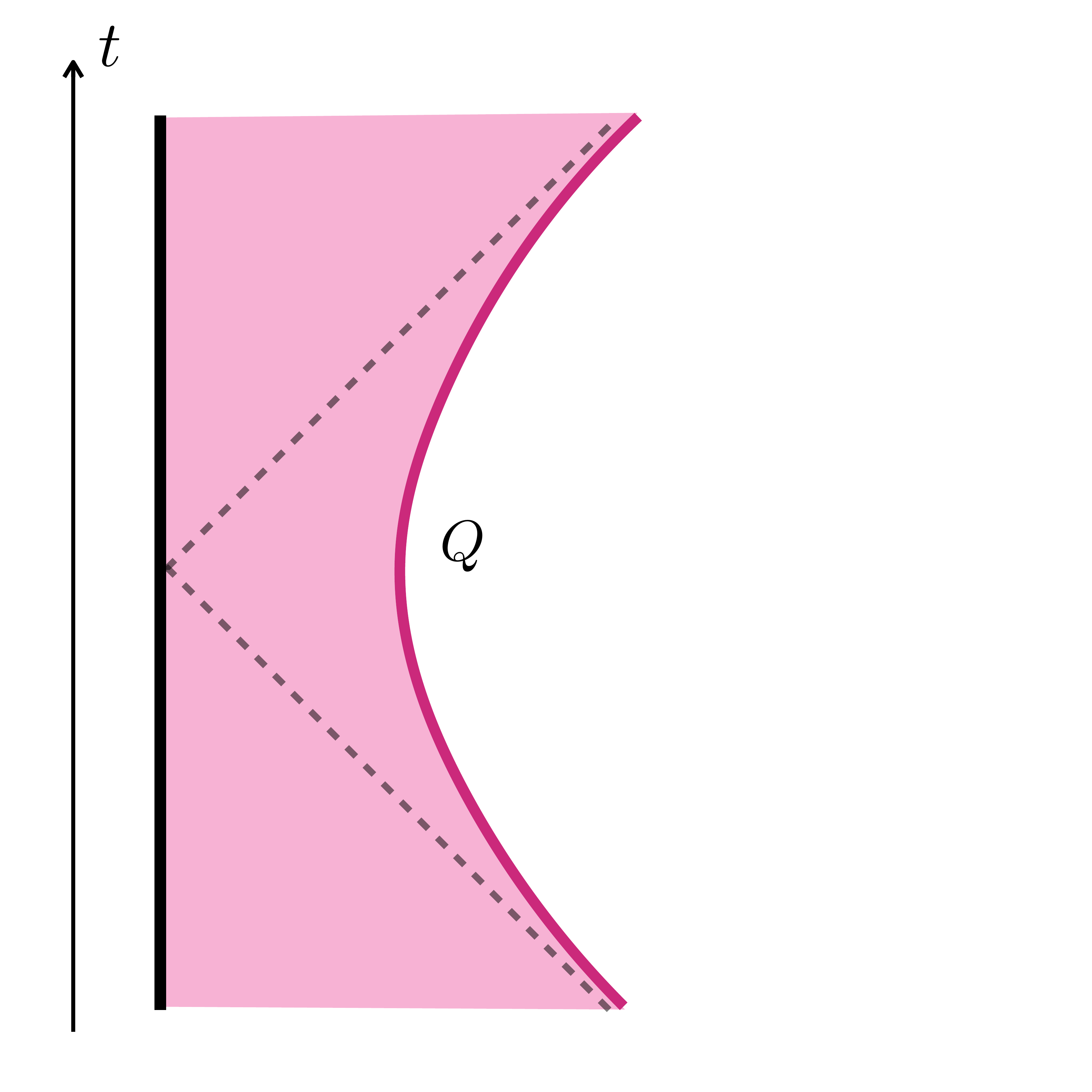}
    \caption{The $x={\rm const.}$ slice of the Euclidean setup and the Lorentzian counterpart at the phase transition point $\Delta \phi = \Delta \phi_c$. The Lorentzian configuration turns out to be a Poincar\'e AdS with an end-of-the-world brane accelerating and falling towards the Poincar\'e horizon. }
    \label{fig:PTpoint}
\end{figure}

As the last case, let us consider the phase transition point $\Delta \phi = \Delta \phi_c = 2K[-1] \simeq 2.6$. The corresponding Euclidean setup has been already studied in section \ref{sec:bdyJanus}. By performing the analytic continuation, one can find that the Lorentzian setup is a Poincare AdS with an end-of-the-world brane accelerating and falling towards the Poincare horizon. See figure \ref{fig:PTpoint} for a sketch. 

\begin{figure}[H]
    \centering
     \includegraphics[width=.5\textwidth,page=1]{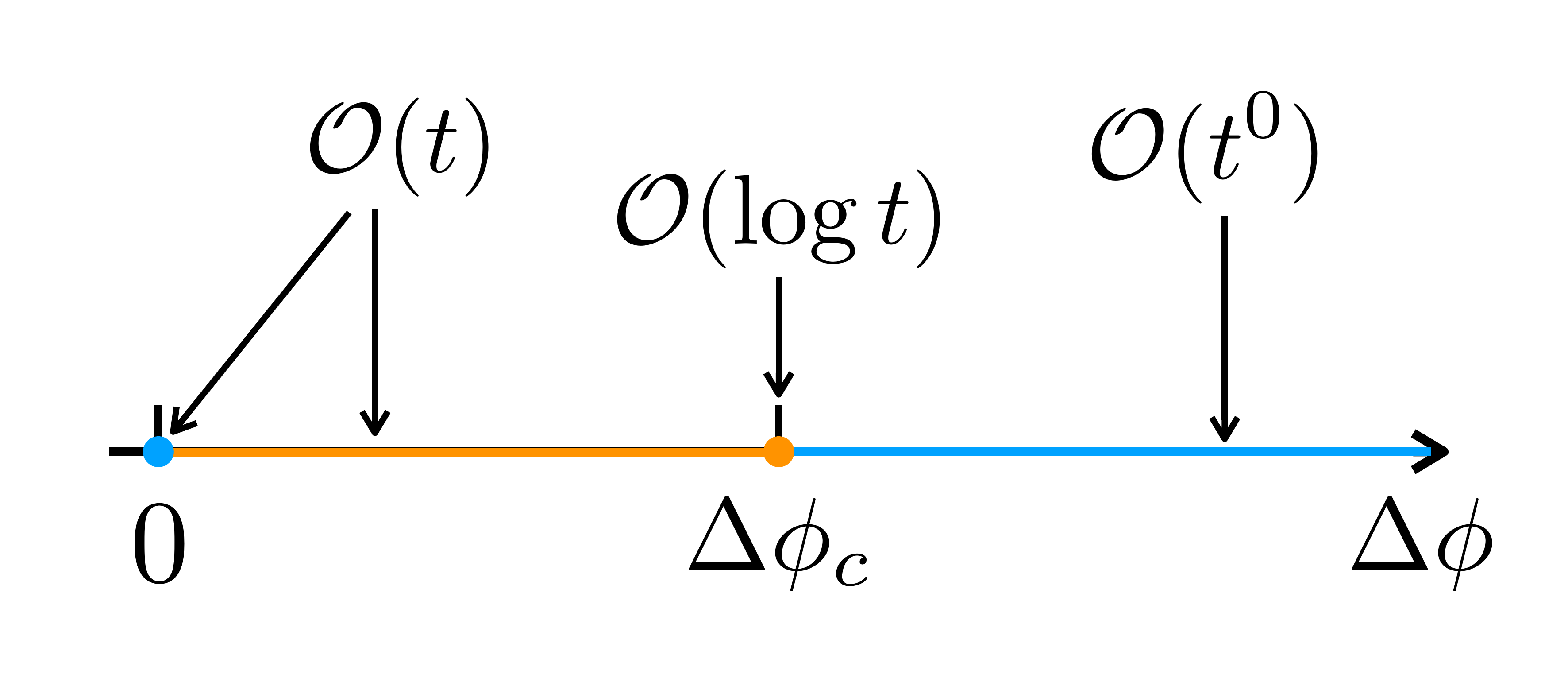}
    \caption{The leading time evolution of the pseudo entropy in the phase transition induced by $\Delta \phi$.}
    \label{fig:PTdiagram}
\end{figure}

If we approximates the holographic pseudo entropy with the minimal geodesic lying in the time slice $t = {\rm const. }$, one can find the time growth of the approximated pseudo entropy is 
\begin{align}
    \frac{c}{6} \log t + \cdots.
\end{align} 
In general, when there is an end-of-the-world brane approaching the speed of light and falling towards the Poincar\'e horizon, the coefficient can be modified by the back reaction\footnote{The HRT surface goes outside of the Poinca\'e patch when the end-of-the-world brane approaches the speed of light and falls towards the Poincar\'e horizon. The segment outside of the Poinca\'e patch can modify the coefficient of the $\log t$ term in the time evolution. See \cite{Shimaji:2018czt,Caputa:2019avh} for details.} of the end-of-the-world brane but the logarithmic behavior is robust. 
\begin{align}
    S(\mathcal{T}_A) = \mathcal{O}(\log t). 
\end{align}
It would be an interesting future direction to determine the coefficient in this case. See figure \ref{fig:PTdiagram} for a sketch of what we have found in this phase transition. 

\subsection{Possible connection to measurement-induced phase transition}
For our transition matrix in a holographic two dimensional CFT, we observed the time evolution of pseudo entropy changes from the linear growth $\mathcal{O}(t)$ (small $\Delta \phi$ phase with BTZ geometry) to the trivial one $\mathcal{O}(t^0)$ (large $\Delta \phi$ phase with thermal AdS geometry) when turning up the parameter $\Delta \phi$, and the behavior at the phase transition point with Poincar\'e geometry is $\mathcal{O}(\log t)$. This looks very similar to the behavior of entanglement entropy under the measurement-induced phase transition in 1D quantum many-body systems \cite{Skinner:2018tjl,Li:2018mcv,Li:2019zju}. Motivated by this, we would like to propose below a possible interpretation of our phase transition of transition matrix as a version of the measurement-induced phase transition. 

To begin with, let us first point out that the gravity side of the current setup effectively realize a classical open system. In the current setup, the BCFT side is a path integral which represents a transition matrix, which means that the BCFT side is not in a standard quantum state, but a combination of two quantum states. On the other hand, the gravity side in the Lorentzian signature can be determined by solving the Einstein equation with the initial condition at $t=0$ and turns out to be a standard state in classical gravity. The only exotic part on the gravity side is that the brane-localized scalar field $\phi$ takes imaginary values and turns out to be a ghost in the Lorentzian signature. This can be easily seen from the equation of motion of the scalar field (\ref{EOMsctd}) with the Wick rotation $\tau=it$ performed. As a result, the energy in the gravity sector leaks to the brane-localized scalar sector under the time evolution, and the current setup effectively realize an open system when viewing from the bulk gravity sector. The ``openness" of the system is characterized by the shift of the scalar field $\Delta \phi$. 

The behavior on the gravity side discussed above motivates us to conjecture that the field theory side effectively realize a {\it single} quantum state under a {\it non-unitary} time evolution which describes a quantum open system\footnote{Note that the canonical explanation on the field theory side is a combination of {\it two} quantum states under {\it unitary} time evolution.}. In this explanation, the quantity computed by the HRT formula turns out to be the entanglement entropy of the single quantum state under a non-unitary time evolution, but not the pseudo entropy of the transition matrix under a unitary time evolution. Again, the ``openness" of the quantum system is characterized by the shift of the scalar field $\Delta \phi$. When $\Delta \phi$ gets larger, the system gets more open to the outside environment which may be interpreted as measurements, and the entanglement growth gets reduced and experiences the phase transition. Indeed, this is very similar to the measurement-induced phase transition.  

To summary, discussions above suggest that our
time-dependent solution with the EOW brane with a ghost field may be a gravity dual of
an open quantum system where a quantum system evolves in a non-unitary way, driven by
e.g. non-hermitian Hamiltonians or projection measurements. Note that if we perform projection measurements in the UV degrees of freedom, then the EOW brane should intersect with the AdS boundary as in the setups studied earlier \cite{Numasawa:2016emc,Antonini:2022sfm}.
Instead in our present construction, we expect non-unitary operations such as projections are acted on the IR degrees of freedom as the EOW brane is inserted in the interior. 
Such non-unitary EOW branes seem to imply a new holographic relation including open systems. We leave more detailed studies of this interpretation for an important future problem.

\section{Conclusions and Discussions}

In this paper, we studied the dynamics of a scalar field localized on the end-of-the-world (EOW) brane as a new extension of the minimal setup of AdS/BCFT. This brane-localized scalar field is dual to a boundary primary operator. As opposed to setups with bulk scalar fields, it is often possible to analytically solve the equation of motion at least partially. Owing to this advantage, we obtained various new solutions which are only possible in the presence of the scalar field. 

Firstly, we found planar solutions which are gravity duals of boundary RG flows. Next, by imposing rotational invariance, we obtained hemisphere, annulus and cone shaped solutions in the Poincar\'{e} AdS$_3$. Among them, annulus and cone shaped ones are impossible in the minimal AdS/BCFT model.  We also give interpretations from the viewpoint of boundary RG flows for the hemisphere and annulus shaped solutions. We leave the field theoretic interpretation of the cone shaped solutions as a future problem. At the intuitive level, they look like poking small holes in the space. Notice that these constructions of above solutions can be done in any dimension in a similar way, though we often presented solutions in AdS$_3/$BCFT$_2$ explicitly. 

We can also get intriguing solutions for BCFTs on strips. Even when the two boundaries of the strip have different values of the scalar potential, we can construct a connected EOW brane in our setup. We find that there is a phase transition point in the value of the difference of the scalar potential $\Delta\phi$.
If $\Delta\phi$ is smaller than the value at the phase transition point, the gravity dual is given by a connected EOW brane in thermal AdS$_3$. If it is larger, then the dual geometry is given by disconnected EOW branes in BTZ.
At the phase transition point and at the zero temperature, the dual is a connected EOW brane in the Poincar\'{e} AdS$_3$.
Our phase diagram Fig.\ref{fig:Compare} is analogous to the Hawking-Page transition with a chemical potential, which describes the physics of confinement/deconfinement phase transition.

We can also regard these solutions as transition matrices via a Wick rotation. In this interpretation, the phase transition we found describes the transition of dynamical evolutions of pseudo entropy from the area law $O(t^0)$ to volume law $O(t)$. At the phase transition point we may expect the logarithmic evolution $O(\log t)$ in two dimensions. It is intriguing to note that in the corresponding Lorentzian gravity dual, the EOW brane does not intersect with the AdS boundary, which can be regarded as a deformation of \cite{Hartman:2013qma}. We may try to interpret this time-dependent geometry as a gravity dual of the time evolution of a certain quantum state in the CFT. However, the new thing in our solution is that the scalar field on the EOW brane takes imaginary values, which is equivalent to a ghost-like scalar on the brane. The amount of such non-unitarity is controlled by the parameter $\Delta \phi$. The above results imply that as the non-unitary effect gets larger, the entanglement growth gets reduced. This is analogous to the measurement-induced phase transition \cite{Skinner:2018tjl,Li:2018mcv,Li:2019zju}. This suggests that our time-dependent solution with the EOW brane with a ghost field may be a gravity dual of an open quantum system where a quantum system evolves in a non-unitary way derived by e.g. non-hermitian Hamiltonians or projection measurements. This obviously deserves future investigations. 

Since our setup is a small extension of the minimal AdS/BCFT, it would be interesting to analyze setups of AdS/BCFT with richer contents of brane-localized fields such as multiple scalar fields and gauge fields. Such further extensions may allow us to study boundary problems in condensed matter physics. Another application may be to build new holographic cosmology models, by taking into account the brane-world dynamics of localized graviton on EOW branes.  Finally, since we relied on a bottom-up construction in this paper, it will be important to consider the string theory realizations of our models. 

\section*{Acknowledgements}

We are grateful to Brian Swingle and Tomonori Ugajin for useful discussions.
This work is supported by the Simons Foundation through the ``It from Qubit'' collaboration
and by MEXT KAKENHI Grant-in-Aid for Transformative Research Areas (A) through the ``Extreme Universe'' collaboration: Grant Number 21H05187.
This work is also supported by Inamori Research Institute for Science and
by JSPS Grant-in-Aid for Scientific Research (A) No.~21H04469.
ZW is supported by Grant-in-Aid for JSPS Fellows No. 20J23116.
TT would like to thank 2022 Simons Collaboration on It from Qubit Annual Meeting where a part of the present work was done.

\appendix
\section{The round-shaped solution of \texorpdfstring{$\dot{\p}^2\geq0$}{dot phi squared geq zero}}\label{ap:round-sol}

We solve $\dot{\p}^2\geq 0$ in \eqref{eq:polar ODE}:
\ba
\dot{\phi}^2=-\frac{\ddot{z}}{2z\s{1+\dot{z}^2}}+\frac{\dot{z}\s{1+\dot{z}^2}}{2zr}\geq 0
\ea
under some boundary conditions. In the considering region, we can simplify into
\ba
\dot{z}\left( 1+\dot{z}^{2}\right) \geq \ddot{z} r
\ea
since $z\geq 0$. Introducing $y(r)=\dot
{z}(r)$, we have
\ba\label{eq:A3}
y(1+y^2)-\dot{y}r\geq 0.
\ea

First, we can easily find soltions for the case of equal sign
\ba
y(r)=\frac{Ar}{\sqrt{1-A^{2} r^{2}}}.
\ea
As mentioned in Chapter \ref{ch:TypeI}, this solution corresponds to a part of an exact sphere. Now we make $A$ dependent of $r$. Then inequality \eqref{eq:A3} can be read as
\ba
-\frac{r^{2}\dot{A}(r)}{\left( 1-r^{2} A^{2}(r)\right)^{3/2}}\geq 0,
\ea
equivalently, $\dot{A}(r)\leq 0$.

Next, We restrict $A(r)$ by boundary conditions. In the Type I, for example, we impose $z(0)>0, \dot{z}(0)=0$ and $\ddot{z}(0)<0$ on the general solution
\ba
z(r)=\int_0^r\frac{A(r)r}{\sqrt{1-A^{2}(r) r^{2}}} + z_0,
\ea
and, therefore, solutions must satisfy $A(0)<0$ and $z_0>0$. The same applies to the Type III. On the other hand, in the Type II, we impose $z(0)=0$ and $\dot{z}(0)>0$. This means
\ba
0\leq\k:=\lim_{r\to0}A(r)r\leq1.
\ea
Therefore, we can conclude $A(r)$ must be displayed by
\ba
A(r)=\frac{\k}{r}-P(r)
\ea
where function $P(r)$ satisfies $\dot{P}(r)\geq-\k/r^2$ and $P(0)$ is finite.

\bibliographystyle{JHEP}
\bibliography{BraneMatter}


\end{document}